\documentclass[12pt]{article}
\usepackage{latexsym,graphicx}
\newcommand{\be}{\begin{equation}}
\newcommand{\ee}{\end{equation}}
\usepackage[left=2.5cm,top=2.5cm,right=2.5cm,bottom=1.5cm]{geometry}
\usepackage{float}
\usepackage{epstopdf}
\usepackage{amsmath}
\begin{document}
\begin{center}
\large{\bf{ A new shape function for wormholes in  $f(R)$  gravity and General Relativity}} \\
\vspace{10mm}
\normalsize{Ambuj Kumar Mishra$^1$ Umesh Kumar Sharma$^2$  } \\
\vspace{5mm}
\normalsize{$^{1,2}$Department of Mathematics, Institute of Applied Sciences and Humanities, GLA University,
Mathura-281 406, Uttar Pradesh, India}\\
\vspace{2mm}
$^1$E-mail: ambuj\_math@rediffmail.com   \\
\vspace{2mm}
$^2$E-mail:sharma.umesh@gla.ac.in  \\

\end{center}
\vspace{10mm}
\begin{abstract}
		In the present work, a new shape function is proposed inside modified $f(R)$ gravity and General Relativity in wormhole (WH) geometry. The shape function obeyed all the desired conditions of WH geometry. The equation of state (EoS) parameter, anisotropy parameter and the energy conditions are computed. The tangential null energy conditions and  the weak energy condition  are validated, as well as the radial energy conditions, which demonstrates the nonappearance of  exotic matter due to modified gravity allied with such a new proposal.
	
 	\end{abstract}

\smallskip
Keywords: $f(R)$ gravity, Wormholes, Shape function, Energy conditions \\

PACS number: 04.50.kd

\section{Introduction}
The geometrical structures as throat like, called wormholes, which interface two isolated and different areas of space-times and have no singularity or horizon. As a bridge model of a particle, the wormholes solutions  were firstly found in 1935 by Einstein and Rosen \cite{ref1}. The investigation of Lorentzian wormholes with regards to General Relativity (GR) was invigorated by the remarkable work of Morris and Thorne in 1988 \cite{ref2,ref3}. The flaring-out state of the throat is the basic component in wormhole physics, which in general relativity involves the infringement of the null energy condition (NEC). The violation of null energy condition represents the presence of exotic matter. A significant issue in wormhole solutions is the setting up of energy conditions. In such manner, different techniques have been suggested in the previous works that manage the scenarios of energy conditions with in wormhole background. Along this line, work on thin-shell wormholes has been done in \cite{ref4} and dynamical wormholes \cite{ref5,ref6,ref7}, where the supporting matter is focused on the wormholes throat.\\

Different autonomous high-precision observational information have acknowledged with alarming proof that the Universe is experiencing a period of accelerating expansion \cite{ref8}. Many proposals have been suggested in the published work to clarify this idea, varying from modified gravity theory to dark energy models. In modified gravity theory, one may suppose that the Einstein's hypothesis of general relativity (GR) does not work at large scales, and a more generalized action depicts the gravitational field. Hilbert derived the Einstein field equation of GR from the action principle, by embracing the scalar curvature as a linear function, $R$, in the Lagrangian gravitational density. Although, there are no presumptive basis to limit the gravitational Lagrangian to this structure, and naturally a few speculations have been proposed. In the context of modified theories of gravity, the presence of higher order terms in curvature would allow for building thin-shell wormholes supported by ordinary matter \cite{ref9}. Recently, many people try to build and study wormhole solutions within the framework of modified theories of gravity in various context \cite{ref10}.\\ 

  The modified $f(R)$ gravity, a popular gravity theory, which modifies the Einstein GR  gravitational action $R$ by an arbitrary function $f(R)$, where $R$ is a well- known Ricci scalar. The accelerated expansion of the Universe  may be  demonstrated by these models  \cite{ref11}. It is observed that, in  $f(R)$ gravity,  the energy conditions can be satisfy by the  higher-order curvature invariants \cite{ref12}. The wormhole solutions in modified $f(R)$ gravity with non constant Ricci scalar are obtained  in \cite{ref13}. Lately, the junction conditions applied to the development of pure double layer bubbles and thin-shell wormholes in $f(R)$ gravity \cite{ref14}. Using the non-commutative geometry, the new static wormholes are built in $f(R)$ gravity \cite{ref15} and the wormhole solutions cosmological evolution is explored in \cite{ref16}.  Wormhole solutions (Lorentzian), which are consistent with the cosmological evolution and solar system observations  were also investigated in viable modified $f(R)$ theories of gravity \cite{ref17}. These wormholes were observed in explicit $f(R)$ models which are considered to repeat sensible situations of cosmological behavior dependent on the accelerated expansion of the Universe. It was demonstrated that the weak energy conditions (WEC) holds at the region of the throat for specific scopes of the free parameters of the hypothesis. The  weak energy condition is fulfilled at a particular point in space with the simple choices of shape function for the asymptotically flat wormhole solutions. A hybrid and exponential shape function in modified gravity are introduced in \cite{ref18,ref19}. Recently, two different shape function for wormhole structure in $f(R)$ gravity were proposed in \cite{ref20,ref21}.\\
  
 Because of the absence of wormholes observations at present, in spite of the considerable number of endeavors and proposition \cite{ref22}, some material and geometrical  highlights of wormholes, for example, the equation of state (EoS) and the shape function, are as yet not exactly known. Especially, a few structures for the shape function $b(r)$ have been suggested and investigated so far, as one may check, for example, \cite{ref23}. As the shape function $b(r)$  have to obey several conditions and not arbitrary. We propose a new shape function for modified $f(R)$ gravity and GR in this work, which must be, normally as per the shape function conditions.\\ 

This paper is composed in the accompanying way: In Sec. II, the field equations and the wormhole metric is presented in reference to $f(R)$ gravity. In Sec. III, energy conditions are defined.  In Sec. IV, wormholes models are described.  In Sec. V, results and discussions are explored. Finally, conclusions are given in Sec. VI.

\section{Field Equations and Wormhole Metric} 
The theory of  $f(R)$ gravity starts from the generalization of the Einstein's theory of relativity. The action in $f(R)$ theory is defined as

\begin{equation}\label{eq1}
S = \frac{1}{2\kappa} \int  \sqrt{-g} \left[f(R)+ \mathcal{L}_m\right]  d^4 x,  
\end{equation}
where $\kappa= 2 \pi G$, $g$ is the metric determinant and $\mathcal{L}_m $ is matter Lagrangian density of the metric $g_{\upsilon\mu}$. For mathematical simplicity $\kappa$ is considered to be unity.

Varying Eq. (\ref{eq1}) with respect to  $g_{\upsilon\mu}$, the fourth-order non-linear field equations are obtained as
\begin{equation}\label{eq2}
F_{R} R_{\upsilon\mu} -\frac{1}{2} f g_{\upsilon\mu} - \bigtriangledown_{\upsilon} \bigtriangledown_{\mu} F_{R} + \Box F_{R} \,g_{\upsilon\mu} = T^m _{\upsilon\mu},   
\end{equation} 
where $\Box= \bigtriangledown_{\upsilon} \bigtriangledown^{\upsilon}$, $\bigtriangledown_{\upsilon}$ denotes, covariant derivative operator. $R$ and $R_{\upsilon\mu}$ denote curvature scalar and Ricci tensor respectively.  $T^m _{\upsilon\mu}$ denotes energy momentum tensor and $F_{R}=\frac{d f}{d R}$.

The effective field equation can be obtained by using Eq. (\ref{eq2}) as 

\begin{equation}\label{eq3}
G_{\upsilon \mu} = T^{eff} _{\upsilon\mu}=R_{\upsilon \mu} -\frac{1}{2}R g_{\upsilon\mu} = T^{g}_{\upsilon\mu} + \frac{8 \pi}{F_{R}} T^{m}_{\upsilon\mu},
\end{equation}
where $G_{\upsilon \mu}$ denotes the Einstein tensor, $T^{eff} _{\upsilon\mu}$ denotes effective energy momentum tensor and $T^{g}_{\upsilon\mu} = \frac{1}{F_{R}}\left[\bigtriangledown_{\upsilon}\bigtriangledown_{\mu}F_{R}-\left(\Box F_{R}+\frac{1}{2}RF_{R}-\frac{1}{2}f\right)g_{\upsilon\mu}\right]$. The energy momentum tensor for the matter source of the wormholes is $T_{\upsilon\mu}=\frac{\partial\mathcal{L}_m}{\partial g^{\upsilon\mu}}$, which can be written as

\begin{equation}\label{eq4}
T_{\upsilon\mu}=\left(\rho + p_{l}\right)u_{\upsilon}u_{\mu} -p_{l}g_{\upsilon\mu} + \left(p_{r}-p_{l}\right)X_{\upsilon}X_{\mu},
\end{equation}
such that $u^{\upsilon}u_{\upsilon}=-1$ and $X^{\upsilon}X_{\upsilon}=1$. Here $rho$, $p_r$ and $p_l$ denote energy density, radial pressure and tangential pressure respectively.

We explore static and spherically symmetric wormhole geometries in $f(R)$ in gravity, so we take metric describing wormhole geometry as
\begin{equation}\label{eq5}
ds^{2}= -e^{2\chi(r)}dt^{2}+e^{\psi(r)}dr^{2} +r^{2}\left(d\theta^{2}+ sin^{2}\theta^{2}\phi^{2}\right),
\end{equation}  
where $\chi(r)$ is arbitrary function of $r$, indicates gravitational redshift (called redshift function). $\psi(r) = \left(1-\frac{b(r)}{r}\right)^{-1}$ and $r_{0}<r<\infty$. The minimum value $r_{0}$ denotes throat radius of wormhole, $b(r)$ describes geometry of the wormhole and is known as the shape function. This shape function  is responsible for shape of the wormhole and it must have the following properties:$\, (i)\, b(r_{0})=r_{0}, (ii)\, b'(r_0)<1, (iii)\, \frac{b(r)-b'(r)r}{b(r)^{2}}>0, (iv)\, \frac{b(r)}{r}\rightarrow 0\,\, as\,\, r\rightarrow\infty, (v)\,\frac{b(r)}{r}<1\,\, for\,\, r>r_{0}$.  Moreover, to avoid event horizon (non-traversable condition) $\chi(r)$ should be every where finite. However, In this article, we consider constant redshift function for mathematical simplicity.

Here $R$ (Ricci scalar tensor) is  given by $R=\frac{2b'(r)}{r^{2}}$.
This yields Einstein's fields equations for the action Eq.(\ref{eq1}) in $f(R)$ gravity as:

 \begin{equation}\label{eq6}
 \rho = \frac{ b'(r) F_{R}}{r^{2}}- H,
 \end{equation}

\begin{equation}\label{eq67}
p_{r} = -\frac{ b(r) F_{R}}{r^{3}} - \left(1- \frac{b(r)}{r}\right)\left[F_{R}'' + \frac{F_{R}'\left(r b'
	(r)-b(r)\right)}{2r^{2}\left(1-\frac{b(r)}{r}\right)}\right]+ H,
\end{equation}

\begin{equation}\label{eq8}
p_{l} = \frac{ \left(b(r)-rb'(r)\right) F_{R}}{2r^{3}}-\frac{F_{R}'}{r}\left(1-\frac{b(r)}{r}\right)+ H,
\end{equation}
where $H = \frac{1}{4} \left(R F_{R} + \Box F_{R} + T\right)$  and prime upon the function indicates derivative with respect to radial coordinate $r$. 

The radial state parameter i.e EOS parameter in terms of radial pressure is defined as
\begin{equation}\label{eq9}
\omega = \frac{p_{r}}{\rho}.
\end{equation}

The anisotropy parameter $\Delta$ in terms of radial pressure $p_{r}$ and tangential pressure $p_{l}$ is defined as

\begin{equation}\label{eq10}
\Delta = p_{l}-p_{r}.
\end{equation}
 The negative and positive sign of $\delta$ indicate that geometry is attractive or repulsive, $\delta = 0$ indicates geometry has an isotropic pressure.

\section{Energy Conditions}
The energy conditions play important roll to analyze physical existence of wormhole geometry. The important energy conditions are NEC (null energy condition), WEC (weak energy condition), SEC (strong energy condition), and DEC (dominant energy condition). NEC for any null vector is defined as $T_{\upsilon\mu} k^{\upsilon}k^{\mu} \geq 0\Leftrightarrow NEC$. Alternatively, NEC can also be defined in terms of principal pressures as $\forall i, \rho + p_{i}\geq 0 \Leftrightarrow NEC$. WEC for a timelike vector is defined as $T_{\upsilon\mu} V^{\upsilon}V^{\mu} \geq 0\Leftrightarrow WEC$. This WEC in terms of principal pressures can also be defined as $\rho\geq 0$ and $\forall i, \rho + p_{i}\geq 0 \Leftrightarrow WEC $. For a time-like vector, SEC is defined as $(T_{\upsilon\mu}-\frac{T}{2}g_{\upsilon\mu}) V^{\upsilon}V^{\mu} \geq 0\Leftrightarrow SEC$, where $T$ denotes trace of energy momentum tensor. SEC in terms of principal pressures is defined as $T = -\rho + \sum_{j} p_{j}$ and  $\forall j, \rho + p_{j}\geq 0,  \rho +\sum_{j} p_{j}\geq 0  \Leftrightarrow SEC$. DEC for a time-like vector is defined as $DEC \Leftrightarrow T_{\upsilon\mu} V^{\upsilon}V^{\mu} \geq 0$ and $T_{\upsilon\mu}$ is not space like. 
Alternately, DEC is also defined in terms of principal pressures $\rho \geq 0$ and $\forall i, p_{i}\in \left[-\rho, +\rho\right] \Leftrightarrow DEC$.

In the present article, These energy conditions are investigated in terms of radial pressure $p_{r}$ and tangential pressure $p_{l}$ as follows:
\begin{itemize}
	\item $\rho + p_{r}\geq 0, \rho + p_{l}\geq 0$ (NEC),
	\item $\rho\geq0, \rho + p_{r}\geq 0, \rho + p_{l}\geq 0$ (WEC),
	\item $ \rho + p_{r}\geq 0, \rho + p_{l}\geq 0,\rho+p_{r}+2p_{l}\geq 0$ (SEC),
	\item $\rho \geq 0, \rho -|p_{r}| \geq 0, \rho -|p_{l}| \geq 0 $ (DEC).
\end{itemize}


  \begin{figure}
	\centering 
	(a)\includegraphics[width=12cm, height=8cm, angle=0]{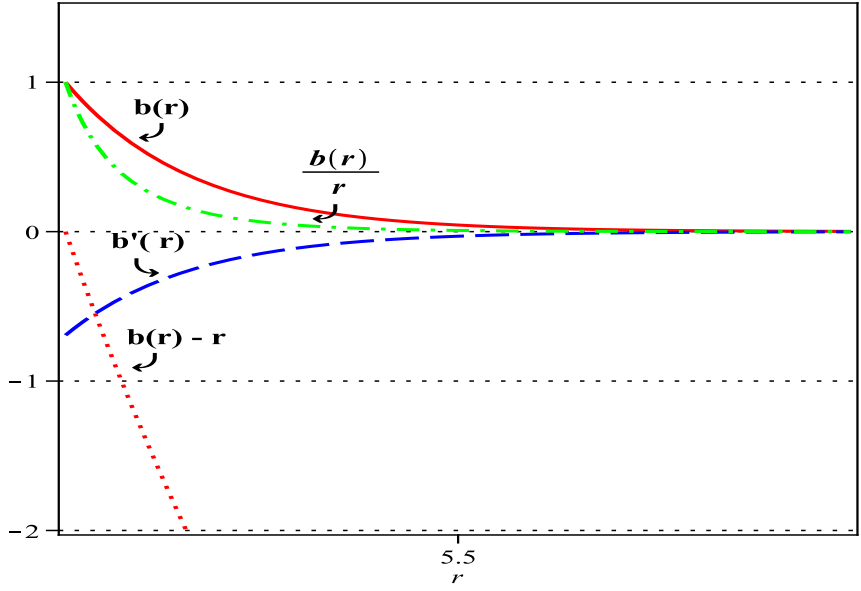}
	\caption {Nature of $b(r)$, throat condition $b(r)<r$, flaring out condition $b'(r)<1$ and asymptotically flatness $\lim\limits_{r\rightarrow \infty} \frac{b(r)}{r}=0$ for $r_0 = 1$.}
\end{figure} 


\section{Models of Wormholes}

In the present paper, $f(R)$ model is taken, which is defined by Nojiri and Odintsov \cite{Nojiri2003} as 
\begin{equation}\label{eq11}
f(R)= R + \alpha R^{m}-\beta R^{-n},
\end{equation}
where $\alpha, \beta, m $ and $n$ are taken as positive constants. The term $R^{m}$ dominates when $R^{-n}$ vanishes in the case of curvature increases and it represent evolution of inflation at early times. Similarly when curvature decrease the term   $R^{-n}$ dominates and $R^{m}$  vanishes that shows the production of present acceleration. In the framework of $FRW$ universe such forms of $f(R)$ are considered by Cao $ et\, al.$ \cite{Cao2017} for exploration of the probabilities of late time acceleration. Moradpour \cite{Moradpour2017} investigated traversable wormholes in the background of Lyra manifold and general relativity. He studies energy conditions by using hyperbolic shape function. Godani and Samanta \cite{Godani2019} investigated non violation of energy conditions in $f(R)$ gravity by using the same hyperbolic shape function. 

In this section, we will obtain wormhole models that for their matter content are constructed from different hypotheses in viable $f(R)$ gravity. Here, a new shape function is defined as 
 
\begin{equation}\label{eq12}
b(r)= {r_{{0}}}\left(\frac {{a}^{r}}{{a}^{r_{0}}}\right),
\end{equation}
where $a$ is base and  $0<a<1$. We can see directly in Fig.1 that the proposed shape function satisfies all the conditions i.e. basic requirements of shape functions, such as throat condition, flaring out condition and typical asymptotically flatness condition for WH geometry.

This new shape function  is taken into account to investigate the wormhole solutions. The energy density $\left(\rho\right)$, radial pressure $\left(p_{r}\right)$, tangential pressure  $\left(p_{l}\right)$ can be obtained by using field equations (\ref{eq6})- (\ref{eq8}) as

\begin{eqnarray}
\rho &=&\frac{1}{r^{2}} \left[ r_{{0}}{a}^{r-r_{{0}}}\ln  \left( a \right) +{2}^{-1+m}\alpha\, \left( {\frac {r_{{0}}{a}^{r-r_{{0}}}\ln  \left( a \right) }{{r}^{2}}} \right) ^{m}m{r}^{2}\right.\nonumber\\
&+&
\left.{2}^{-1-n}\beta\, \left( {\frac {r_{{0}}{a}^{r-r_{{0}}}\ln  \left( a \right) }{{r}^{2}}} \right) ^{-n}n{r}^{2}\right],
 \end{eqnarray} 
 
 \begin{eqnarray} 
p_{r}&=& -\frac{1}{4  {r}^{3}{r_{{0}}}\left(\ln \left( a \right)\right)}\left[ 4\,{r_{{0}}}^{2}{a}^{r-r_{{0}}}\ln  \left( a \right) +4\,\alpha\,{2}^{m} \left( {\frac {r_{{0}}{a}^{r-r_{{0}}}\ln  \left( a\right) }{{r}^{2}}} \right)^{m}m{a}^{r_{{0}}-r}{r}^{3}\right. \nonumber\\
&-&
\left.12\,\alpha\,{	m}^{2} \left( {\frac {r_{{0}}{a}^{r-r_{{0}}}\ln  \left( a \right) }{{r}^{2}}} \right)^{m}{2}^{m}{a}^{r_{{0}}-r}{r}^{3}+12\,\beta\,{n}^{2}\left( {\frac {r_{{0}}{a}^{r-r_{{0}}}\ln  \left( a \right) }{{r}^{2}}
} \right) ^{-n}{2}^{-n}{a}^{r_{{0}}-r}{r}^{3}\right. \nonumber\\
&+&
\left.4\,\beta\,{2}^{-n}\left( {\frac {r_{{0}}{a}^{r-r_{{0}}}\ln  \left( a \right) }{{r}^{2}}} \right)^{-n}n{a}^{r_{{0}}-r}{r}^{3}+4\,{2}^{1+m}\alpha\, \left( {
	\frac {r_{{0}}{a}^{r-r_{{0}}}\ln  \left( a \right) }{{r}^{2}}}
\right) ^{m}{m}^{3}{a}^{r_{{0}}-r}{r}^{3}\right. \nonumber\\
&+&
\left.4\,{2}^{1-n}\beta\, \left( 
{\frac {r_{{0}}{a}^{r-r_{{0}}}\ln  \left( a \right) }{{r}^{2}}}
\right) ^{-n}{n}^{3}{a}^{r_{{0}}-r}{r}^{3}-4\,{2}^{1-n}\beta\,
\left( {\frac {r_{{0}}{a}^{r-r_{{0}}}\ln  \left( a \right) }{{r}^{2}}
} \right) ^{-n}{n}^{3}{r}^{2}r_{{0}}\right. \nonumber\\
&-&
\left.4\,{2}^{1+m}\alpha\, \left({\frac {r_{{0}}{a}^{r-r_{{0}}}\ln  \left( a \right) }{{r}^{2}}}\right)^{m}{m}^{3}{r}^{2}r_{{0}}-13\,\alpha\,{m}^{2} \left( {\frac {	r_{{0}}{a}^{r-r_{{0}}}\ln  \left( a \right) }{{r}^{2}}} \right) ^{m}{2}^{m}{r}^{3}r_{{0}}\ln  \left( a \right) \right. \nonumber\\
&+&
\left.13\,\beta\,{n}^{2} \left({\frac {r_{{0}}{a}^{r-r_{{0}}}\ln  \left( a \right) }{{r}^{2}}}\right) ^{-n}{2}^{-n}{r}^{3}r_{{0}}\ln  \left( a \right) +4\,{2}^{1-n}\beta\, \left( {\frac{r_{{0}}{a}^{r-r_{{0}}}\ln  \left( a \right) }{{r}^{2}}} \right) ^{-n}{n}^{3}{r}^{3}r_{{0}}\ln  \left( a \right)\right. \nonumber\\
&+&
\left.4\,{2}^{1+m}\alpha\, \left( {\frac {r_{{0}}{a}^{r-r_{{0}}}\ln  \left( a\right) }{{r}^{2}}} \right) ^{m}{m}^{3}{r}^{3}r_{{0}}\ln  \left( a\right) -4\,{2}^{-1-n}\beta\, \left( {\frac {r_{{0}}{a}^{r-r_{{0}}}	\ln  \left( a \right) }{{r}^{2}}} \right) ^{-n}{n}^{3} \left( \ln \left( a \right)  \right) ^{2}{r}^{4}r_{{0}}\right. \nonumber\\
&-&
\left.3\,\beta\,{n}^{2}{r}^{4}\left( \ln  \left( a \right)  \right) ^{2} \left( {\frac {r_{{0}}{a}^	{r-r_{{0}}}\ln  \left( a \right) }{{r}^{2}}} \right) ^{-n}{2}^{-n}r_{{0}}-4\,{2}^{-1+m}\alpha\, \left( {\frac {r_{{0}}{a}^{r-r_{{0}}}\ln \left( a \right) }{{r}^{2}}} \right) ^{m}{m}^{3} \left( \ln  \left( a\right)  \right) ^{2}{r}^{4}r_{{0}}\right. \nonumber
\end{eqnarray}

\begin{eqnarray}
&+&
\left.3\,\alpha\,{m}^{2}{r}^{4} \left(\ln  \left( a \right)  \right) ^{2} \left( {\frac {r_{{0}}{a}^{r-r_{{0}}}\ln  \left( a \right) }{{r}^{2}}} \right) ^{m}{2}^{m}r_{{0}}-4\,{2}^{1+m}\alpha\, \left( {\frac {r_{{0}}{a}^{r-r_{{0}}}\ln  \left( a	\right) }{{r}^{2}}} \right) ^{m}m{a}^{r_{{0}}-r}{r}^{4}\ln  \left( a\right)\right. \nonumber\\
&-&
\left.4\,{2}^{1-n}\beta\, \left( {\frac {r_{{0}}{a}^{r-r_{{0}}}
\ln  \left( a \right) }{{r}^{2}}} \right) ^{-n}n{a}^{r_{{0}}-r}{r}^{4}
\ln  \left( a \right) -4\,{2}^{1-n}\beta\, \left( {\frac {r_{{0}}{a}^{
r-r_{{0}}}\ln  \left( a \right) }{{r}^{2}}} \right) ^{-n}{n}^{3}{a}^{r
_{{0}}-r}{r}^{4}\ln  \left( a \right)\right. \nonumber\\ 
&-&
\left.4\,{2}^{1+m}\alpha\, \left({\frac {r_{{0}}{a}^{r-r_{{0}}}\ln  \left(a\right)}{{r}^{2}}}\right)^{m}{m}^{3}{a}^{r_{{0}}-r}{r}^{4}\ln  \left( a \right) +16\,\alpha\,{m}^{2}{r}^{4} \left( {\frac {r_{{0}}{a}^{r-r_{{0}}}\ln	\left( a \right) }{{r}^{2}}} \right) ^{m}{2}^{m}{a}^{r_{{0}}-r}\ln \left( a \right)\right. \nonumber\\
 &-&
\left.16\,\beta\,{n}^{2}{r}^{4} \left( {\frac {r_{{0}}{a}
^{r-r_{{0}}}\ln  \left( a \right) }{{r}^{2}}} \right) ^{-n}{2}^{-n}{a}
^{r_{{0}}-r}\ln  \left( a \right) -4\,\alpha\,{m}^{2}{r}^{5} \left( 
\ln  \left( a \right)  \right) ^{2} \left( {\frac {r_{{0}}{a}^{r-r_{{0
}}}\ln  \left( a \right) }{{r}^{2}}} \right) ^{m}{2}^{m}{a}^{r_{{0}}-r
}\right. \nonumber\\
&+&
\left.4\,{2}^{-1+m}\alpha\, \left( {\frac {r_{{0}}{a}^{r-r_{{0}}}\ln 
\left( a \right) }{{r}^{2}}} \right) ^{m}m{a}^{r_{{0}}-r}{r}^{5}
\left( \ln  \left( a \right)  \right) ^{2}+4\,{2}^{-1-n}\beta\,
\left( {\frac {r_{{0}}{a}^{r-r_{{0}}}\ln  \left( a \right) }{{r}^{2}}
} \right) ^{-n}n{a}^{r_{{0}}-r}{r}^{5} \left( \ln  \left( a \right) 
\right) ^{2}\right. \nonumber\\
&+&
\left.4\,{2}^{-1+m}\alpha\, \left( {\frac {r_{{0}}{a}^{r-r_{{0
}}}\ln  \left( a \right) }{{r}^{2}}} \right) ^{m}{m}^{3}{a}^{r_{{0}}-r
} \left( \ln  \left( a \right)  \right) ^{2}{r}^{5}+4\,{2}^{-1-n}\beta
\, \left( {\frac {r_{{0}}{a}^{r-r_{{0}}}\ln  \left( a \right) }{{r}^{2
}}} \right) ^{-n}{n}^{3}{a}^{r_{{0}}-r} \left( \ln  \left( a \right) 
\right) ^{2}{r}^{5}\right. \nonumber\\
&+&
\left.4\,\beta\,{n}^{2}{r}^{5} \left( \ln  \left( a\right)  \right) ^{2} \left( {\frac {r_{{0}}{a}^{r-r_{{0}}}\ln \left( a \right) }{{r}^{2}}}\right)^{-n}{2}^{-n}{a}^{r_{{0}}-r}-20\,{2}^{-1-n}\beta\,{n}^{2} \left( {\frac {r_{{0}}{a}^{r-r_{{0}}}\ln	\left( a \right) }{{r}^{2}}} \right) ^{-n}{r}^{2}r_{{0}}\right. \nonumber\\
&+&
\left.20\,{2}^{-1+m}\alpha\,{m}^{2} \left( {\frac{r_{{0}}{a}^{r-r_{{0}}}\ln  \left( a\right) }{{r}^{2}}} \right) ^{m}{r}^{2}r_{{0}}+5\,{2}^{m}{r}^{3}r_{{0}}\alpha\, \left( {\frac {r_{{0}}{a}^{r-r_{{0}}}\ln  \left( a \right)}{{r}^{2}}} \right) ^{m}m\ln  \left( a \right) \right. \nonumber\\
&-&
\left.{2}^{m}\alpha\,\left( {\frac {r_{{0}}{a}^{r-r_{{0}}}\ln  \left( a \right) }{{r}^{2}}} \right) ^{m}m{r}^{4} \left( \ln  \left( a \right)  \right) ^{2}r_{{0}}+5\,{2}^{-n}{r}^{3}r_{{0}}\beta\, \left( {\frac {r_{{0}}{a}^{r-r_{{0	}}}\ln  \left( a \right) }{{r}^{2}}} \right) ^{-n}n\ln  \left( a\right) \right. \nonumber\\
&-&
\left.{2}^{-n}\beta\, \left( {\frac {r_{{0}}{a}^{r-r_{{0}}}\ln\left( a \right) }{{r}^{2}}} \right) ^{-n}n{r}^{4} \left( \ln \left( a \right)  \right) ^{2}r_{{0}} \right],
  \end{eqnarray}

 \begin{eqnarray}
p_{l}&=& -\frac{1}{4 {r}^{3}{r_{{0}}}	\left( \ln  \left( a \right)  \right)} \left[ 4\,{r}^{4}{2}^{-1+m}\alpha\,{m}^{2} \left( {\frac {r_{{0}}{a}^{r-r_{{0}}}\ln  \left( a \right) }{{r}^{2}}} \right) ^{m}{a}^{r_	{{0}}-r}\ln  \left( a \right)\right. \nonumber\\
&-&
\left.4\,{r}^{3}r_{{0}}{2}^{-1+m}\alpha\,{m}^{2} \left( {\frac {r_{{0}}{a}^{r-r_{{0}}}\ln  \left( a \right) }{{r}^{2}}} \right) ^{m}\ln  \left( a \right) -4\,\alpha\,{m}^{2} \left( {\frac {r_{{0}}{a}^{r-r_{{0}}}\ln  \left( a \right) }{{r}^{2}}}
\right) ^{m}{2}^{m}{a}^{r_{{0}}-r}{r}^{3}\right. \nonumber\\
&+&
\left.4\,\alpha\,{m}^{2} \left( {\frac {r_{{0}}{a}^{r-r_{{0}}}\ln  \left( a \right) }{{r}^{2}}}\right) ^{m}{2}^{m}{r}^{2}r_{{0}}-4\,{r}^{4}{2}^{-1+m}\alpha\,
\left( {\frac {r_{{0}}{a}^{r-r_{{0}}}\ln  \left( a \right) }{{r}^{2}}
} \right) ^{m}m{a}^{r_{{0}}-r}\ln  \left( a \right)\right. \nonumber\\ 
&+&
\left.3\,{2}^{m}{r}^{3}r_{{0}}\alpha\, \left( {\frac {r_{{0}}{a}^{r-r_{{0}}}\ln  \left( a\right) }{{r}^{2}}} \right) ^{m}m\ln  \left( a \right) +4\,\alpha\,{2}^{m} \left( {\frac {r_{{0}}{a}^{r-r_{{0}}}\ln  \left( a \right) }{{r}^{2}}} \right) ^{m}m{a}^{r_{{0}}-r}{r}^{3}\right. \nonumber\\
&-&
\left.5\,\alpha\,{2}^{m} \left( {\frac {r_{{0}}{a}^{r-r_{{0}}}\ln  \left(a\right)}{{r}^{2}}}\right)^{m}m{r}^{2}r_{{0}}-4\,{r}^{4}{2}^{-1-n}\beta\,{n}^{2}\left( {\frac {r_{{0}}{a}^{r-r_{{0}}}\ln  \left( a \right) }{{r}^{2}}} \right) ^{-n}{a}^{r_{{0}}-r}\ln  \left( a \right)\right. \nonumber\\ 
&+&
\left.4\,{r}^{3}r_{{0}}{2}^{-1-n}\beta\,{n}^{2} \left( {\frac {r_{{0}}{a}^{r-r_{{0}}}\ln 	\left( a \right) }{{r}^{2}}} \right) ^{-n}\ln  \left( a \right) +4\,\beta\,{n}^{2} \left( {\frac {r_{{0}}{a}^{r-r_{{0}}}\ln  \left( a\right) }{{r}^{2}}} \right) ^{-n}{2}^{-n}{a}^{r_{{0}}-r}{r}^{3}\right. \nonumber\\
&-&
\left.4\,\beta\,{n}^{2} \left( {\frac {r_{{0}}{a}^{r-r_{{0}}}\ln  \left(a\right)}{{r}^{2}}}\right)^{-n}{2}^{-n}{r}^{2}r_{{0}}-4\,{r}^{4}{2}^{-1-n}\beta\, \left( {\frac {r_{{0}}{a}^{r-r_{{0}}}\ln  \left( a
\right) }{{r}^{2}}} \right) ^{-n}n{a}^{r_{{0}}-r}\ln  \left( a
\right)\right. \nonumber\\ 
&+&
\left.3\,{2}^{-n}{r}^{3}r_{{0}}\beta\, \left( {\frac {r_{{0}}{a}^{
r-r_{{0}}}\ln  \left( a \right) }{{r}^{2}}} \right) ^{-n}n\ln  \left( 
a \right) +4\,\beta\,{2}^{-n} \left( {\frac {r_{{0}}{a}^{r-r_{{0}}}
\ln  \left( a \right) }{{r}^{2}}} \right) ^{-n}n{a}^{r_{{0}}-r}{r}^{3}
\right. \nonumber\\
&-&
\left.5\,\beta\,{2}^{-n} \left( {\frac {r_{{0}}{a}^{r-r_{{0}}}\ln  \left(a\right)}{{r}^{2}}}\right)^{-n}n{r}^{2}r_{{0}}-2\,{r_{{0}}}^{2}{a}^{r-r_{{0}}}\ln  \left( a \right) +2\,{r_{{0}}}^{2} \left( \ln  \left( a \right)  \right) ^{2}r{a}^{r-r_{{0}}} \right],
\end{eqnarray} 

\begin{eqnarray}
\rho + p_{r}&=& \frac{1}{4 {r}^{3}{r_{{0}}}	\left( \ln \left( a \right) \right)} \left[ 4\,{r_{{0}}}^{2} \left( \ln  \left( a \right)  \right)^{2}r{a}^{r-r_{{0}}}-4\,{r_{{0}}}^{2}{a}^{r-r_{{0}}}\ln  \left( a\right)\right.\nonumber\\ 
&-&
\left.4\,\alpha\,{2}^{m} \left( {\frac {r_{{0}}{a}^{r-r_{{0}}}\ln 
\left( a \right) }{{r}^{2}}} \right) ^{m}m{a}^{r_{{0}}-r}{r}^{3}+12\,
\alpha\,{m}^{2} \left( {\frac {r_{{0}}{a}^{r-r_{{0}}}\ln  \left( a
\right) }{{r}^{2}}} \right) ^{m}{2}^{m}{a}^{r_{{0}}-r}{r}^{3}\right.\nonumber\\
&-&
\left. 12\,\beta\,{n}^{2} \left( {\frac {r_{{0}}{a}^{r-r_{{0}}}\ln  \left(a\right)}{{r}^{2}}}\right)^{-n}{2}^{-n}{a}^{r_{{0}}-r}{r}^{3}-4\,\beta\,{2}^{-n} \left( {\frac {r_{{0}}{a}^{r-r_{{0}}}\ln  \left( a
\right) }{{r}^{2}}} \right) ^{-n}n{a}^{r_{{0}}-r}{r}^{3}\right.\nonumber\\
&-&
\left.4\,{2}^{1+m}\alpha\, \left( {\frac {r_{{0}}{a}^{r-r_{{0}}}\ln  \left(a\right)}{{r}^{2}}}\right)^{m}{m}^{3}{a}^{r_{{0}}-r}{r}^{3}-4\,{2}^{1-n}\beta\, \left( {\frac {r_{{0}}{a}^{r-r_{{0}}}\ln  \left( a \right) }{{r}^{2}}} \right) ^{-n}{n}^{3}{a}^{r_{{0}}-r}{r}^{3}\right.\nonumber\\
&+&
\left.4\,{2}^{1-n}\beta\,\left( {\frac {r_{{0}}{a}^{r-r_{{0}}}\ln  \left( a \right) }{{r}^{2}}} \right) ^{-n}{n}^{3}{r}^{2}r_{{0}}+4\,{2}^{1+m}\alpha\, \left( {
\frac {r_{{0}}{a}^{r-r_{{0}}}\ln  \left( a \right) }{{r}^{2}}}
\right) ^{m}{m}^{3}{r}^{2}r_{{0}}\right.\nonumber\\
&+&
\left.13\,\alpha\,{m}^{2} \left( {\frac {r_{{0}}{a}^{r-r_{{0}}}\ln
 \left( a \right) }{{r}^{2}}} \right) ^{m}{2}^{m}{r}^{3}r_{{0}}\ln  \left( a \right) -13\,\beta\,{n}^{2} \left( {\frac {r_{{0}}{a}^{r-r_{{0}}}\ln  \left( a \right) }{{r}^{2}}}
\right) ^{-n}{2}^{-n}{r}^{3}r_{{0}}\ln  \left( a \right)\right.\nonumber\\
&-&
\left.4\,{2}^{1-n}\beta\, \left( {\frac {r_{{0}}{a}^{r-r_{{0}}}\ln  \left( a \right) }{	{r}^{2}}} \right) ^{-n}{n}^{3}{r}^{3}r_{{0}}\ln  \left( a \right) -4\,{2}^{1+m}\alpha\, \left( {\frac {r_{{0}}{a}^{r-r_{{0}}}\ln  \left( a\right) }{{r}^{2}}} \right) ^{m}{m}^{3}{r}^{3}r_{{0}}\ln  \left( a\right)\right.\nonumber\\ 
&+&
\left.4\,{2}^{-1-n}\beta\, \left( {\frac {r_{{0}}{a}^{r-r_{{0}}}
\ln  \left( a \right) }{{r}^{2}}} \right) ^{-n}{n}^{3} \left( \ln 
\left( a \right)  \right) ^{2}{r}^{4}r_{{0}}+3\,\beta\,{n}^{2}{r}^{4}
\left( \ln  \left( a \right)  \right) ^{2} \left( {\frac {r_{{0}}{a}^
{r-r_{{0}}}\ln  \left( a \right) }{{r}^{2}}} \right) ^{-n}{2}^{-n}r_{{0}}\right.\nonumber\\
&+&
\left.4\,{2}^{-1+m}\alpha\, \left( {\frac {r_{{0}}{a}^{r-r_{{0}}}\ln 
\left( a \right) }{{r}^{2}}} \right) ^{m}{m}^{3} \left( \ln  \left( a
\right) \right) ^{2}{r}^{4}r_{{0}}-3\,\alpha\,{m}^{2}{r}^{4} \left( 
\ln  \left( a \right)  \right) ^{2} \left( {\frac {r_{{0}}{a}^{r-r_{{0
}}}\ln  \left( a \right) }{{r}^{2}}} \right) ^{m}{2}^{m}r_{{0}}\right.\nonumber\\
&+&
\left.4\,{2}^{1+m}\alpha\, \left( {\frac {r_{{0}}{a}^{r-r_{{0}}}\ln  \left( a\right) }{{r}^{2}}} \right) ^{m}m{a}^{r_{{0}}-r}{r}^{4}\ln  \left( a\right)+4\,{2}^{1-n}\beta\,\left({\frac{r_{{0}}{a}^{r-r_{{0}}}
\ln  \left( a \right) }{{r}^{2}}} \right) ^{-n}n{a}^{r_{{0}}-r}{r}^{4}
\ln  \left( a \right)\right.\nonumber\\ 
&+&
\left.4\,{2}^{1-n}\beta\, \left( {\frac {r_{{0}}{a}^{r-r_{{0}}}\ln  \left( a \right) }{{r}^{2}}} \right) ^{-n}{n}^{3}{a}^{r	_{{0}}-r}{r}^{4}\ln  \left( a \right) +4\,{2}^{1+m}\alpha\, \left( {
\frac {r_{{0}}{a}^{r-r_{{0}}}\ln  \left( a \right) }{{r}^{2}}}
\right) ^{m}{m}^{3}{a}^{r_{{0}}-r}{r}^{4}\ln  \left( a \right)\right.\nonumber\\ 
&-&
\left.16\,\alpha\,{m}^{2}{r}^{4} \left( {\frac {r_{{0}}{a}^{r-r_{{0}}}\ln 	\left( a \right) }{{r}^{2}}} \right) ^{m}{2}^{m}{a}^{r_{{0}}-r}\ln\left(a\right)+16\,\beta\,{n}^{2}{r}^{4} \left( {\frac {r_{{0}}{a}^{r-r_{{0}}}\ln  \left( a \right) }{{r}^{2}}} \right) ^{-n}{2}^{-n}{a}^{r_{{0}}-r}\ln  \left( a \right)\right.\nonumber\\ 
&+&
\left.4\,\alpha\,{m}^{2}{r}^{5} \left( \ln  \left( a \right)  \right) ^{2} \left( {\frac {r_{{0}}{a}^{r-r_{{0	}}}\ln  \left( a \right) }{{r}^{2}}} \right) ^{m}{2}^{m}{a}^{r_{{0}}-r}-4\,{2}^{-1+m}\alpha\, \left( {\frac {r_{{0}}{a}^{r-r_{{0}}}\ln\left( a \right) }{{r}^{2}}} \right) ^{m}\right.\nonumber\\
&\times&
\left.m{a}^{r_{{0}}-r}{r}^{5}
\left( \ln  \left( a \right)  \right) ^{2}-4\,{2}^{-1-n}\beta\,
\left( {\frac {r_{{0}}{a}^{r-r_{{0}}}\ln  \left( a \right) }{{r}^{2}}
} \right) ^{-n}n{a}^{r_{{0}}-r}{r}^{5} \left( \ln  \left( a \right) 
\right) ^{2}\right.\nonumber\\
&-&
\left.4\,{2}^{-1+m}\alpha\, \left( {\frac {r_{{0}}{a}^{r-r_{{0
}}}\ln  \left( a \right) }{{r}^{2}}} \right) ^{m}{m}^{3}{a}^{r_{{0}}-r
} \left( \ln  \left( a \right)  \right) ^{2}{r}^{5}\right.\nonumber\\
&-&
\left.4\,{2}^{-1-n}\beta
\, \left( {\frac {r_{{0}}{a}^{r-r_{{0}}}\ln  \left( a \right) }{{r}^{2
}}} \right) ^{-n}{n}^{3}{a}^{r_{{0}}-r} \left( \ln  \left( a \right) 
\right) ^{2}{r}^{5} \right.\nonumber\\
&-&
\left.4\,\beta\,{n}^{2}{r}^{5} \left( \ln  \left( a
\right)  \right) ^{2} \left( {\frac {r_{{0}}{a}^{r-r_{{0}}}\ln 
\left( a \right) }{{r}^{2}}} \right) ^{-n}{2}^{-n}{a}^{r_{{0}}-r}+20
\,{2}^{-1-n}\beta\,{n}^{2} \left( {\frac {r_{{0}}{a}^{r-r_{{0}}}\ln 
\left( a \right) }{{r}^{2}}} \right) ^{-n}{r}^{2}r_{{0}}\right.\nonumber\\
&-&
\left.20\,{2}^{-1+m}\alpha\,{m}^{2} \left( {\frac {r_{{0}}{a}^{r-r_{{0}}}\ln  \left( a\right) }{{r}^{2}}} \right) ^{m}{r}^{2}r_{{0}}-3\,{2}^{m}{r}^{3}r_{{0}}\alpha\, \left( {\frac {r_{{0}}{a}^{r-r_{{0}}}\ln  \left( a \right)}{{r}^{2}}} \right) ^{m}m\ln  \left( a \right) \right.\nonumber\\
&+&
\left.{2}^{m}\alpha\,
\left( {\frac {r_{{0}}{a}^{r-r_{{0}}}\ln  \left( a \right) }{{r}^{2}}
} \right) ^{m}m{r}^{4} \left( \ln  \left( a \right)  \right) ^{2}r_{{0
}}-3\,{2}^{-n}{r}^{3}r_{{0}}\beta\, \left( {\frac {r_{{0}}{a}^{r-r_{{0
}}}\ln  \left( a \right) }{{r}^{2}}} \right) ^{-n}n\ln  \left( a
\right)\right.\nonumber\\ 
&+&
\left.{2}^{-n}\beta\, \left( {\frac {r_{{0}}{a}^{r-r_{{0}}}\ln 
\left( a \right) }{{r}^{2}}} \right) ^{-n}n{r}^{4} \left( \ln 
\left( a \right)  \right) ^{2}r_{{0}} \right], 
\end{eqnarray}

\begin{eqnarray}
\rho + p_{l}&=&\frac{1}{4 {r}^{3}{r_{{0}}} \left( \ln 
\left( a \right)  \right)}\, \left[ 2\,{r_{{0}}}^{2} \left( \ln  \left(a\right)\right)^{2}r{a}^{r-r_{{0}}}+4\,{r}^{3}r_{{0}}{2}^{-1+m}\alpha\,{m}^{2} \left( {	\frac {r_{{0}}{a}^{r-r_{{0}}}\ln  \left( a \right) }{{r}^{2}}}\right) ^{m}\ln  \left( a \right)\right. \nonumber\\
 &-&
\left.{2}^{m}{r}^{3}r_{{0}}\alpha\,\left({\frac{r_{{0}}{a}^{r-r_{{0}}}\ln  \left( a \right) }{{r}^{2}}} \right) ^{m}m\ln  \left( a \right) -4\,{r}^{3}r_{{0}}{2}^{-1-n}\beta\,{n}^{2} \left( {\frac {r_{{0}}{a}^{r-r_{{0}}}\ln  \left( a \right) }{{r}^{2}}} \right) ^{-n}\ln  \left( a \right)\right. \nonumber\\ 
&-&
\left.{2}^{-n}{r}^{3}r_{{0}}\beta\,\left({\frac{r_{{0}}{a}^{r-r_{{0}}}\ln  \left( a \right) }{{r}^{2}}} \right) ^{-n}n\ln  \left( a \right) +2\,{r_{{0}}}^{2}{a}^{r-r_{{0}}}\ln  \left( a \right)\right. \nonumber\\
 &+&
\left.5\,\beta\,{2}^{-n} \left( {\frac {r_{{0}}{a}^{r-r_{{0}}}\ln  \left(a\right)}{{r}^{2}}}\right)^{-n}n{r}^{2}r_{{0}}+5\,\alpha\,{2}^{m} \left( {\frac {r_{{0}}{a}^{r-r_{{0}}}\ln \left( a \right) }{{r}^{2}}} \right) ^{m}m{r}^{2}r_{{0}}\right. \nonumber\\
&-&
\left.4\,\alpha\,{2}^{m} \left( {\frac {r_{{0}}{a}^{r-r_{{0}}}\ln  \left(a\right)}{{r}^{2}}}\right)^{m}m{a}^{r_{{0}}-r}{r}^{3}+4\,\alpha\,{m}^{2} \left({\frac {r_{{0}}{a}^{r-r_{{0}}}\ln  \left( a \right) }{{r}^{2}}}\right)^{m}{2}^{m}{a}^{r_{{0}}-r}{r}^{3}\right. \nonumber\\
&-&
\left.4\,\beta\,{n}^{2} \left( {\frac {r_{{0}}{a}^{r-r_{{0}}}\ln  \left(a\right)}{{r}^{2}}}\right)^{-n}{2}^{-n}{a}^{r_{{0}}-r}{r}^{3}-4\,\beta\,{2}^{-n}\left( {\frac {r_{{0}}{a}^{r-r_{{0}}}\ln  \left( a \right) }{{r}^{2}}} \right) ^{-n}n{a}^{r_{{0}}-r}{r}^{3}\right. \nonumber\\
&+&
\left.4\,\beta\,{n}^{2} \left( {\frac {r_{{0}}{a}^{r-r_{{0}}}\ln  \left( a \right) }{{r}^{2}}}\right) ^{-n}{2}^{-n}{r}^{2}r_{{0}}-4\,\alpha\,{m}^{2} \left( {\frac 
{r_{{0}}{a}^{r-r_{{0}}}\ln  \left( a \right) }{{r}^{2}}} \right) ^{m}{
2}^{m}{r}^{2}r_{{0}}\right. \nonumber\\
&-&
\left.4\,{r}^{4}{2}^{-1+m}\alpha\,{m}^{2} \left({\frac {r_{{0}}{a}^{r-r_{{0}}}\ln  \left( a \right) }{{r}^{2}}}
\right) ^{m}{a}^{r_{{0}}-r}\ln  \left( a \right) \right. \nonumber\\
&+&
\left.4\,{r}^{4}{2}^{-1+m}\alpha\, \left( {\frac {r_{{0}}{a}^{r-r_{{0}}}\ln  \left( a \right) }{{r}^{2}}} \right) ^{m}m{a}^{r_{{0}}-r}\ln  \left( a \right) \right. \nonumber\\
&+&
\left.4\,{r}^{4}{2}^{-1-n}\beta\,{n}^{2} \left( {\frac {r_{{0}}{a}^{r-r_{{0}}}\ln\left( a \right) }{{r}^{2}}} \right) ^{-n}{a}^{r_{{0}}-r}\ln  \left( a \right)\right. \nonumber\\
&+&
\left.4\,{r}^{4}{2}^{-1-n}\beta\, \left( {\frac {r_{{0}}{a}^{r-r_
{{0}}}\ln  \left( a \right) }{{r}^{2}}} \right) ^{-n}n{a}^{r_{{0}}-r}
\ln  \left( a \right)  \right],
\end{eqnarray}

\begin{eqnarray}
\rho + p_{r}+ 2\,p_{l}&=&-\frac{1}{4 {r}{r_{{0}}} \left( \ln  \left( a
\right)  \right)}\, \left[ -20\,\alpha\,{m}^{2}r \left( {\frac {r_{{0}}{a}^{r-r_{{0}}}\ln  \left( a \right) }{{r}^{2}}} \right) ^{m}{2}^{m}{a}^{r_{{0}}-r}\right. \nonumber\\
&+&
\left.4\,{2}^{1-n}\beta\, \left( {\frac {r_{{0}}{a}^{r-r_{{0}}}\ln 
\left( a \right) }{{r}^{2}}} \right) ^{-n}{n}^{3}{a}^{r_{{0}}-r}r+20
\,\beta\,{n}^{2}r \left( {\frac {r_{{0}}{a}^{r-r_{{0}}}\ln  \left( a
\right) }{{r}^{2}}} \right) ^{-n}{2}^{-n}{a}^{r_{{0}}-r}\right. \nonumber\\
&+&
\left.4\,{2}^{1+m}\alpha\, \left( {\frac {r_{{0}}{a}^{r-r_{{0}}}\ln  \left(a\right)}{{r}^{2}}}\right)^{m}{m}^{3}{a}^{r_{{0}}-r}r+12\,\alpha\,{2}^{m}\left( {\frac {r_{{0}}{a}^{r-r_{{0}}}\ln  \left( a \right) }{{r}^{2}}} \right) ^{m}m{a}^{r_{{0}}-r}r\right. \nonumber\\
&+&
\left.12\,\beta\,{2}^{-n} \left( {\frac {r_{{0}}{a}^{r-r_{{0}}}\ln  \left(a\right)}{{r}^{2}}}\right)^{-n}n{a}^{r_{{0}}-r}r-
4\,{2}^{1-n}\beta\, \left( {\frac {r_{{0}}{a}^{r-r_{{0}}}\ln  \left( a \right) }{{r}^{2}}} \right) ^{-n}{n}^{3}r_{{0}}\right. \nonumber\\
&-&
\left. 4\,{2}^{1+m}\alpha\, \left( {\frac {r_{{0}}{a}^{r-r_{{0}}}\ln  \left( a\right) }{{r}^{2}}} \right)^{m}{m}^{3}r_{{0}}+36\,{2}^{-1+m}\alpha\,{m}^{2} \left( {\frac {r_{{0}}{a}^{r-r_{{0}}}\ln  \left( a \right) }{{r}^{2}}} \right) ^{m}r_{{0}}\right. \nonumber\\
&-&
\left.20\,{2}^{-1+m}\alpha\, \left( {\frac {r_{{0}}{a}^{r-r_{{0}}}\ln  \left( a \right) }{{r}^{2}}} \right) ^{m}mr_{{0}}-36\,{2}^{-1-n}\beta\,{n}^{2} \left( {\frac {r_{{0}}{a}^{r-r_{{0}}}	\ln  \left( a \right) }{{r}^{2}}} \right) ^{-n}r_{{0}}\right. \nonumber
\end{eqnarray}

\begin{eqnarray}
&-&
\left.20\,{2}^{-1-n}\beta\, \left( {\frac {r_{{0}}{a}^{r-r_{{0}}}\ln  \left( a \right) }{{r}^{2}}} \right) ^{-n}nr_{{0}}-{2}^{m}\alpha\, \left( {\frac {r_{{0}}{a}^{r-r_{{0}}}\ln  \left( a \right) }{{r}^{2}}} \right) ^{m}m{r}^{2}\left( \ln  \left( a \right)  \right) ^{2}r_{{0}}\right. \nonumber\\
&-&
\left.{2}^{-n}\beta\,
\left( {\frac {r_{{0}}{a}^{r-r_{{0}}}\ln  \left( a \right) }{{r}^{2}}
} \right) ^{-n}n{r}^{2} \left( \ln  \left( a \right)  \right) ^{2}r_{{0}}+20\,\alpha\,{m}^{2}{r}^{2}\ln  \left( a \right)  \left( {\frac {r_{{0}}{a}^{r-r_{{0}}}\ln  \left( a \right) }{{r}^{2}}} \right) ^{m}{2}^{m}{a}^{r_{{0}}-r}\right. \nonumber\\
&-&
\left.20\,\beta\,{n}^{2}{r}^{2}\ln  \left( a \right) 
\left( {\frac {r_{{0}}{a}^{r-r_{{0}}}\ln  \left( a \right) }{{r}^{2}}
} \right) ^{-n}{2}^{-n}{a}^{r_{{0}}-r}-17\,\alpha\,{m}^{2} \left( {
\frac {r_{{0}}{a}^{r-r_{{0}}}\ln  \left( a \right) }{{r}^{2}}}
\right) ^{m}{2}^{m}rr_{{0}}\ln  \left( a \right)\right. \nonumber\\
& +&
\left.17\,\beta\,{n}^{2}\left( {\frac {r_{{0}}{a}^{r-r_{{0}}}\ln  \left( a \right) }{{r}^{2}}} \right) ^{-n}{2}^{-n}rr_{{0}}\ln  \left( a \right) +4\,{2}^{1-n}\beta\, \left( {\frac{r_{{0}}{a}^{r-r_{{0}}}\ln  \left( a \right) }{{
r}^{2}}} \right) ^{-n}{n}^{3}rr_{{0}}\ln  \left( a \right)\right. \nonumber\\ 
&+&
\left.4\,{2}^{1+m}\alpha\, \left( {\frac {r_{{0}}{a}^{r-r_{{0}}}\ln  \left( a \right) }{{r}^{2}}} \right) ^{m}{m}^{3}rr_{{0}}\ln  \left( a \right) -4\,{2}^{-1-n}\beta\, \left( {\frac {r_{{0}}{a}^{r-r_{{0}}}\ln  \left( a\right) }{{r}^{2}}} \right) ^{-n}\right. \nonumber\\
&\times&
\left.{n}^{3} \left( \ln  \left( a\right)  \right)^{2}{r}^{2}r_{{0}}-3\,\beta\,{n}^{2}{r}^{2} \left(\ln  \left( a \right)  \right) ^{2} \left( {\frac {r_{{0}}{a}^{r-r_{{0
}}}\ln  \left( a \right) }{{r}^{2}}} \right) ^{-n}{2}^{-n}r_{{0}}\right. \nonumber\\
&-&
\left.4\,{
2}^{-1+m}\alpha\, \left( {\frac {r_{{0}}{a}^{r-r_{{0}}}\ln  \left( a
\right) }{{r}^{2}}} \right) ^{m}{m}^{3} \left( \ln  \left( a \right) 
\right) ^{2}{r}^{2}r_{{0}}\right. \nonumber\\
&+&
\left.3\,\alpha\,{m}^{2}{r}^{2} \left( \ln \left( a \right)  \right) ^{2} \left( {\frac {r_{{0}}{a}^{r-r_{{0}}}\ln  \left( a \right) }{{r}^{2}}} \right) ^{m}{2}^{m}r_{{0}}-4\,{2}^{1
-n}\beta\, \left( {\frac {r_{{0}}{a}^{r-r_{{0}}}\ln  \left( a \right) 
	}{{r}^{2}}} \right) ^{-n}\right. \nonumber\\
&\times&
\left.{n}^{3}{a}^{r_{{0}}-r}{r}^{2}\ln  \left( a\right) -4\,{2}^{1+m}\alpha\, \left( {\frac {r_{{0}}{a}^{r-r_{{0}}}
\ln  \left( a \right) }{{r}^{2}}} \right) ^{m}{m}^{3}{a}^{r_{{0}}-r}{r
}^{2}\ln  \left( a \right)\right. \nonumber\\ 
&-&
\left.4\,\alpha\,{m}^{2}{r}^{3} \left( \ln\left( a \right)  \right) ^{2} \left( {\frac {r_{{0}}{a}^{r-r_{{0}}}\ln  \left( a \right) }{{r}^{2}}} \right) ^{m}{2}^{m}{a}^{r_{{0}}-r}+4\,{2}^{-1+m}\alpha\, \left( {\frac {r_{{0}}{a}^{r-r_{{0}}}\ln  \left(a \right) }{{r}^{2}}} \right) ^{m}\right. \nonumber\\
&\times&
\left.m{a}^{r_{{0}}-r}{r}^{3} \left( \ln \left( a \right)  \right) ^{2}+\,{2}^{-1-n}\beta\, \left( {\frac {r_	{{0}}{a}^{r-r_{{0}}}\ln  \left( a \right) }{{r}^{2}}} \right) ^{-n}n{a}^{r_{{0}}-r}{r}^{3} \left( \ln  \left( a \right)  \right) ^{2}\right. \nonumber\\
&+&
\left.4\,{2}^{-1+m}\alpha\, \left( {\frac {r_{{0}}{a}^{r-r_{{0}}}\ln  \left( a\right) }{{r}^{2}}} \right) ^{m}{m}^{3}{a}^{r_{{0}}-r} \left( \ln \left( a \right)  \right) ^{2}{r}^{3}+\,{2}^{-1-n}\beta\, \left( {\frac {r_{{0}}{a}^{r-r_{{0}}}\ln  \left( a \right) }{{r}^{2}}}\right) ^{-n}\right. \nonumber\\
&\times&
\left.{n}^{3}{a}^{r_{{0}}-r} \left( \ln  \left( a \right) \right) ^{2}{r}^{3}+ 4\,\beta\,{n}^{2}{r}^{3} \left( \ln  \left( a
\right)  \right) ^{2} \left( {\frac {r_{{0}}{a}^{r-r_{{0}}}\ln 
\left( a \right) }{{r}^{2}}} \right) ^{-n}{2}^{-n}{a}^{r_{{0}}-r}\right. \nonumber
\end{eqnarray}
\begin{eqnarray}
&-&
\left.12\,{2}^{m}\alpha\, \left( {\frac {r_{{0}}{a}^{r-r_{{0}}}\ln  \left( a\right) }{{r}^{2}}} \right) ^{m}m{a}^{r_{{0}}-r}{r}^{2}\ln  \left( a\right) -12\,{2}^{-n}\beta\, \left( {\frac {r_{{0}}{a}^{r-r_{{0}}}	\ln  \left( a \right) }{{r}^{2}}} \right) ^{-n}n{a}^{r_{{0}}-r}{r}^{2}\ln  \left( a \right) \right. \nonumber\\
&+&
\left.9\,{2}^{m}\alpha\, \left( {\frac {r_{{0}}{a}^{r-r_{{0}}}\ln  \left( a \right) }{{r}^{2}}} \right) ^{m}mrr_{{0}}\ln 
\left( a \right) +9\,{2}^{-n}\beta\, \left( {\frac {r_{{0}}{a}^{r-r_{
{0}}}\ln  \left( a \right) }{{r}^{2}}} \right) ^{-n}nrr_{{0}}\ln 
\left( a \right)  \right],
\end{eqnarray}

\begin{eqnarray}
\rho-\left|p_{r}\right|&=&\frac{1}{4 r^{2}}\, \left[4\,r_{{0}}{a}^{r-r_{{0}}}\ln  \left( a \right) +4\,{2}^{-
1+m}\alpha\, \left( {\frac {r_{{0}}{a}^{r-r_{{0}}}\ln  \left( a
\right) }{{r}^{2}}} \right) ^{m}m{r}^{2}\right. \nonumber\\
&+&
\left.4\,{2}^{-1-n}\beta\, \left( {\frac {r_{{0}}{a}^{r-r_{{0}}}\ln  \left( a \right) }{{r}^{2}}}\right) ^{-n}n{r}^{2}- \left|  \left( -{2}^{m}\alpha\, \left( {\frac 	{r_{{0}}{a}^{r-r_{{0}}}\ln  \left( a \right) }{{r}^{2}}} \right) ^{m}m
{r}^{4} \left( \ln  \left( a \right)  \right) ^{2}r_{{0}}\right.\right.\right. \nonumber\\
&-&
\left.\left.\left.{2}^{-n}\beta\, \left( {\frac {r_{{0}}{a}^{r-r_{{0}}}\ln  \left( a \right) }{{
r}^{2}}} \right) ^{-n}n{r}^{4} \left( \ln  \left( a \right)  \right) ^
{2}r_{{0}}-{2}^{1+m}\alpha\, \left( {\frac {r_{{0}}{a}^{r-r_{{0}}}\ln 
\left( a \right) }{{r}^{2}}} \right) ^{m}{m}^{3} \left( \ln  \left( a
\right)  \right) ^{2}{r}^{4}r_{{0}}\right.\right.\right. \nonumber\\
&-&
\left.\left.\left.{2}^{1-n}\beta\, \left( {\frac {r_{{0}}{a}^{r-r_{{0}}}\ln  \left( a \right) }{{r}^{2}}} \right) ^{-n}{n}^{3} \left( \ln  \left( a \right)  \right) ^{2}{r}^{4}r_{{0}}\right. \right.\right.\nonumber\\
&+&
\left.\left.\left.{2}^{1+m}\alpha\,m{a}^{r_{{0}}-r}{r}^{5} \left( \ln  \left( a \right) \right) ^{2} \left( {\frac {r_{{0}}{a}^{r-r_{{0}}}\ln  \left( a\right) }{{r}^{2}}} \right) ^{m}+{2}^{1+m}\alpha\,{m}^{3}{a}^{r_{{0}}-r} \left( \ln  \left( a \right)  \right) ^{2}{r}^{5} \left( {\frac {r	_{{0}}{a}^{r-r_{{0}}}\ln  \left( a \right) }{{r}^{2}}} \right) ^{m}\right. \right.\right.\nonumber\\
&+&
\left.\left.\left.{2}^{1-n}\beta\,n{a}^{r_{{0}}-r}{r}^{5} \left( \ln  \left( a \right)\right) ^{2} \left( {\frac {r_{{0}}{a}^{r-r_{{0}}}\ln  \left( a\right) }{{r}^{2}}} \right) ^{-n}+{2}^{1-n}\beta\,{n}^{3}{a}^{r_{{0}}-r} \left( \ln  \left( a \right)  \right) ^{2}{r}^{5} \left( {\frac {r
_{{0}}{a}^{r-r_{{0}}}\ln  \left( a \right) }{{r}^{2}}} \right) ^{-n}\right. \right.\right.\nonumber\\
&-&
\left.\left.\left.13\,\alpha\,{m}^{2} \left( {\frac {r_{{0}}{a}^{r-r_{{0}}}\ln  \left( a\right) }{{r}^{2}}} \right) ^{m}{2}^{m}{r}^{3}r_{{0}}\ln  \left( a\right) +13\,\beta\,{n}^{2} \left( {\frac {r_{{0}}{a}^{r-r_{{0}}}\ln 
\left( a \right) }{{r}^{2}}} \right) ^{-n}{2}^{-n}{r}^{3}r_{{0}}\ln 
\left( a \right) \right. \right.\right.\nonumber\\
&-&
\left.\left.\left.3\,\beta\,{n}^{2}{r}^{4} \left( \ln  \left( a
\right)  \right) ^{2} \left( {\frac {r_{{0}}{a}^{r-r_{{0}}}\ln 
\left( a \right) }{{r}^{2}}} \right) ^{-n}{2}^{-n}r_{{0}}+3\,\alpha\,
{m}^{2}{r}^{4} \left( \ln  \left( a \right)  \right) ^{2} \left( {
\frac {r_{{0}}{a}^{r-r_{{0}}}\ln  \left( a \right) }{{r}^{2}}}
\right) ^{m}{2}^{m}r_{{0}}\right. \right.\right.\nonumber\\
&+&
\left.\left.\left.16\,\alpha\,{m}^{2}{r}^{4} \left( {\frac {
r_{{0}}{a}^{r-r_{{0}}}\ln  \left( a \right) }{{r}^{2}}} \right) ^{m}{2
}^{m}{a}^{r_{{0}}-r}\ln  \left( a \right)-16\,\beta\,{n}^{2}{r}^{4}
\left( {\frac {r_{{0}}{a}^{r-r_{{0}}}\ln  \left( a \right) }{{r}^{2}}
} \right) ^{-n}{2}^{-n}{a}^{r_{{0}}-r}\ln  \left( a \right)\right. \right.\right.\nonumber\\
& -&
\left.\left.\left.4\,\alpha\,{m}^{2}{r}^{5} \left( \ln  \left( a \right)  \right) ^{2} \left( {\frac {r_{{0}}{a}^{r-r_{{0}}}\ln  \left( a \right) }{{r}^{2}}}\right)^{m}{2}^{m}{a}^{r_{{0}}-r}+4\,\beta\,{n}^{2}{r}^{5} \left(\ln  \left( a \right)  \right) ^{2} \left( {\frac {r_{{0}}{a}^{r-r_{{0}}}\ln  \left( a \right) }{{r}^{2}}} \right) ^{-n}{2}^{-n}{a}^{r_{{0}}-r}\right. \right.\right.\nonumber\\
&+&
\left.\left.\left.8\,{2}^{m}\alpha\, \left( {\frac{r_{{0}}{a}^{r-r_{{0}}}\ln \left( a \right) }{{r}^{2}}} \right) ^{m}{m}^{3}{r}^{3}r_{{0}}\ln\left( a \right) -8\,{2}^{m}\alpha\, \left( {\frac {r_{{0}}{a}^{r-r_{{0}}}\ln  \left( a \right) }{{r}^{2}}} \right) ^{m}m{a}^{r_{{0}}-r}{r}^{4}\ln  \left( a \right)\right. \right.\right.\nonumber\\
&-&
\left.\left.\left.8\,{2}^{m}\alpha\, \left( {\frac {r_{{0}}{a
}^{r-r_{{0}}}\ln  \left( a \right) }{{r}^{2}}} \right) ^{m}{m}^{3}{a}^
{r_{{0}}-r}{r}^{4}\ln  \left( a \right) +8\,{2}^{-n}\beta\, \left( {
	\frac {r_{{0}}{a}^{r-r_{{0}}}\ln  \left( a \right) }{{r}^{2}}}
\right) ^{-n}{n}^{3}{r}^{3}r_{{0}}\ln  \left( a \right)\right. \right.\right.\nonumber\\ 
&-&
\left.\left.\left.8\,{2}^{-n}\beta\, \left( {\frac {r_{{0}}{a}^{r-r_{{0}}}\ln  \left( a \right) }{{r}^{2}}} \right) ^{-n}n{a}^{r_{{0}}-r}{r}^{4}\ln  \left( a \right) -8
\,{2}^{-n}\beta\, \left( {\frac {r_{{0}}{a}^{r-r_{{0}}}\ln  \left( a
\right) }{{r}^{2}}} \right) ^{-n}{n}^{3}{a}^{r_{{0}}-r}{r}^{4}\ln 
\left( a \right) \right. \right.\right.\nonumber\\
&+&
\left.\left.\left.8\,{2}^{m}\alpha\, \left( {\frac {r_{{0}}{a}^{r-r_{
{0}}}\ln  \left( a \right) }{{r}^{2}}} \right) ^{m}{m}^{3}{a}^{r_{{0}}
-r}{r}^{3}-8\,{2}^{m}\alpha\, \left( {\frac {r_{{0}}{a}^{r-r_{{0}}}
\ln  \left( a \right) }{{r}^{2}}} \right) ^{m}{m}^{3}{r}^{2}r_{{0}}\right. \right.\right.\nonumber\\
&+&
\left.\left.\left.8\,{2}^{-n}\beta\, \left( {\frac {r_{{0}}{a}^{r-r_{{0}}}\ln  \left( a\right) }{{r}^{2}}} \right) ^{-n}{n}^{3}{a}^{r_{{0}}-r}{r}^{3}-8\,{2}^{-n}\beta\, \left( {\frac {r_{{0}}{a}^{r-r_{{0}}}\ln  \left( a\right) }{{r}^{2}}} \right) ^{-n}{n}^{3}{r}^{2}r_{{0}}\right. \right.\right.\nonumber\\
&+&
\left.\left.\left.5\,{2}^{-n}{r}^{3}r_{{0}}\beta\, \left( {\frac {r_{{0}}{a}^{r-r_{{0}}}\ln  \left( a\right) }{{r}^{2}}} \right) ^{-n}n\ln  \left( a \right) +5\,{2}^{m}{r}^{3}r_{{0}}\alpha\, \left( {\frac {r_{{0}}{a}^{r-r_{{0}}}\ln  \left(a \right) }{{r}^{2}}} \right) ^{m}m\ln  \left( a \right) \right. \right.\right.\nonumber\\
&+&
\left.\left.\left.4\,\alpha\,{2}^{m} \left( {\frac{r_{{0}}{a}^{r-r_{{0}}}\ln  \left( a \right) }{{r
}^{2}}} \right) ^{m}m{a}^{r_{{0}}-r}{r}^{3}-12\,\alpha\,{m}^{2}
\left( {\frac {r_{{0}}{a}^{r-r_{{0}}}\ln  \left( a \right) }{{r}^{2}}
} \right) ^{m}{2}^{m}{a}^{r_{{0}}-r}{r}^{3}\right. \right.\right.\nonumber\\
&+&
\left.\left.\left.12\,\beta\,{n}^{2} \left( {\frac {r_{{0}}{a}^{r-r_{{0}}}\ln  \left( a \right) }{{r}^{2}}}
\right) ^{-n}{2}^{-n}{a}^{r_{{0}}-r}{r}^{3}+4\,\beta\,{2}^{-n}
\left( {\frac {r_{{0}}{a}^{r-r_{{0}}}\ln  \left( a \right) }{{r}^{2}}
} \right) ^{-n}n{a}^{r_{{0}}-r}{r}^{3}\right. \right.\right.\nonumber\\
&-&
\left.\left.\left.10\,\beta\,{n}^{2} \left( {\frac {r_{{0}}{a}^{r-r_{{0}}}\ln  \left( a \right) }{{r}^{2}}}
\right) ^{-n}{2}^{-n}{r}^{2}r_{{0}}+10\,\alpha\,{m}^{2} \left( {
\frac {r_{{0}}{a}^{r-r_{{0}}}\ln  \left( a \right) }{{r}^{2}}}
\right) ^{m}{2}^{m}{r}^{2}r_{{0}}\right. \right.\right.\nonumber\\
&+&
\left.\left.\left.4\,{r_{{0}}}^{2}{a}^{r-r_{{0}}}\ln 
\left( a \right)  \right) {r}^{-3}{r_{{0}}}^{-1} \left( \ln  \left( a
\right)  \right) ^{-1} \right| {r}^{2} \right],
\end{eqnarray}
\begin{eqnarray}
\rho-\left|p_{l}\right|&=&\frac{1}{4}r^{2}\, \left[ 4\,r_{{0}}{a}^{r-r_{{0}}}\ln  \left( a \right) +4\,{2}^{-
1+m}\alpha\, \left( {\frac {r_{{0}}{a}^{r-r_{{0}}}\ln  \left( a
\right) }{{r}^{2}}} \right) ^{m}m{r}^{2}\right.\nonumber\\
&+&
\left.4\,{2}^{-1-n}\beta\, \left({\frac {r_{{0}}{a}^{r-r_{{0}}}\ln  \left( a \right) }{{r}^{2}}}\right) ^{-n}n{r}^{2}- \left|  \left( {2}^{1+m}\alpha\,{m}^{2}{r}^{4}{a}^{r_{{0}}-r}\ln  \left( a \right)  \left( {\frac {r_{{0}}{a}^{r-r_{{0}}}\ln  \left( a \right) }{{r}^{2}}} \right) ^{m}\right. \right.\right.\nonumber\\
&-&
\left.\left.\left.{2}^{1+m}\alpha\,{	m}^{2} \left( {\frac {r_{{0}}{a}^{r-r_{{0}}}\ln  \left( a \right) }{{r}^{2}}} \right) ^{m}{r}^{3}r_{{0}}\ln  \left( a \right) -4\,\alpha\,{m
}^{2} \left( {\frac {r_{{0}}{a}^{r-r_{{0}}}\ln  \left( a \right) }{{r}
^{2}}} \right) ^{m}{2}^{m}{a}^{r_{{0}}-r}{r}^{3}\right. \right.\right.\nonumber\\
&+&
\left.\left.\left.4\,\alpha\,{m}^{2}\left( {\frac {r_{{0}}{a}^{r-r_{{0}}}\ln  \left( a \right) }{{r}^{2}}
} \right) ^{m}{2}^{m}{r}^{2}r_{{0}}-{2}^{1+m}\alpha\, \left( {\frac {r
_{{0}}{a}^{r-r_{{0}}}\ln  \left( a \right) }{{r}^{2}}} \right) ^{m}m{a
}^{r_{{0}}-r}{r}^{4}\ln  \left( a \right) \right. \right.\right.\nonumber\\
&+&
\left.\left.\left.3\,{2}^{m}{r}^{3}r_{{0}}\alpha\, \left( {\frac {r_{{0}}{a}^{r-r_{{0}}}\ln  \left( a \right) }{{r}^{2}}} \right) ^{m}m\ln  \left( a \right) +4\,\alpha\,{2}^{m}\left( {\frac {r_{{0}}{a}^{r-r_{{0}}}\ln  \left( a \right) }{{r}^{2}}
} \right) ^{m}m{a}^{r_{{0}}-r}{r}^{3}\right. \right.\right.\nonumber\\
&-&
\left.\left.\left.5\,\alpha\,{2}^{m} \left( {
\frac {r_{{0}}{a}^{r-r_{{0}}}\ln  \left( a \right) }{{r}^{2}}}
\right) ^{m}m{r}^{2}r_{{0}}-{2}^{1-n}\beta\,{n}^{2}{r}^{4}{a}^{r_{{0}
}-r}\ln  \left( a \right)  \left( {\frac {r_{{0}}{a}^{r-r_{{0}}}\ln 
\left( a \right) }{{r}^{2}}} \right) ^{-n}\right. \right.\right.\nonumber\\
&+&
\left.\left.\left.{2}^{1-n}\beta\,{n}^{2}
\left( {\frac {r_{{0}}{a}^{r-r_{{0}}}\ln  \left( a \right) }{{r}^{2}}
} \right) ^{-n}{r}^{3}r_{{0}}\ln  \left( a \right) +4\,\beta\,{n}^{2}
\left( {\frac {r_{{0}}{a}^{r-r_{{0}}}\ln  \left( a \right) }{{r}^{2}}
} \right) ^{-n}{2}^{-n}{a}^{r_{{0}}-r}{r}^{3}\right. \right.\right.\nonumber\\
&-&
\left.\left.\left.4\,\beta\,{n}^{2}
\left( {\frac {r_{{0}}{a}^{r-r_{{0}}}\ln  \left( a \right) }{{r}^{2}}
} \right) ^{-n}{2}^{-n}{r}^{2}r_{{0}}-{2}^{1-n}\beta\, \left( {\frac {
r_{{0}}{a}^{r-r_{{0}}}\ln  \left( a \right) }{{r}^{2}}} \right) ^{-n}n
{a}^{r_{{0}}-r}{r}^{4}\ln  \left( a \right)\right. \right.\right.\nonumber\\ 
&+&
\left.\left.\left.3\,{2}^{-n}{r}^{3}r_{{0}}
\beta\, \left( {\frac {r_{{0}}{a}^{r-r_{{0}}}\ln  \left( a \right) }{{
r}^{2}}} \right) ^{-n}n\ln  \left( a \right) +4\,\beta\,{2}^{-n}
\left( {\frac {r_{{0}}{a}^{r-r_{{0}}}\ln  \left( a \right) }{{r}^{2}}
} \right) ^{-n}n{a}^{r_{{0}}-r}{r}^{3}\right. \right.\right.\nonumber\\
&-&
\left.\left.\left.5\,\beta\,{2}^{-n} \left( {
\frac {r_{{0}}{a}^{r-r_{{0}}}\ln  \left( a \right) }{{r}^{2}}}
\right) ^{-n}n{r}^{2}r_{{0}}-2\,{r_{{0}}}^{2}{a}^{r-r_{{0}}}\ln 
\left( a \right)\right. \right.\right.\nonumber\\ 
&+&
\left.\left.\left.2\,{r_{{0}}}^{2} \left( \ln  \left( a \right) 
\right) ^{2}r{a}^{r-r_{{0}}} \right) {r}^{-3}{r_{{0}}}^{-1} \left( 
\ln  \left( a \right)  \right) ^{-1} \right| {r}^{2} \right].
\end{eqnarray}


    
 \begin{figure}
 	\centering 
  	(a)\includegraphics[width=4cm, height=4cm, angle=0]{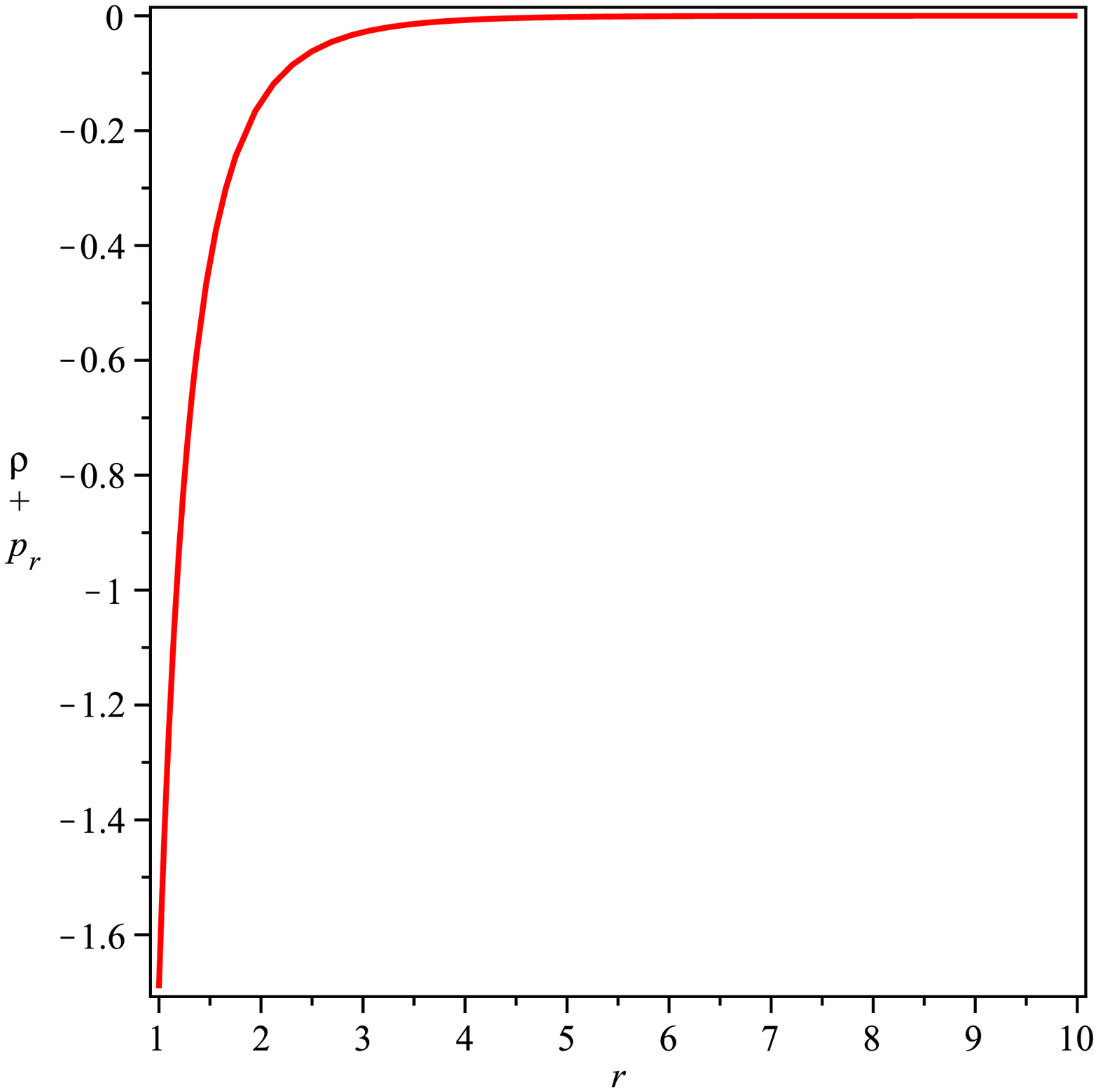}
  	(b)\includegraphics[width=4cm, height=4cm, angle=0]{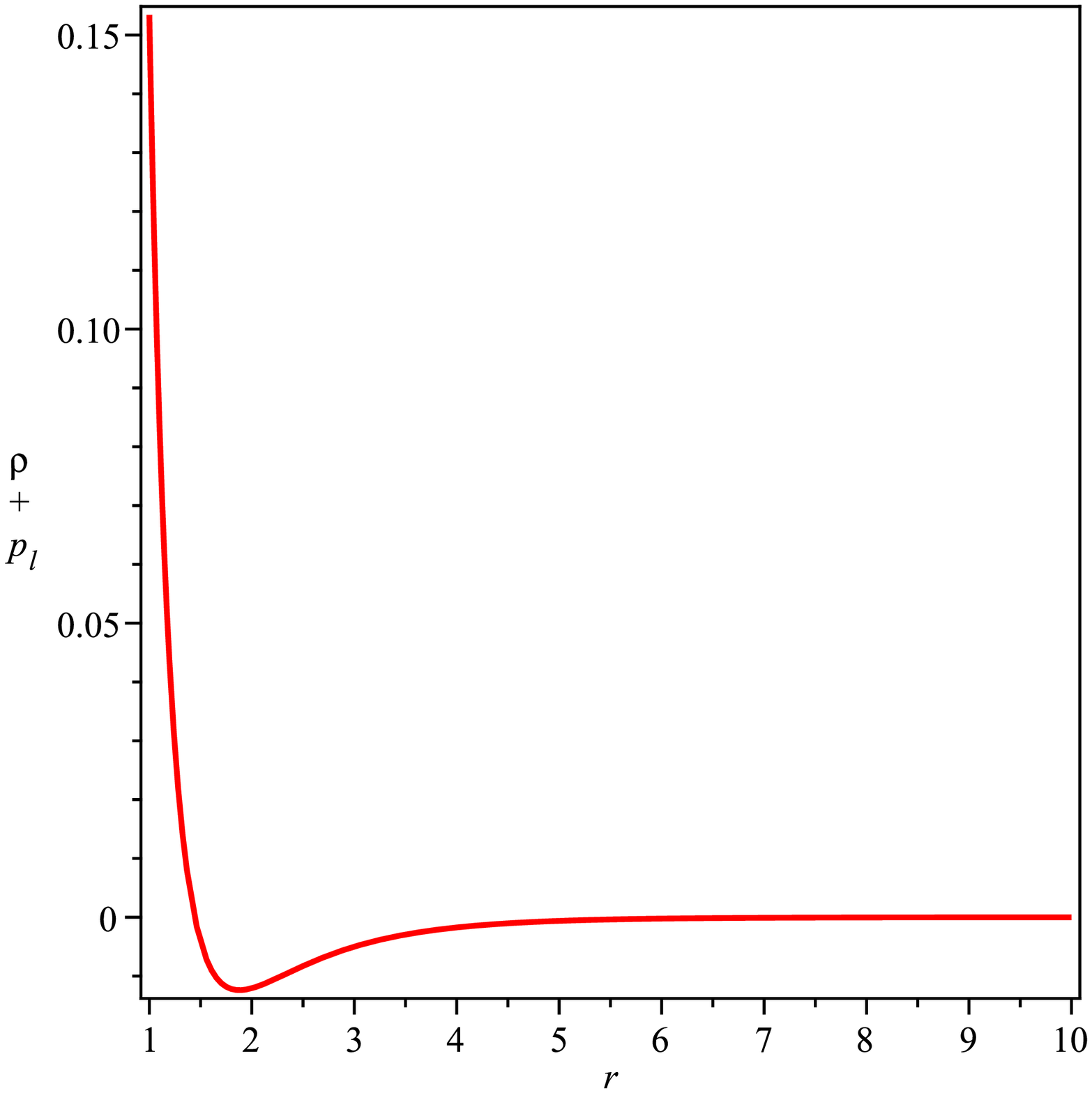}
  	(c)\includegraphics[width=4cm, height=4cm, angle=0]{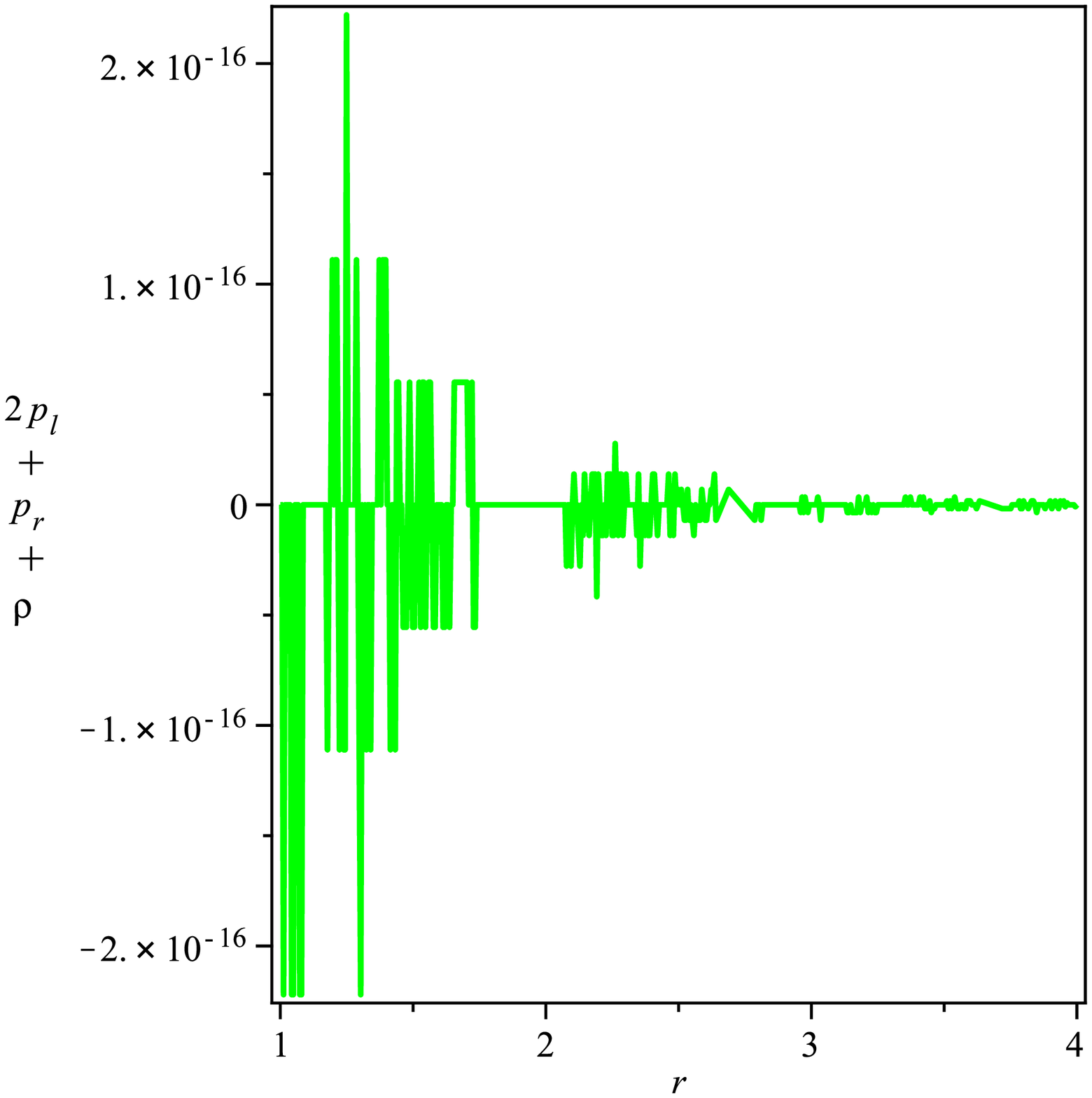}
  	\caption {(a) NEC, $\rho + p_r$ for $\alpha=0, \beta=0$, (b) NEC, $\rho + p_l$ for $\alpha=0, \beta=0$,(c) SEC, $\rho + p_r + 2p_l$ for for $\alpha=0, \beta=0$ .}
\end{figure}
  \begin{figure}
  (a)\includegraphics[width=3.5cm, height=4cm, angle=0]{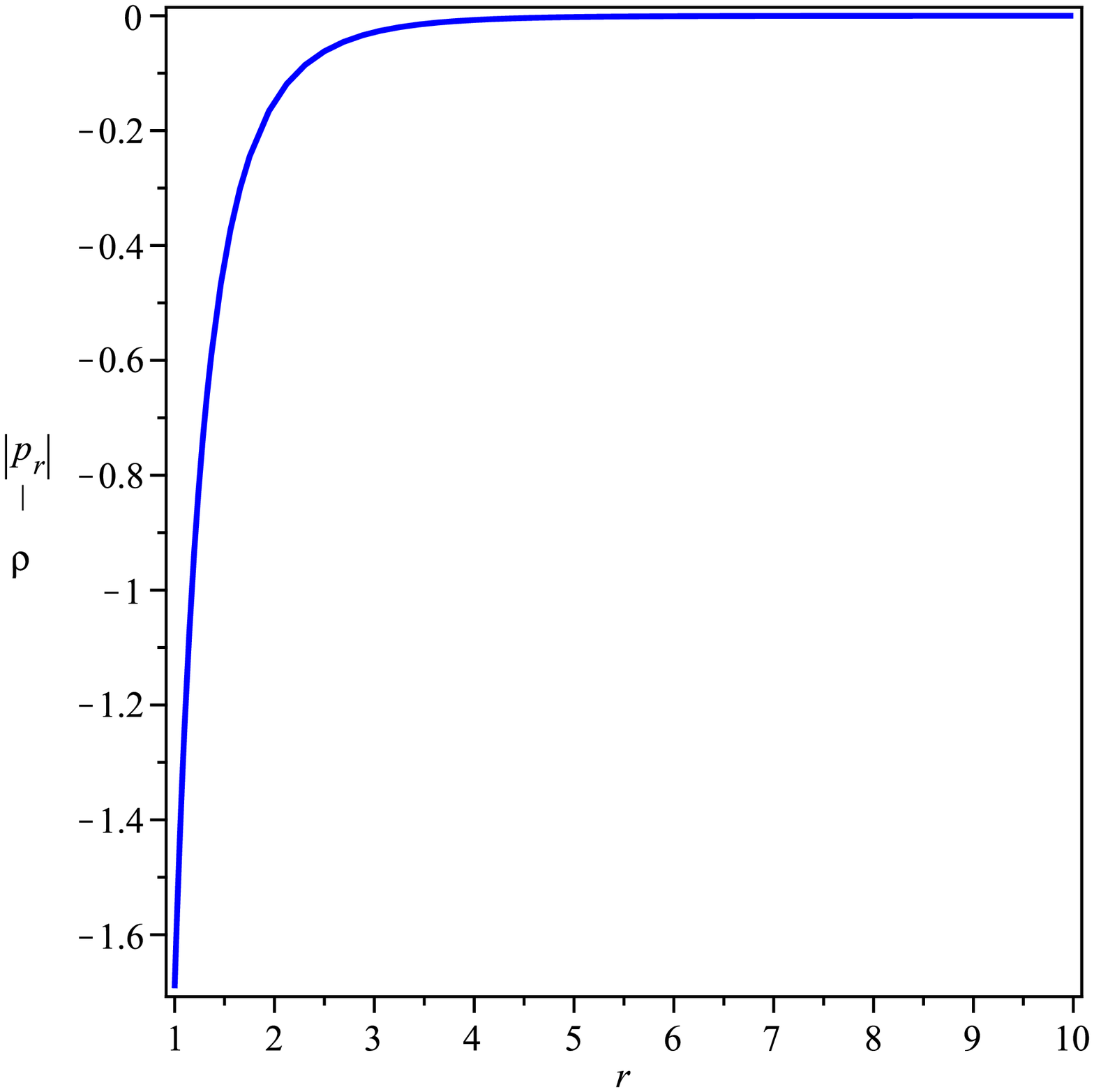}
  (b)\includegraphics[width=3.5cm, height=4cm, angle=0]{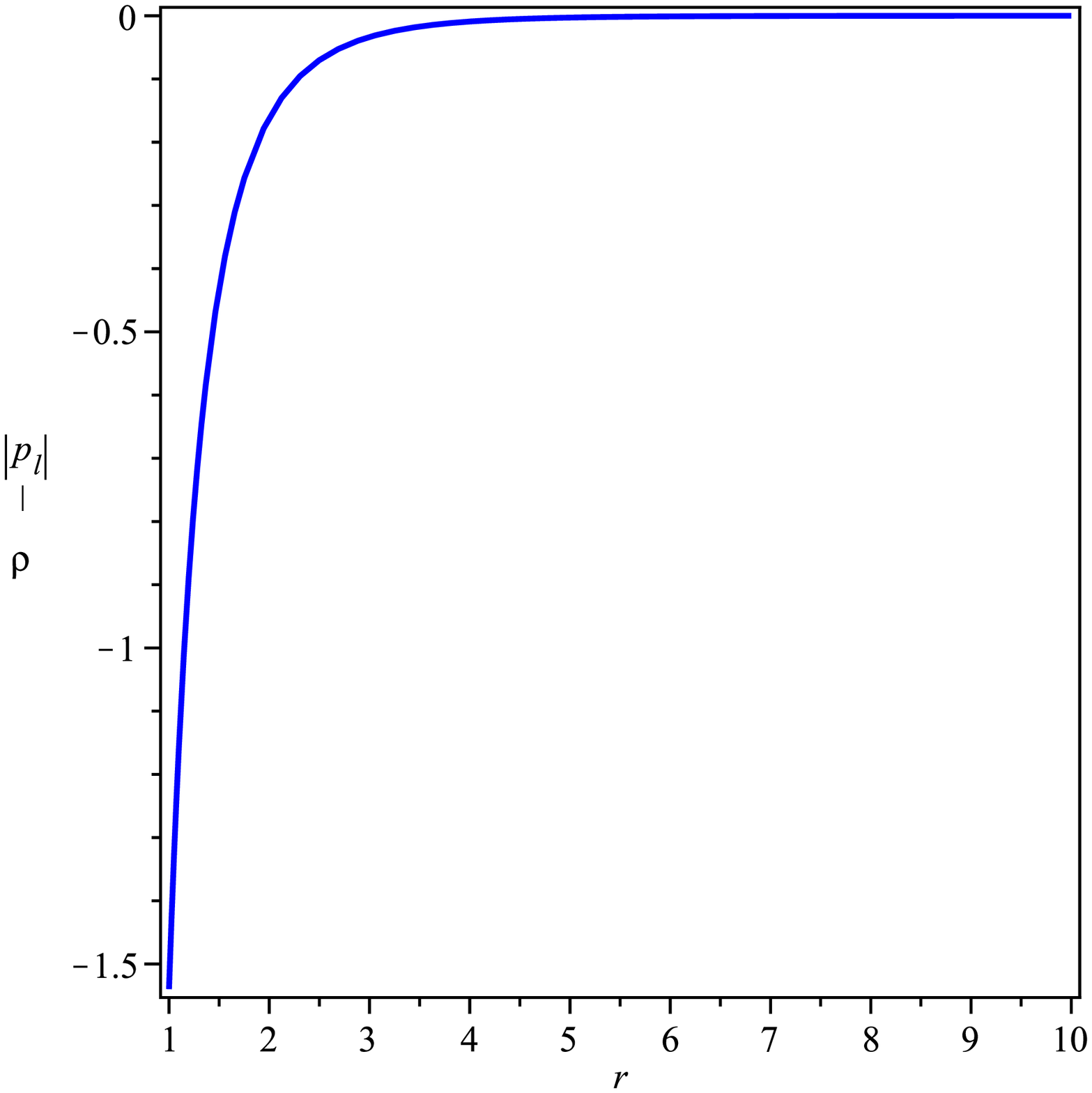}
  (c)\includegraphics[width=3.5cm, height=4cm, angle=0]{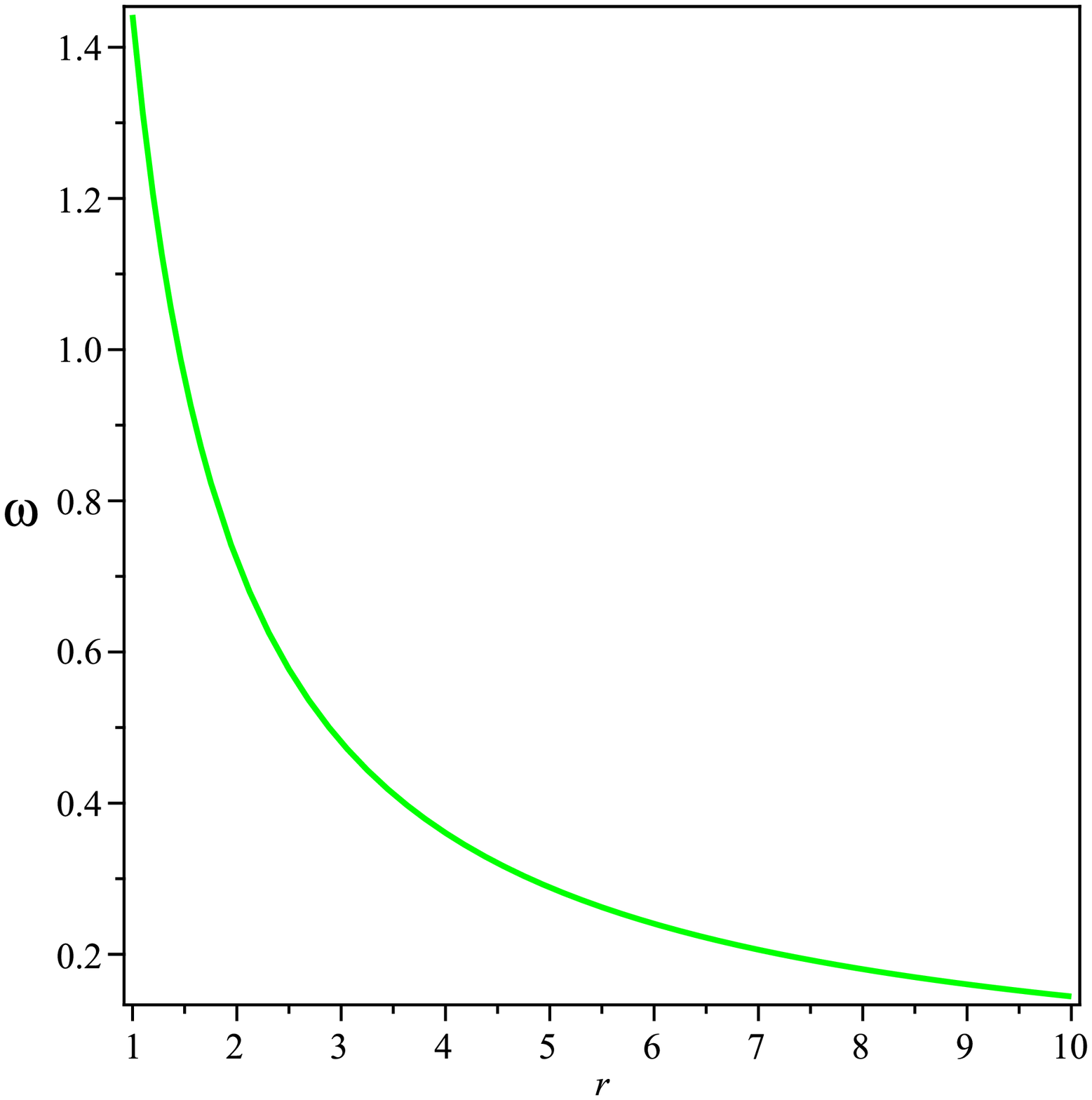}
  (d)\includegraphics[width=3.5cm, height=4cm, angle=0]{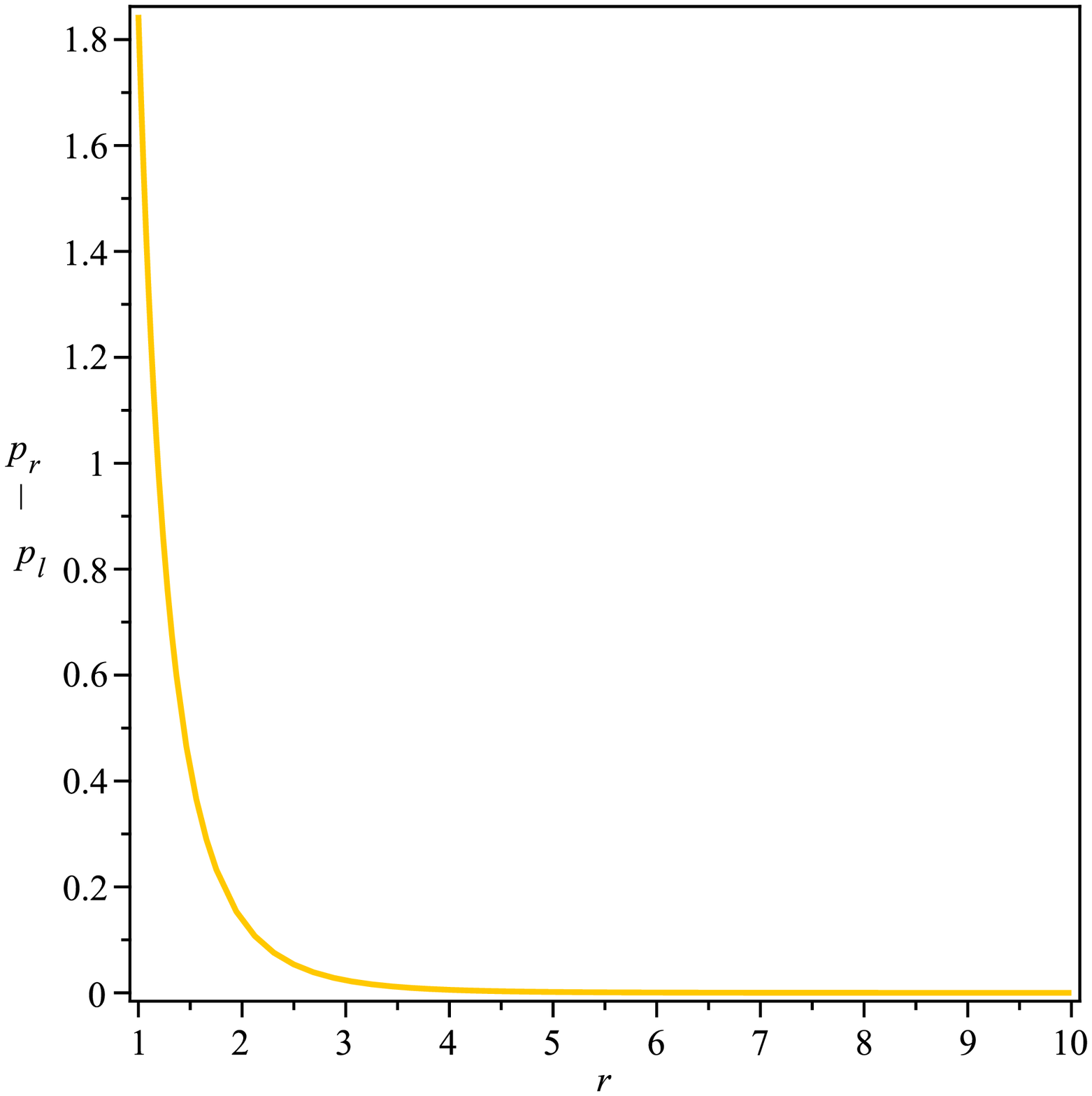}
  \caption {(a) DEC, $\rho - p_r$ for $\alpha=0, \beta=0$, (b) DEC, $\rho - p_l$ for $\alpha=0, \beta=0$, (c) EOS,for $\alpha=0,\beta=0$ (d)$\bigtriangleup =p_l-p_r$ for $\alpha=0,\beta=0$ .} 
  \end{figure}
 \begin{figure}
 \centering 
 (a)\includegraphics[width=3.5cm, height=4cm, angle=0]{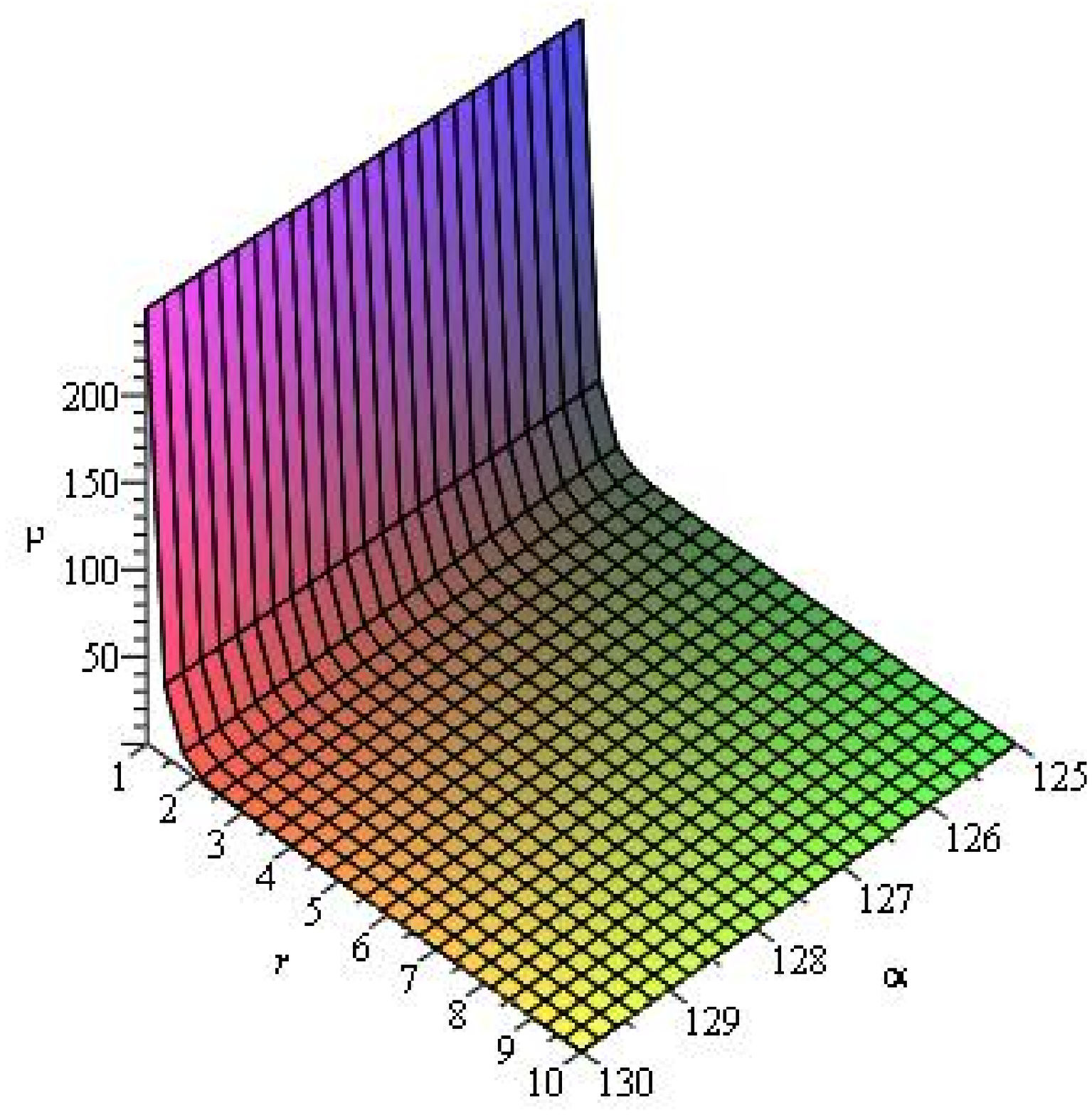}
 (b)\includegraphics[width=3.5cm, height=4cm, angle=0]{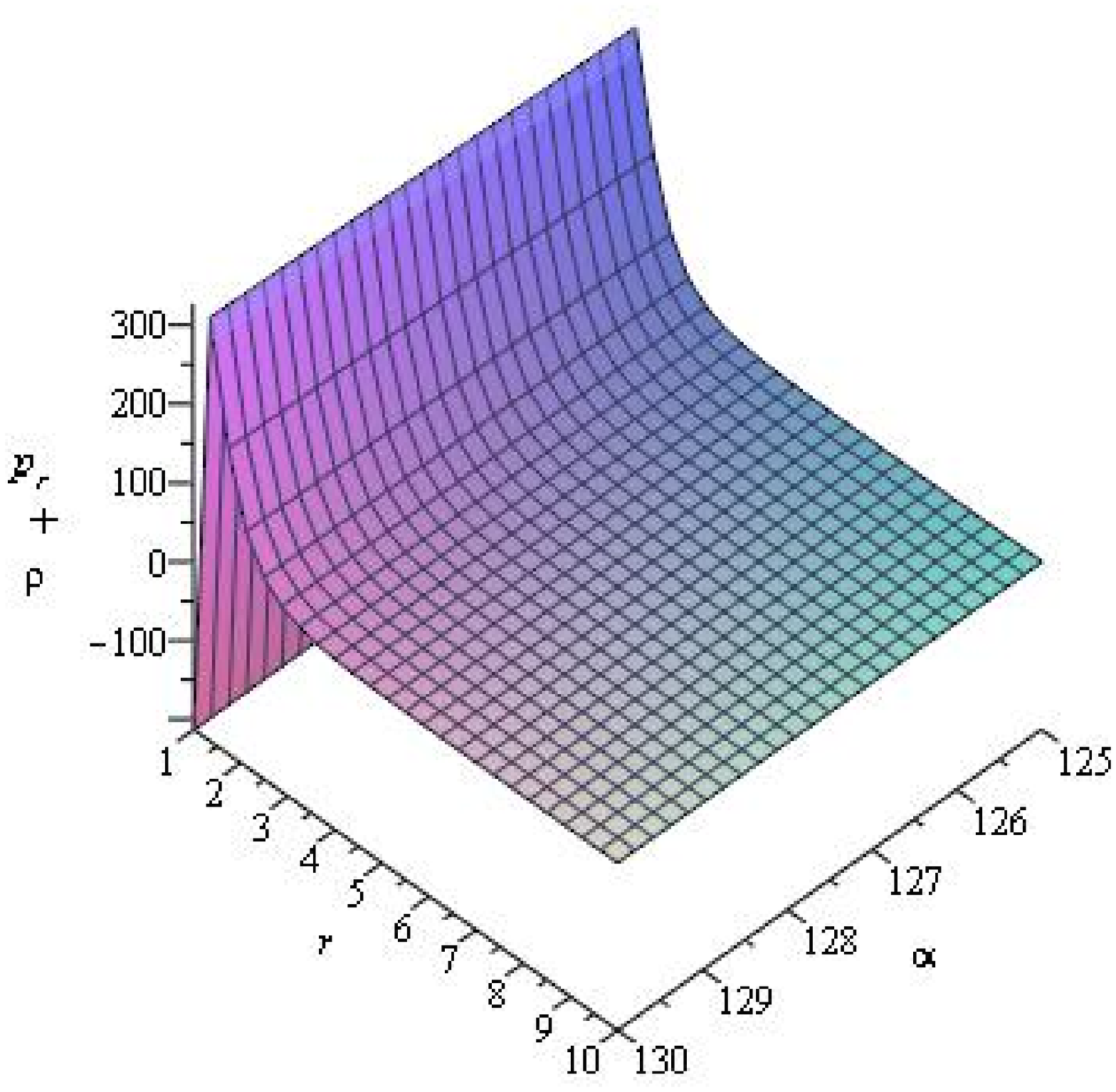}
 (c)\includegraphics[width=3.5cm, height=4cm, angle=0]{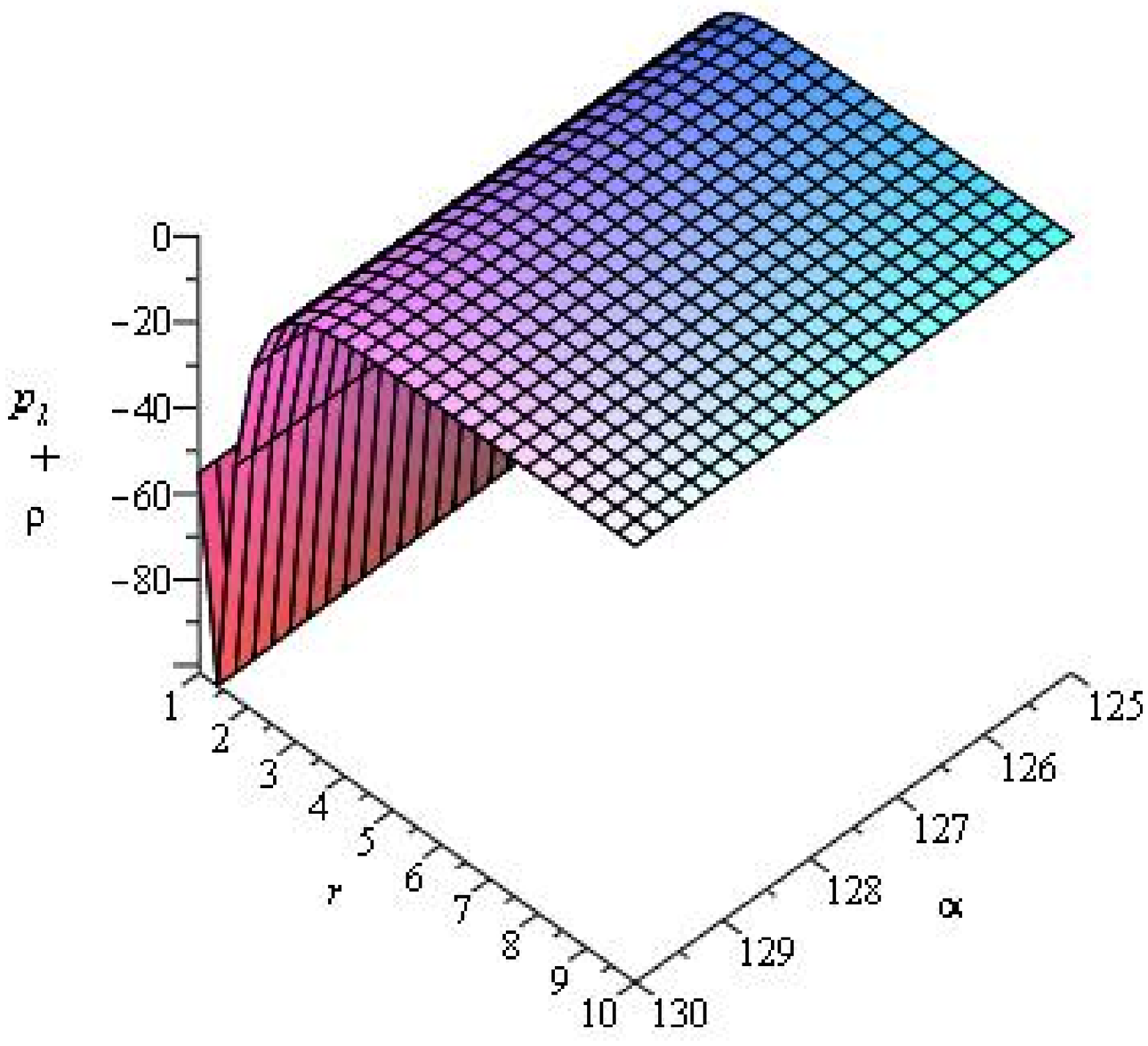}
 (d)\includegraphics[width=3.5cm, height=4cm, angle=0]{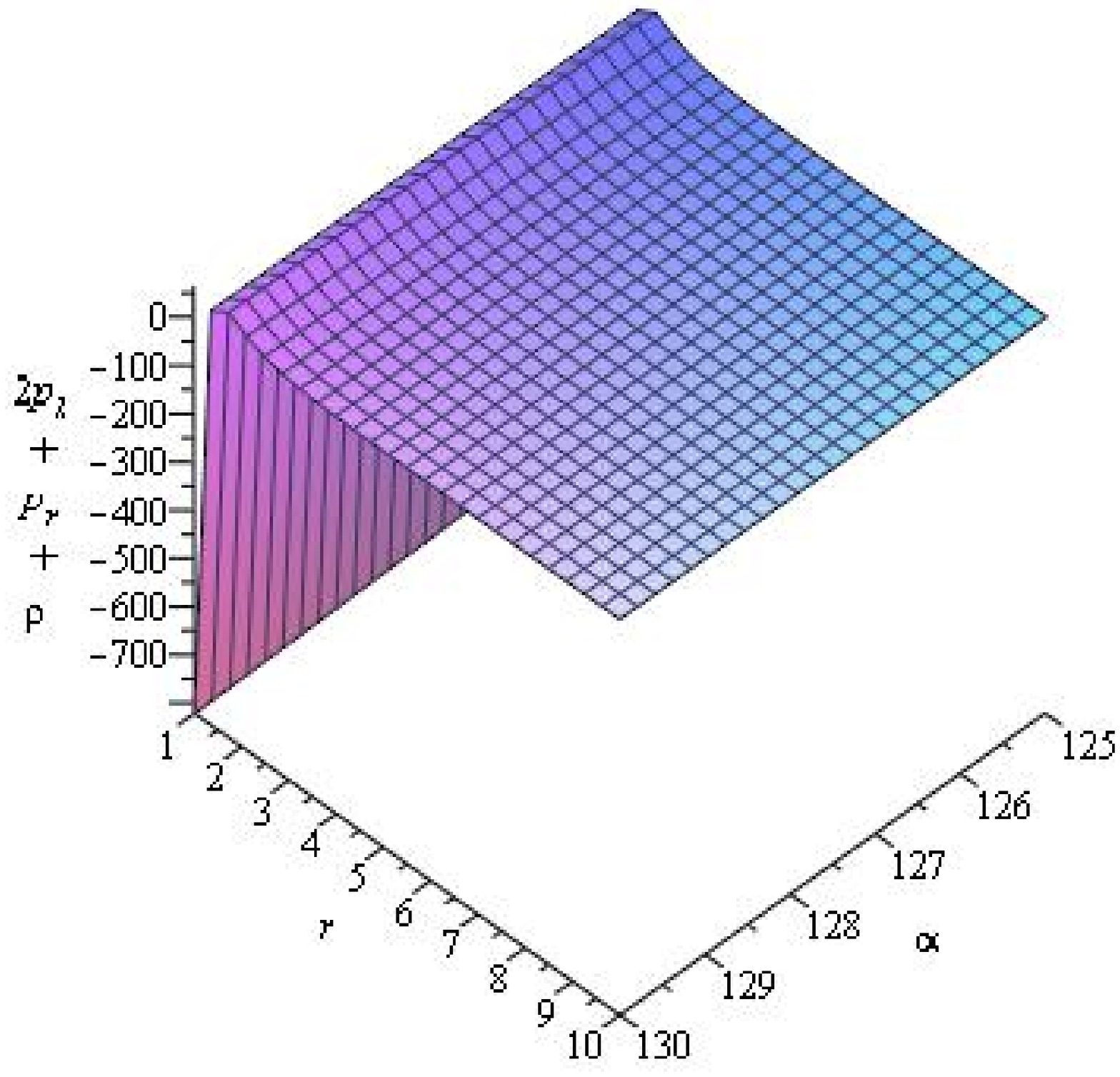}
 \caption {(a) WEC, $\rho$ for $\beta=0, a=0.5, m=2$ (b) NEC, $\rho + p_r$ for $\beta=0, a=0.5, m=2$, (c) NEC, $\rho + p_l$ for$\beta=0, a=0.5, m=2$,(d) SEC, $\rho + p_r + 2p_l$ for$\beta=0, a=0.5, m=2$.}
\end{figure}
  \begin{figure}
   \centering 
  (a)\includegraphics[width=3.5cm, height=4cm, angle=0]{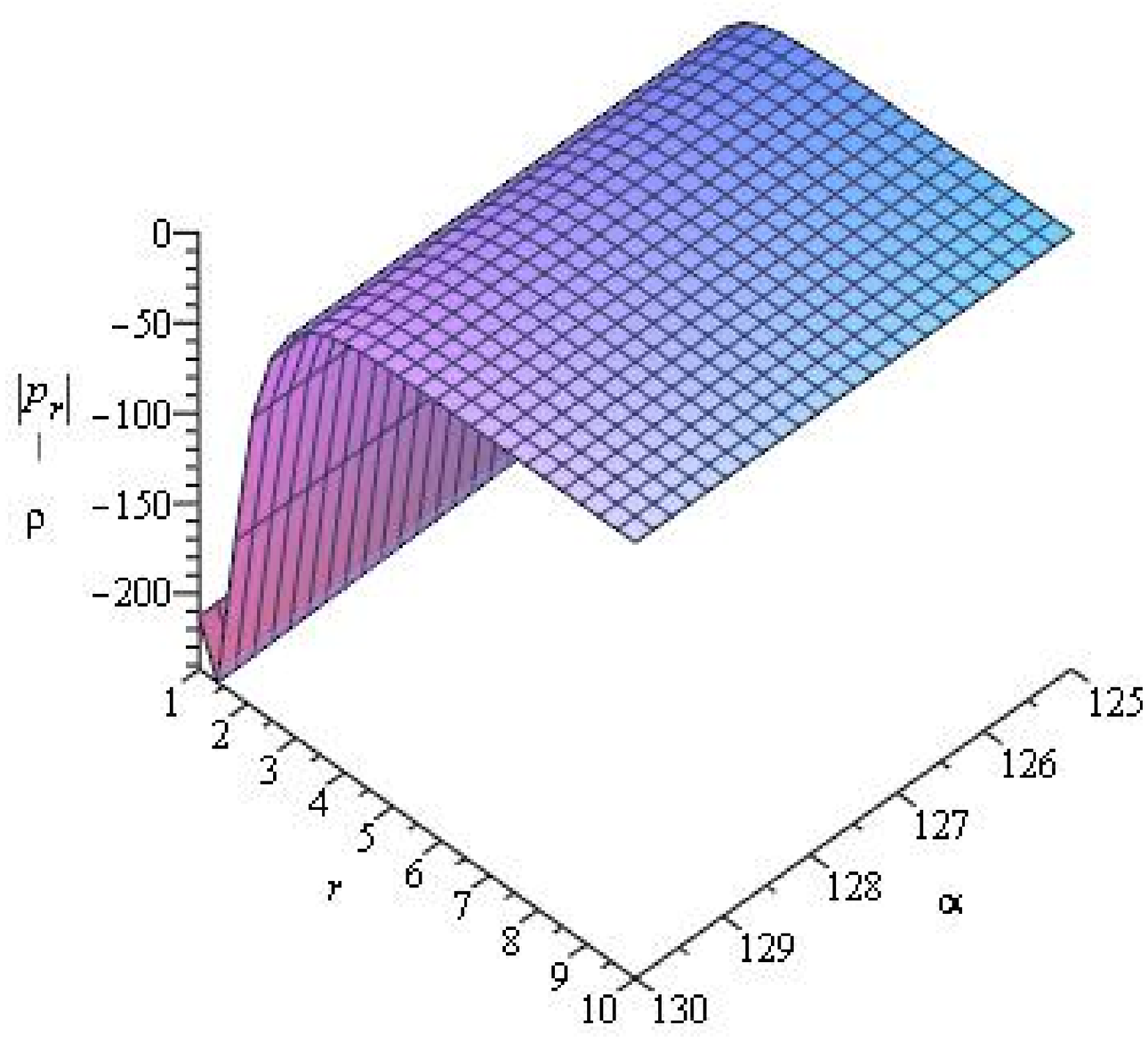}
  (b)\includegraphics[width=3.5cm, height=4cm, angle=0]{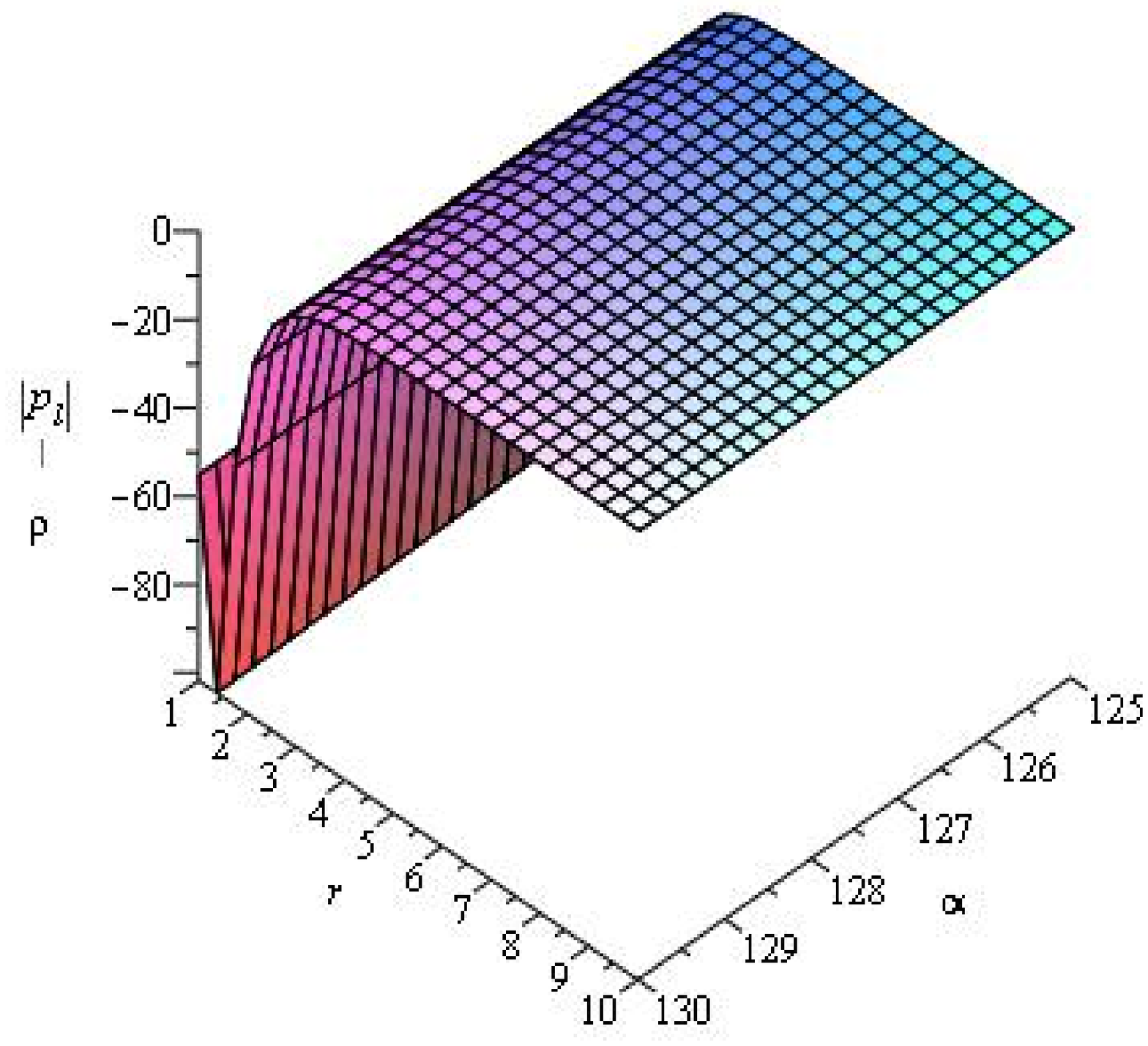}
  (c)\includegraphics[width=3.5cm, height=4cm, angle=0]{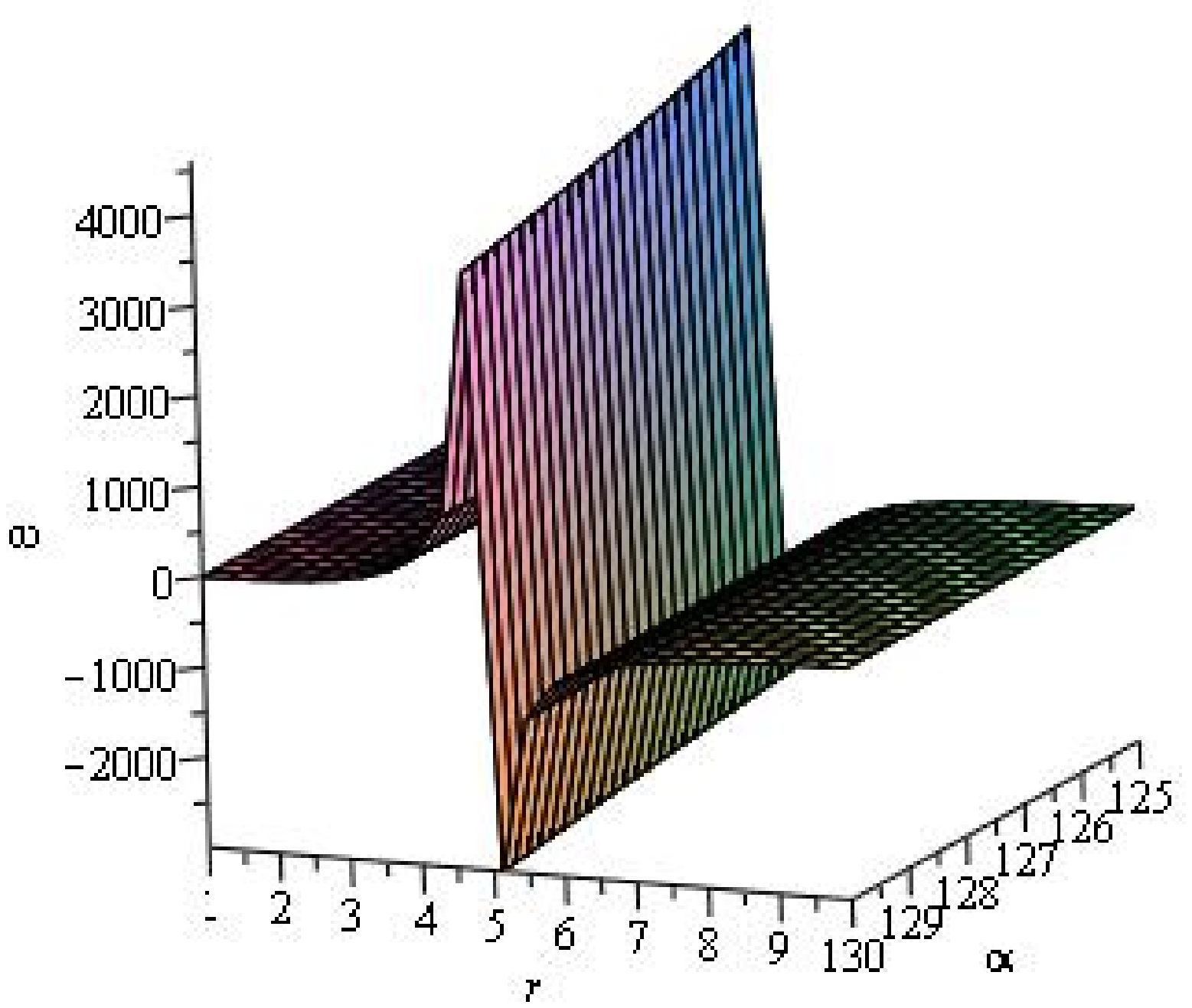}
  (d)\includegraphics[width=3.5cm, height=4cm, angle=0]{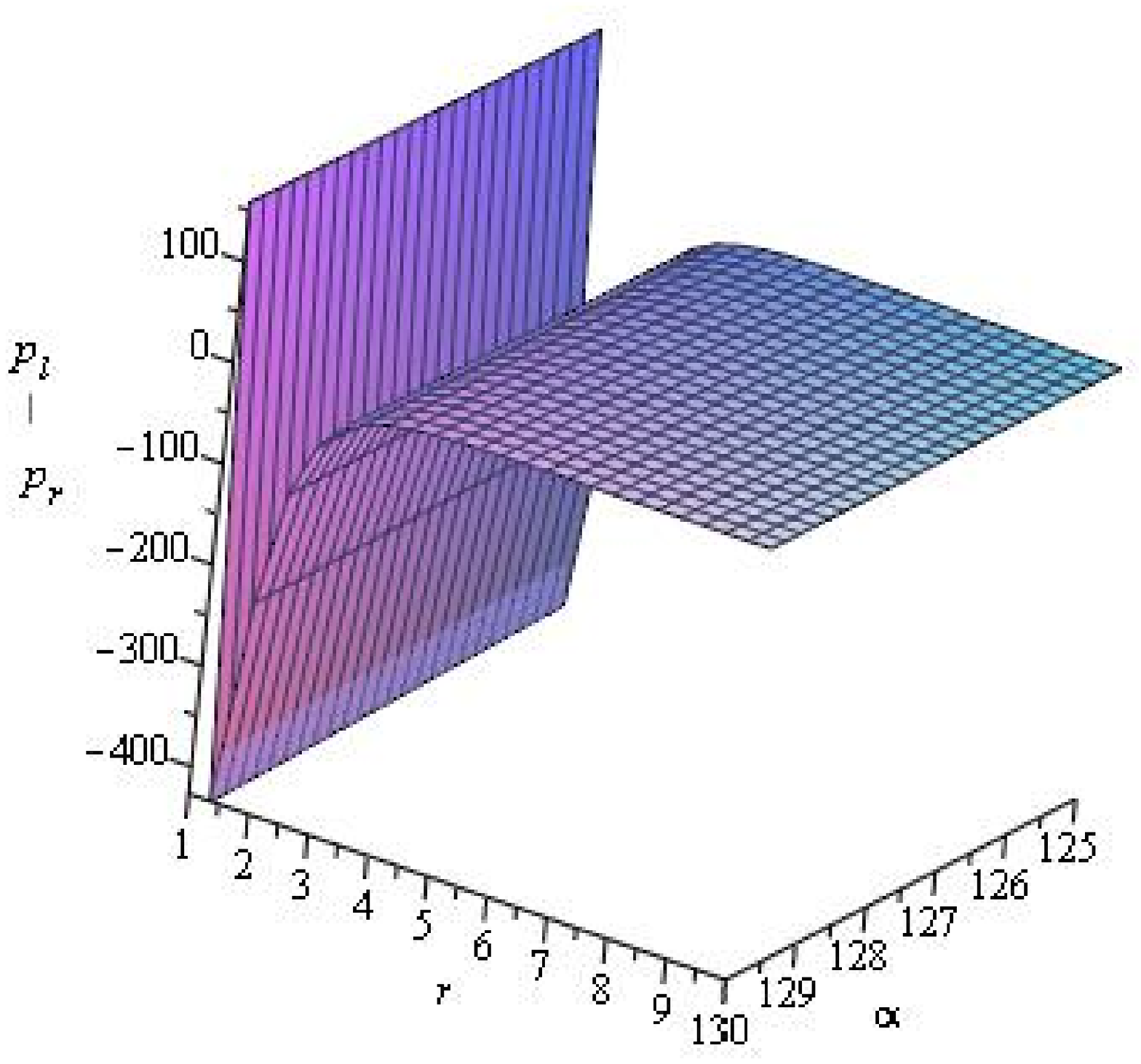}
  \caption {(a) DEC, $\rho - p_r$ for $\beta=0, a=0.5, m=2$, (b) DEC, $\rho - p_l$ for$\beta=0, a=0.5, m=2$, (c) EOS,for $\beta=0, a=0.5, m=2$ (d)$\bigtriangleup=p_l-p_r$ or $\beta=0, a=0.5, m=2$.}
\end{figure} 
  \begin{figure}
 	\centering 
 	(a)\includegraphics[width=3.5cm, height=4cm, angle=0]{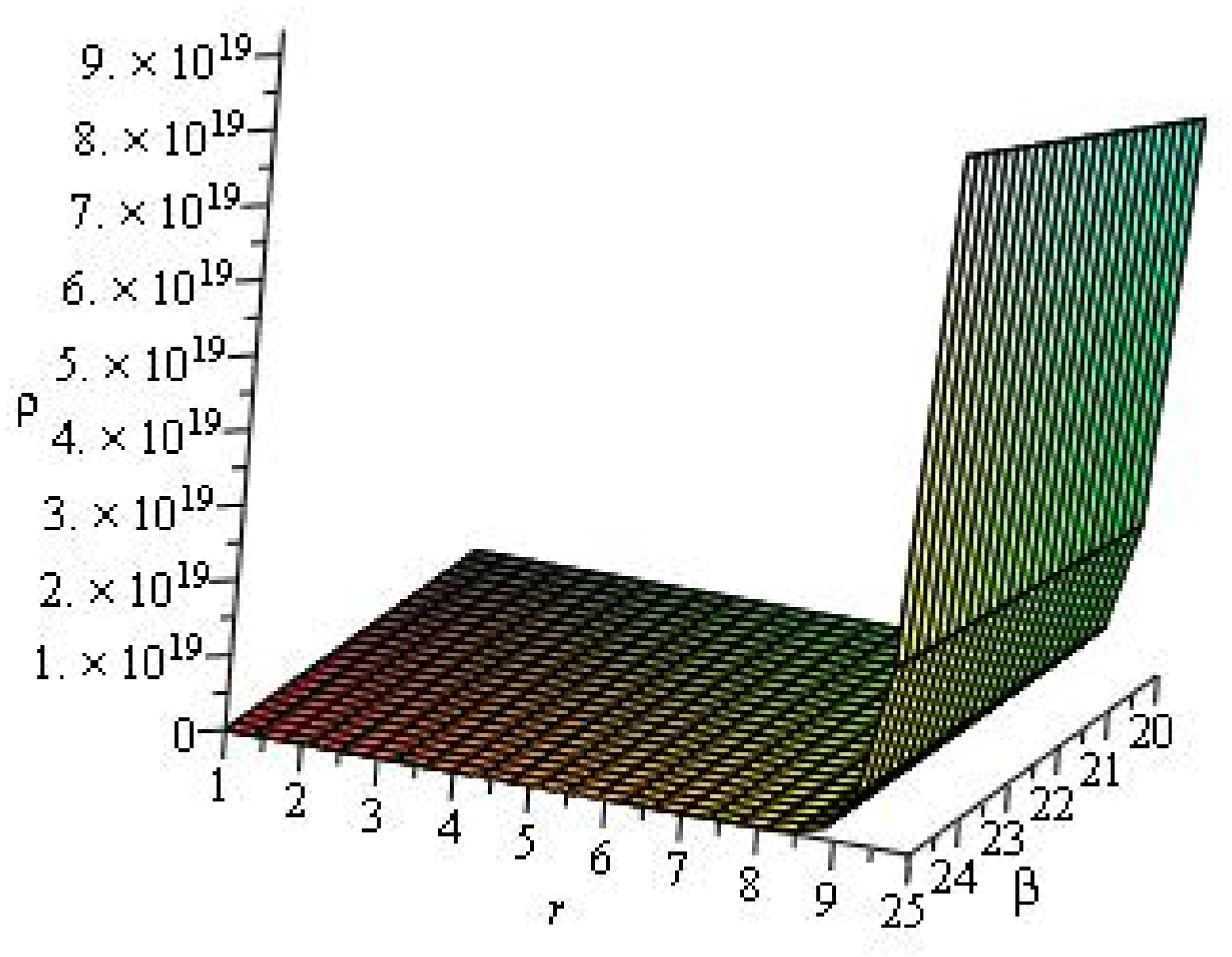}
 	(b)\includegraphics[width=3.5cm, height=4cm, angle=0]{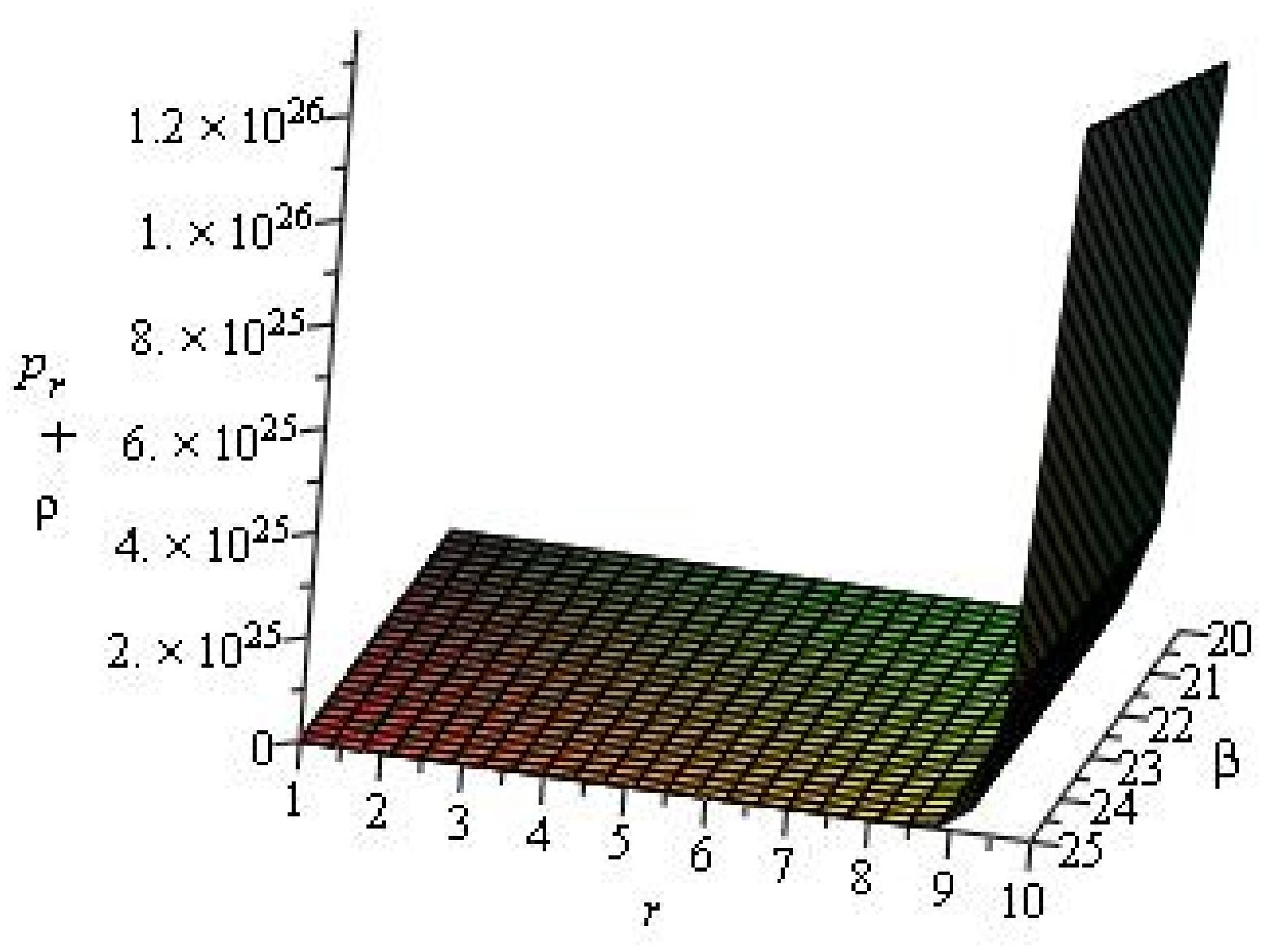}
 	(c)\includegraphics[width=3.5cm, height=4cm, angle=0]{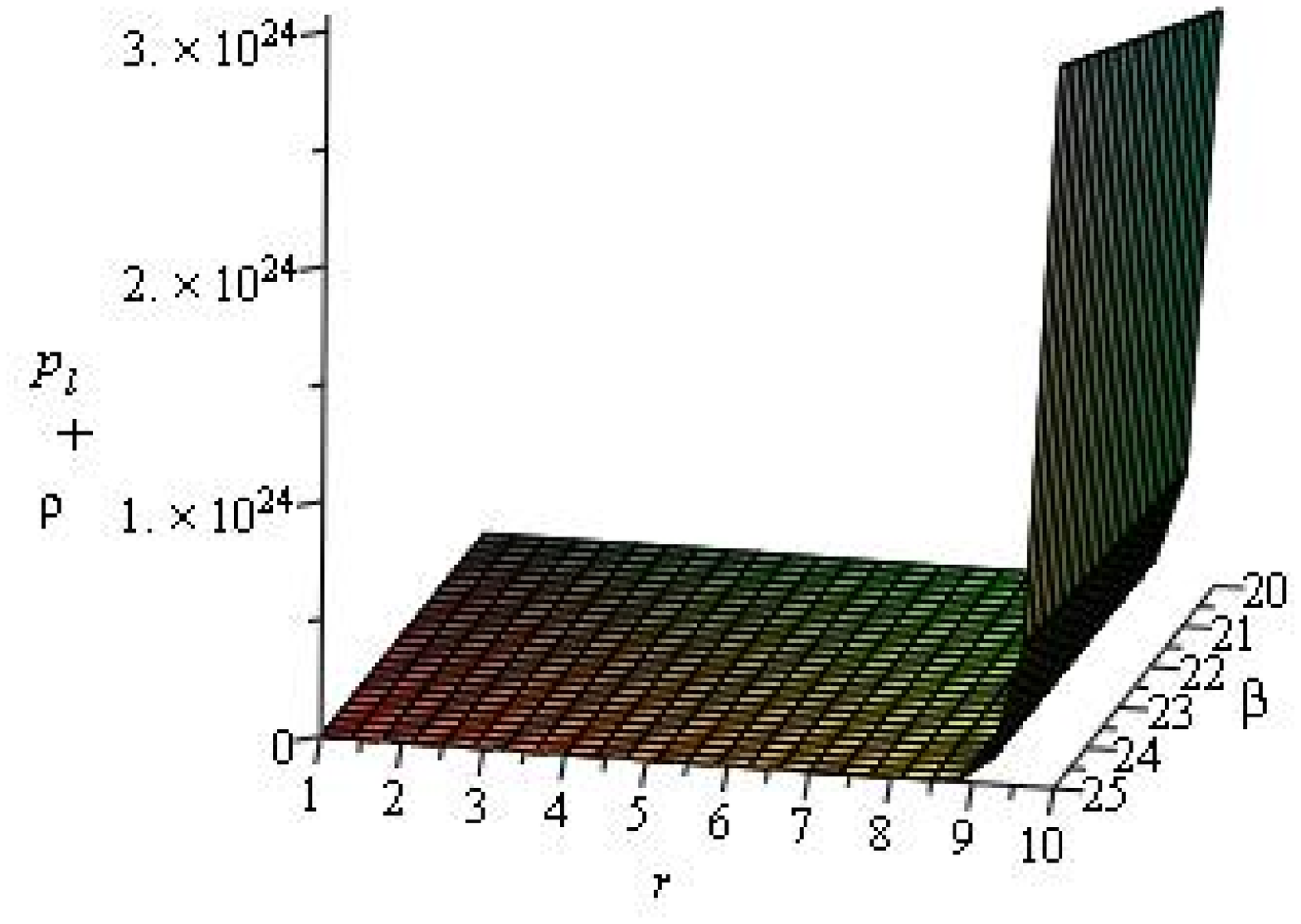}
 	(d)\includegraphics[width=3.5cm, height=4cm, angle=0]{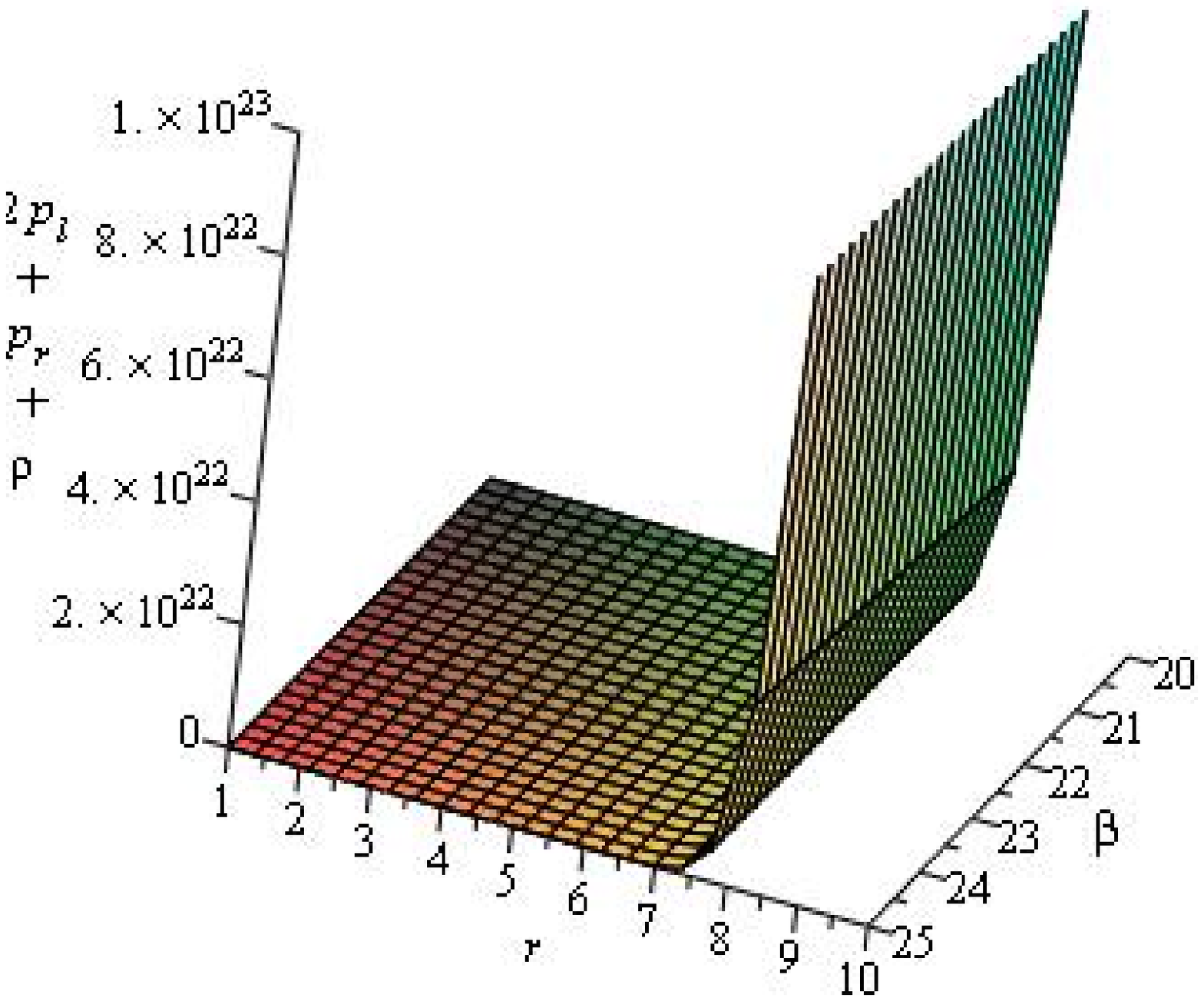}
 	\caption {(a) WEC, $\rho$ for $\alpha=0, a=0.5, n=4$ (b) NEC, $\rho + p_r$ for $\alpha=0, a=0.5, n=4$, (c) NEC, $\rho + p_l$ for$\alpha=0, a=0.5, n=4$,(d) SEC, $\rho + p_r + 2p_l$ for$\alpha=0, a=0.5, n=4$.}
 \end{figure}
 \begin{figure}
	\centering 
	(a)\includegraphics[width=3.5cm, height=4cm, angle=0]{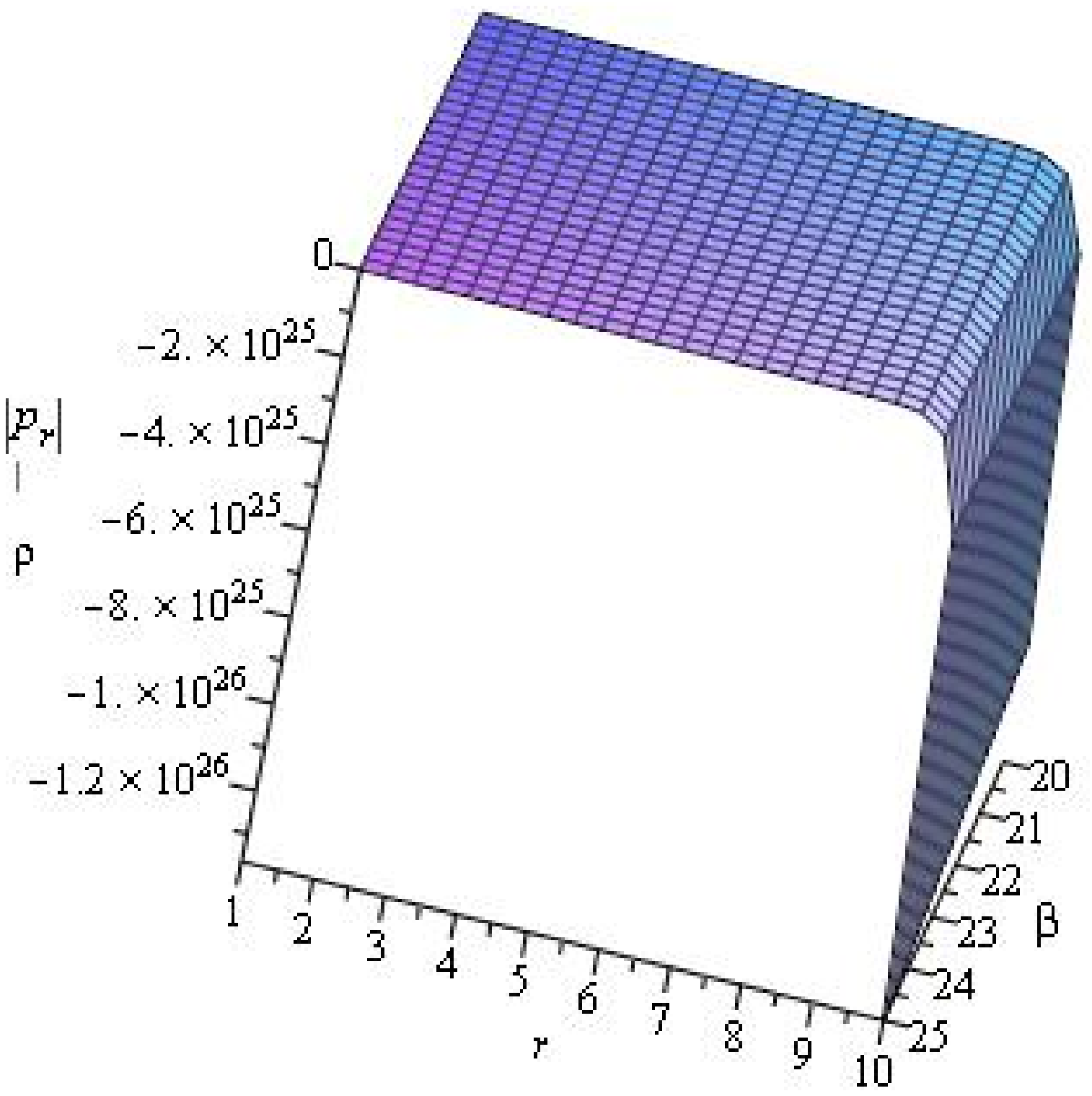}
	(b)\includegraphics[width=3.5cm, height=4cm, angle=0]{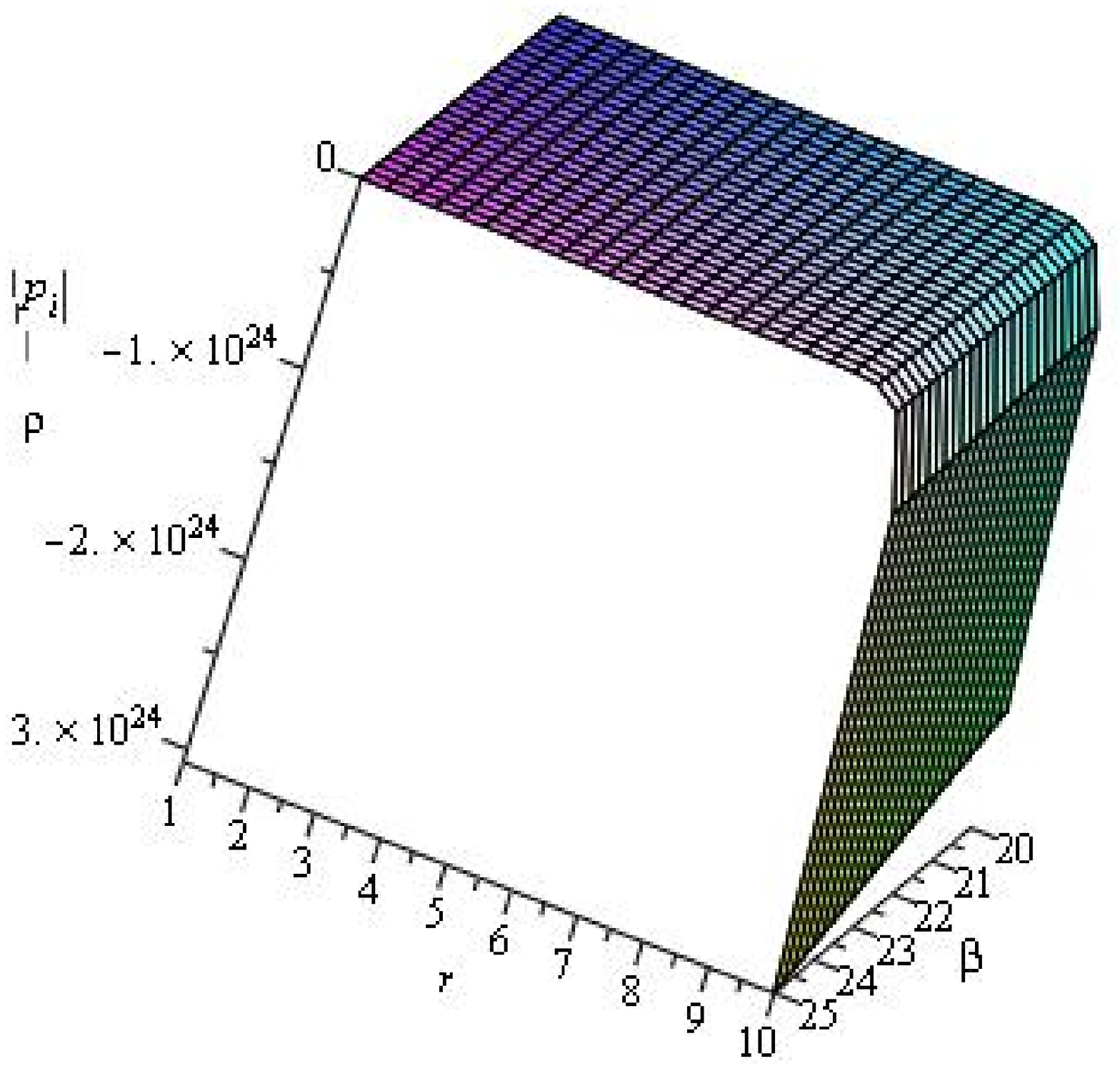}
	(c)\includegraphics[width=3.5cm, height=4cm, angle=0]{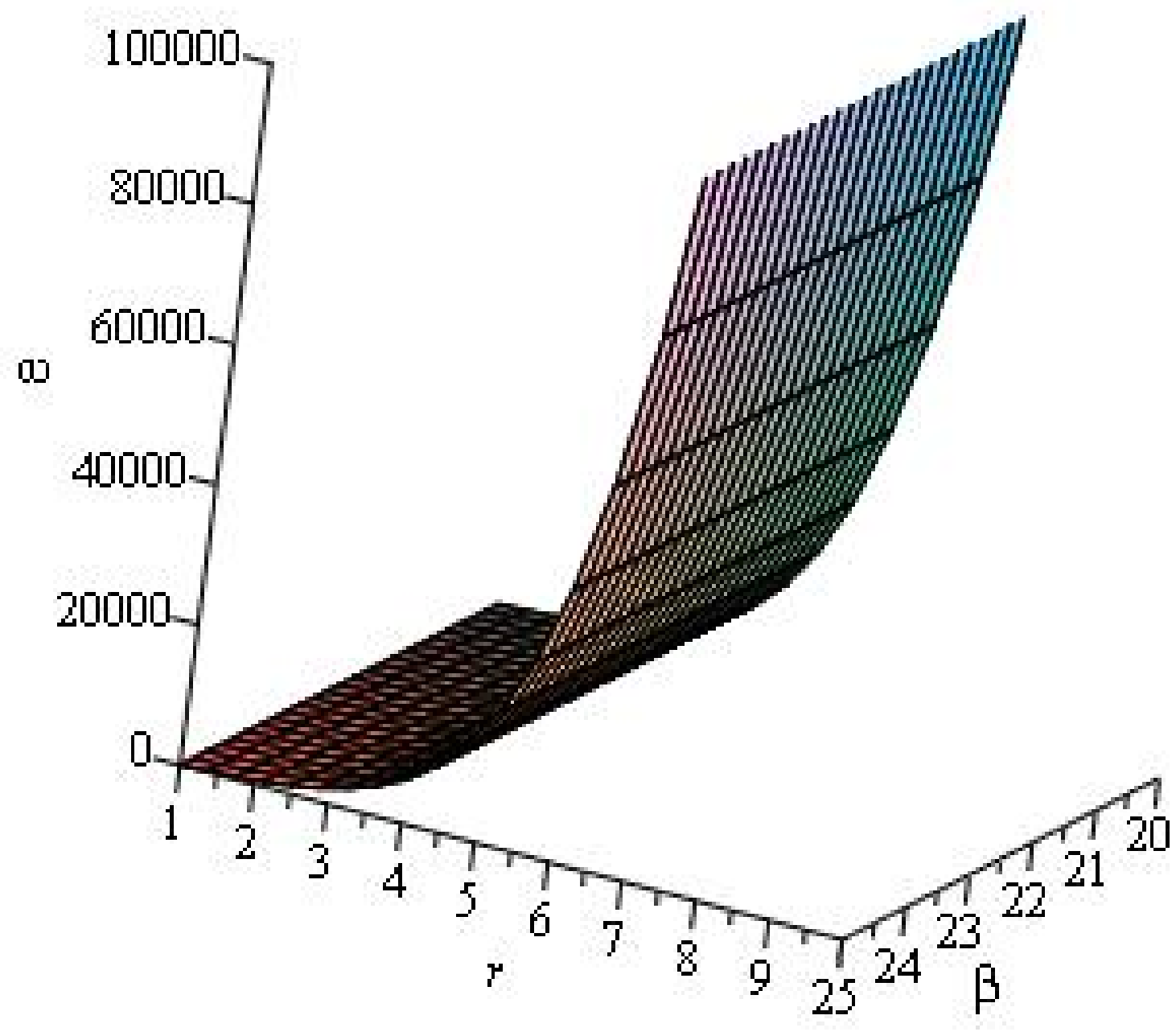}
	(d)\includegraphics[width=3.5cm, height=4cm, angle=0]{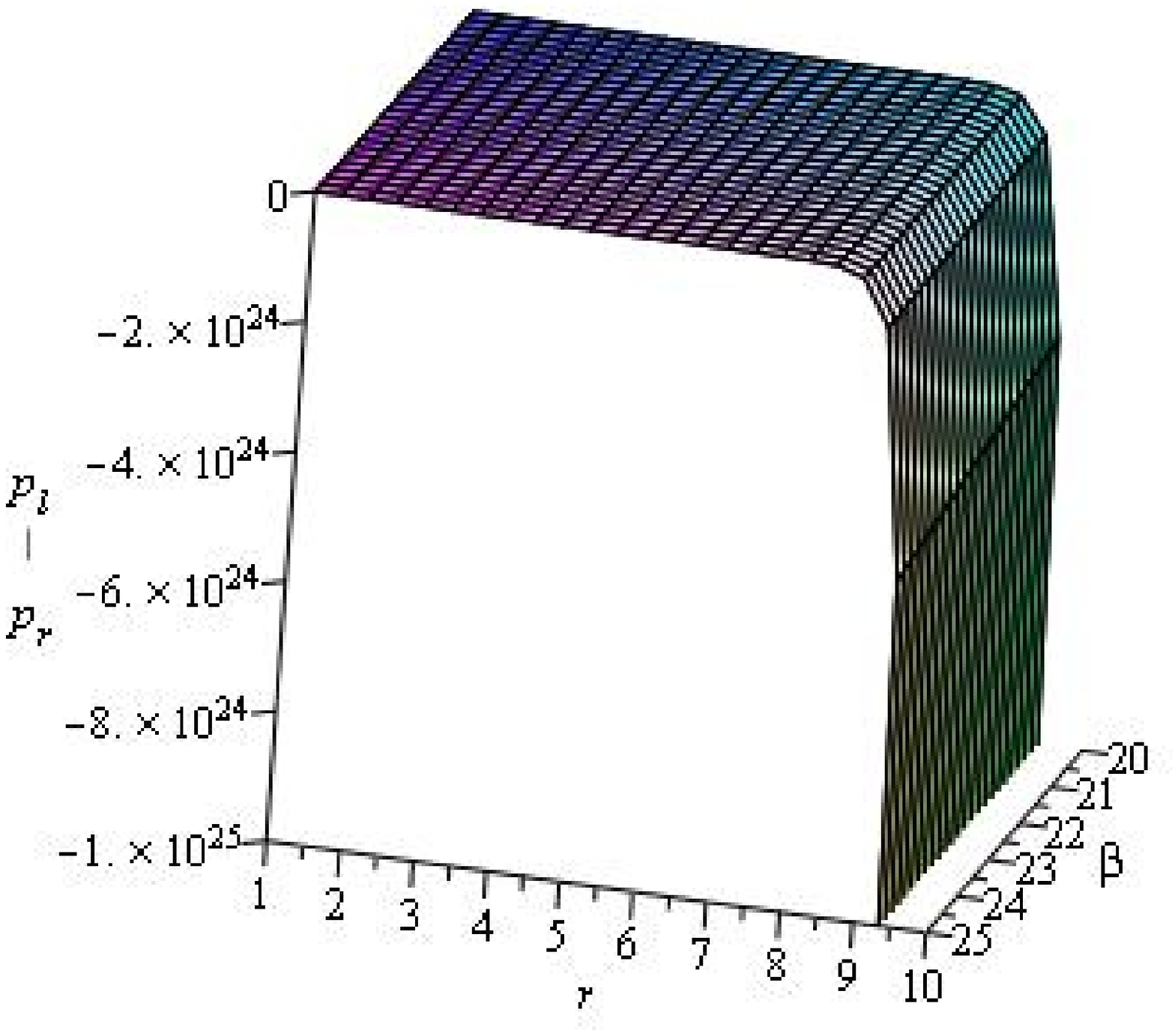}
	\caption {(a) DEC, $\rho - p_r$ for $\alpha=0, a=0.5, n=4$, (b) DEC, $\rho - p_l$ for$\beta=0, a=0.5, n=4$, (c) EOS,for $\alpha=0, a=0.5, n=4$ (d)$\bigtriangleup=p_l-p_r$ for $\alpha=0, a=0.5, n=4$.} 
 \end{figure}	
  \begin{figure}
 	\centering 
 	(a)\includegraphics[width=3.5cm, height=4cm, angle=0]{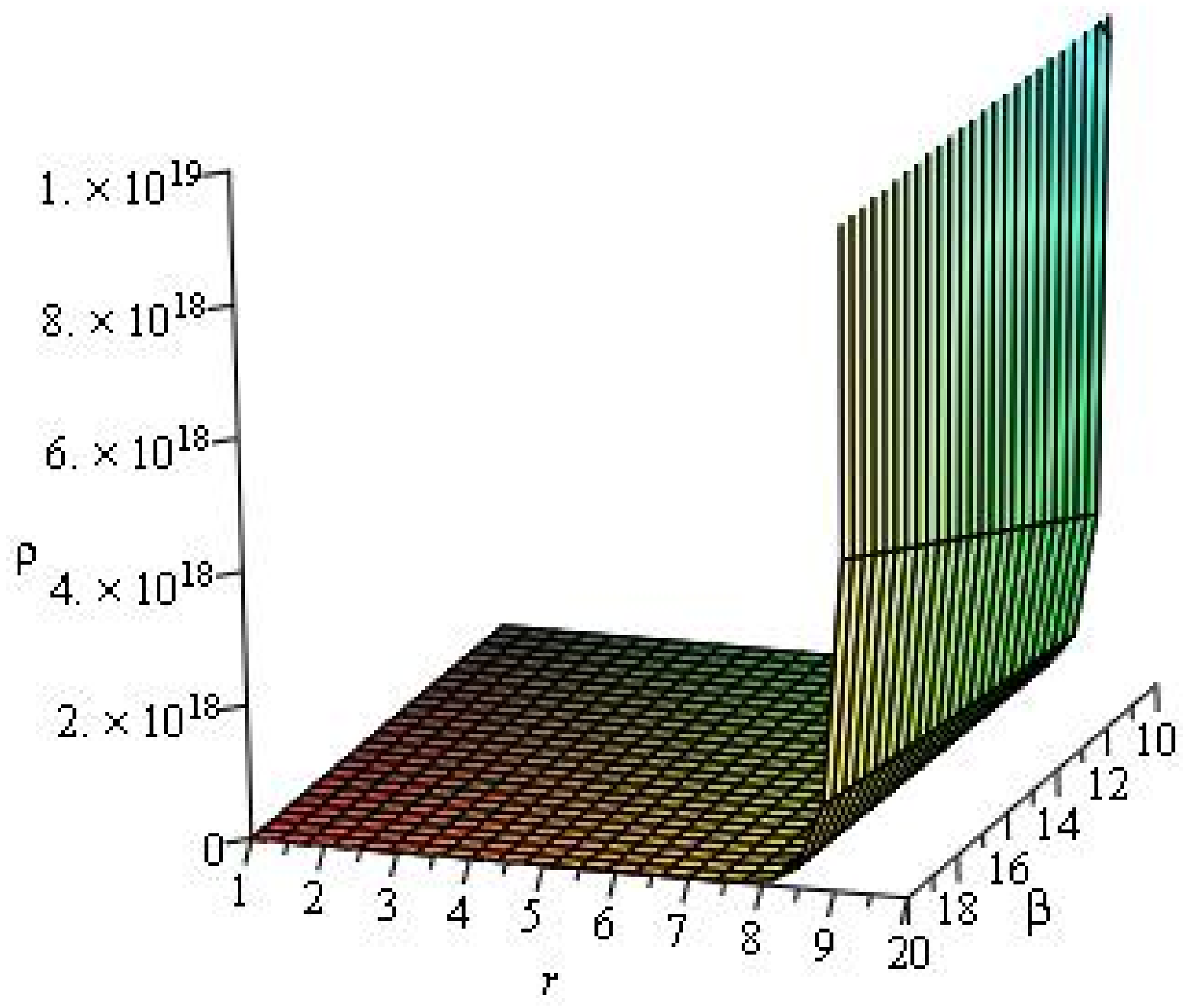}
 	(b)\includegraphics[width=3.5cm, height=4cm, angle=0]{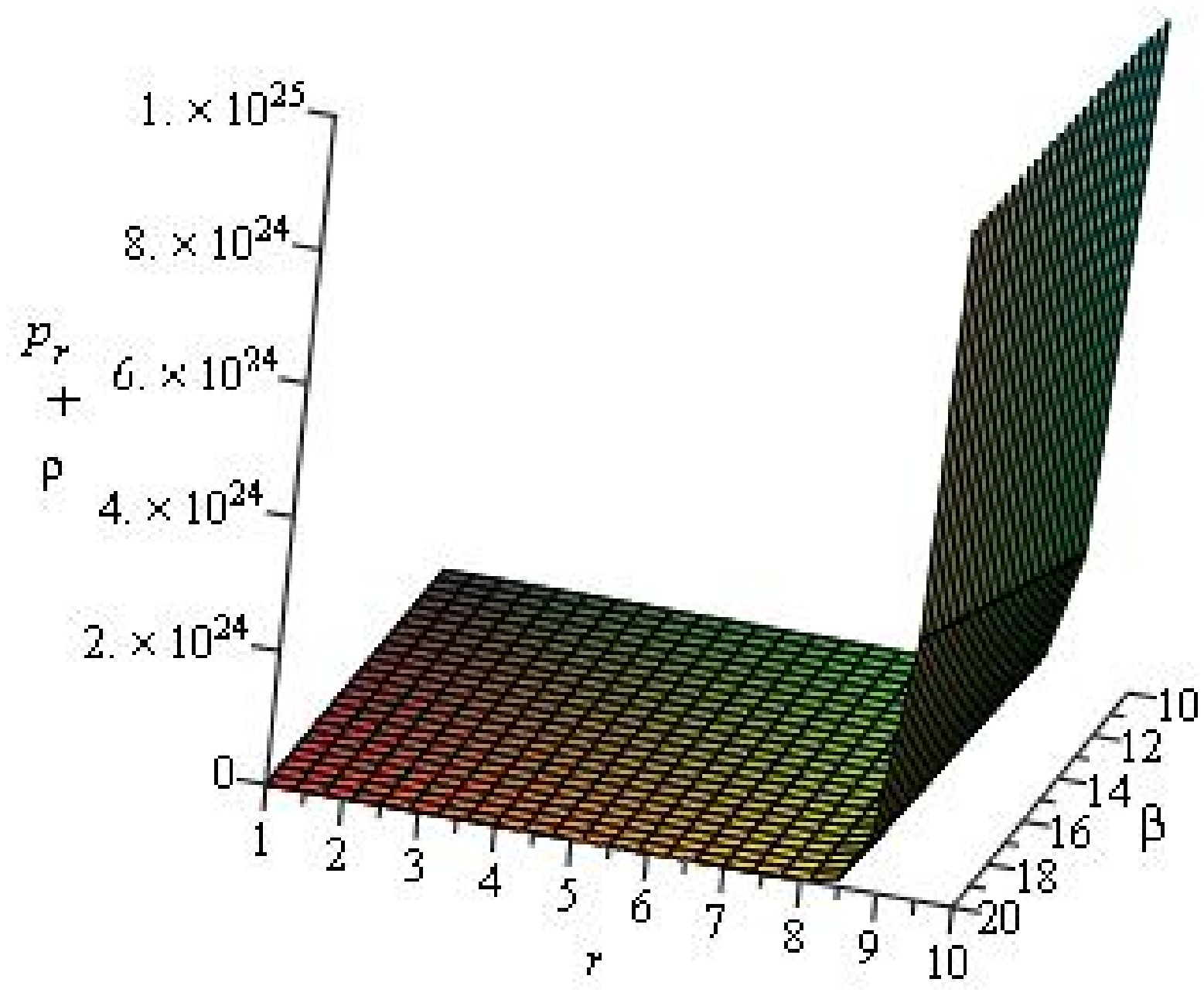}
 	(c)\includegraphics[width=3.5cm, height=4cm, angle=0]{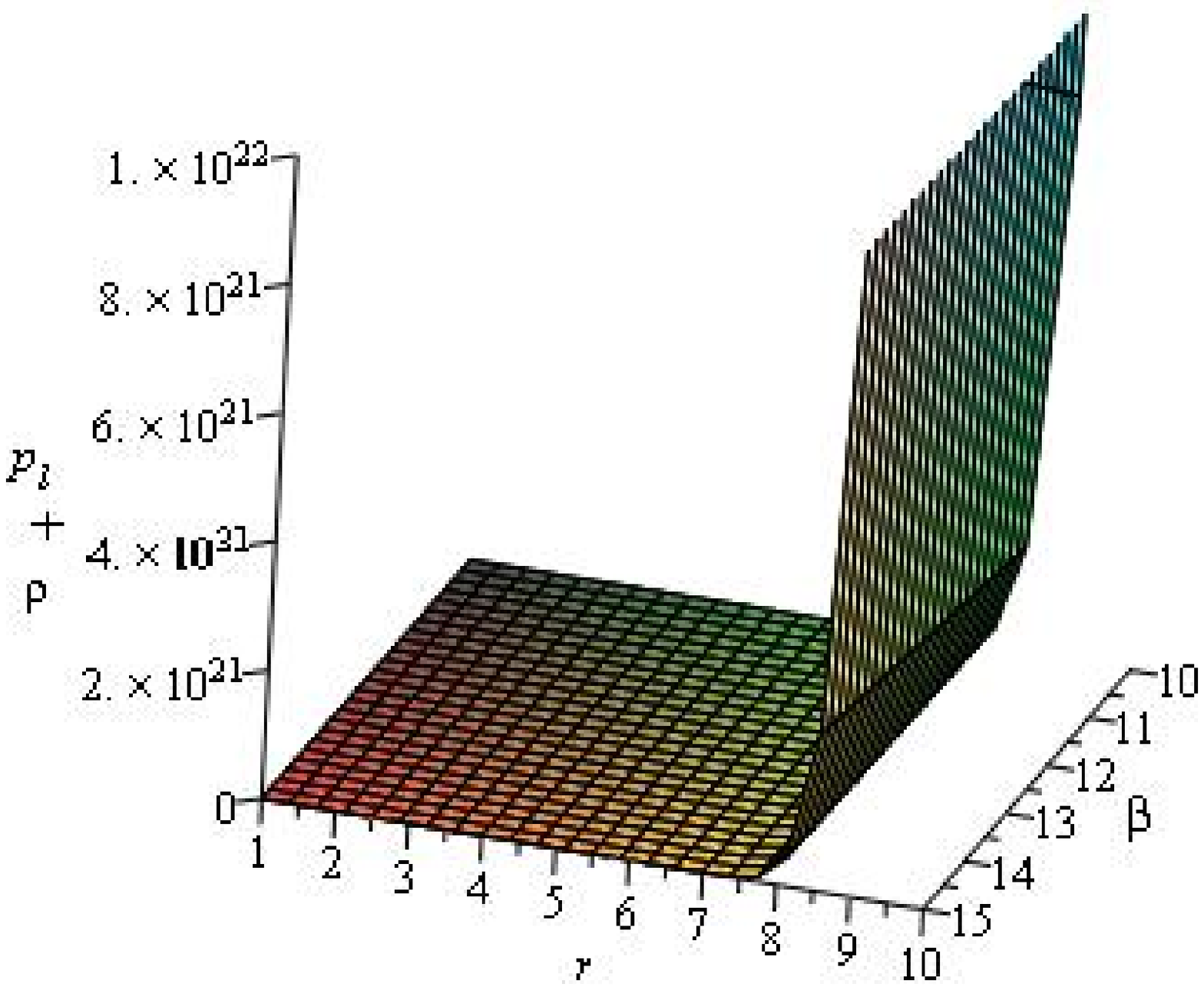}
 	(d)\includegraphics[width=3.5cm, height=4cm, angle=0]{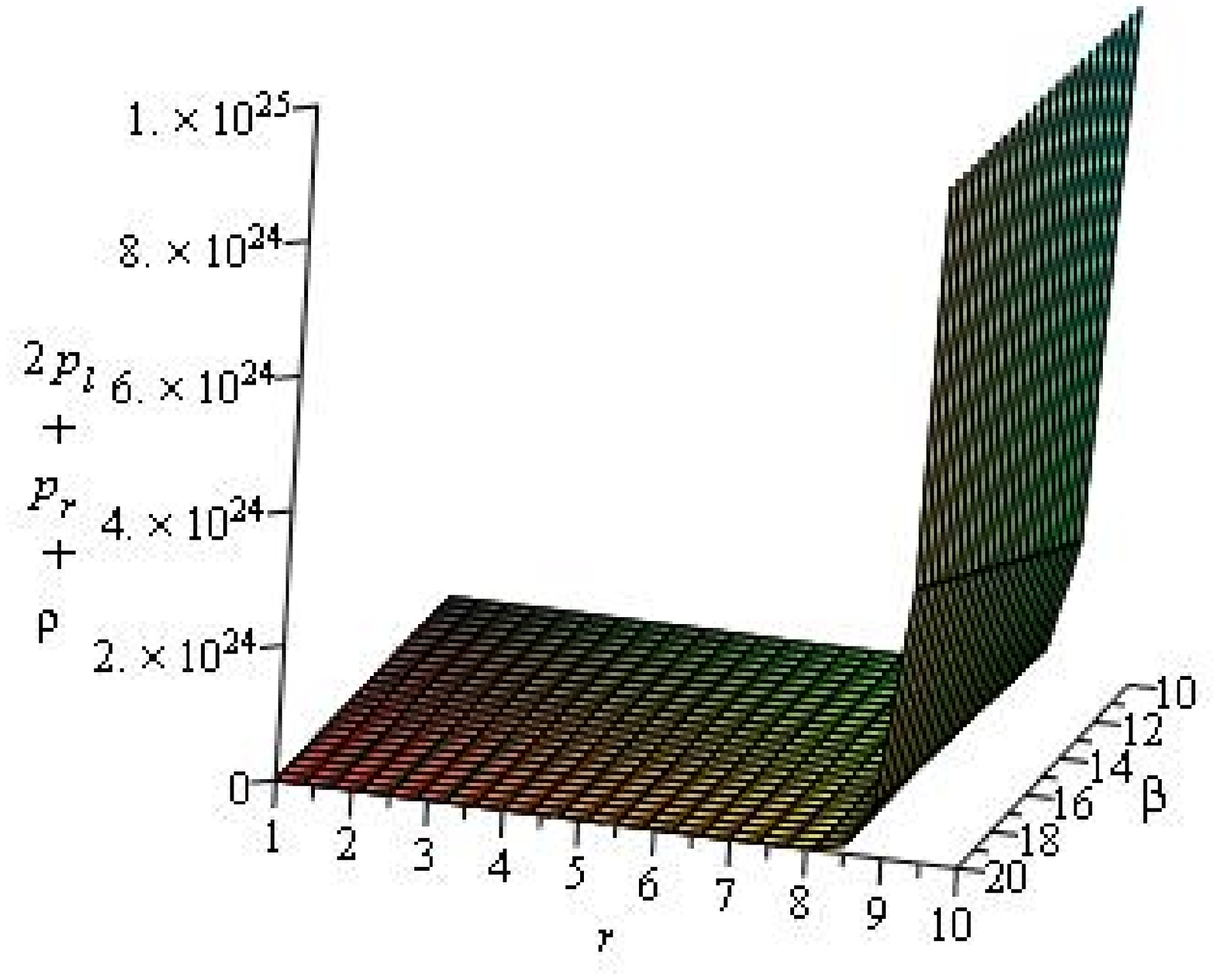}
 	\caption {(a) WEC, $\rho$ for $\alpha=5, a=0.5, m=2, n=4$ (b) NEC, $\rho + p_r$ for $\alpha=5, a=0.5, m=2, n=4$, (c) NEC, $\rho + p_l$ for$\alpha=5, a=0.5, m=2, n=4$,(d) SEC, $\rho + p_r + 2p_l$ for$\alpha=5, a=0.5, m=2, n=4$.}
 \end{figure}
 \begin{figure}
 	\centering 
 	(a)\includegraphics[width=3.5cm, height=4cm, angle=0]{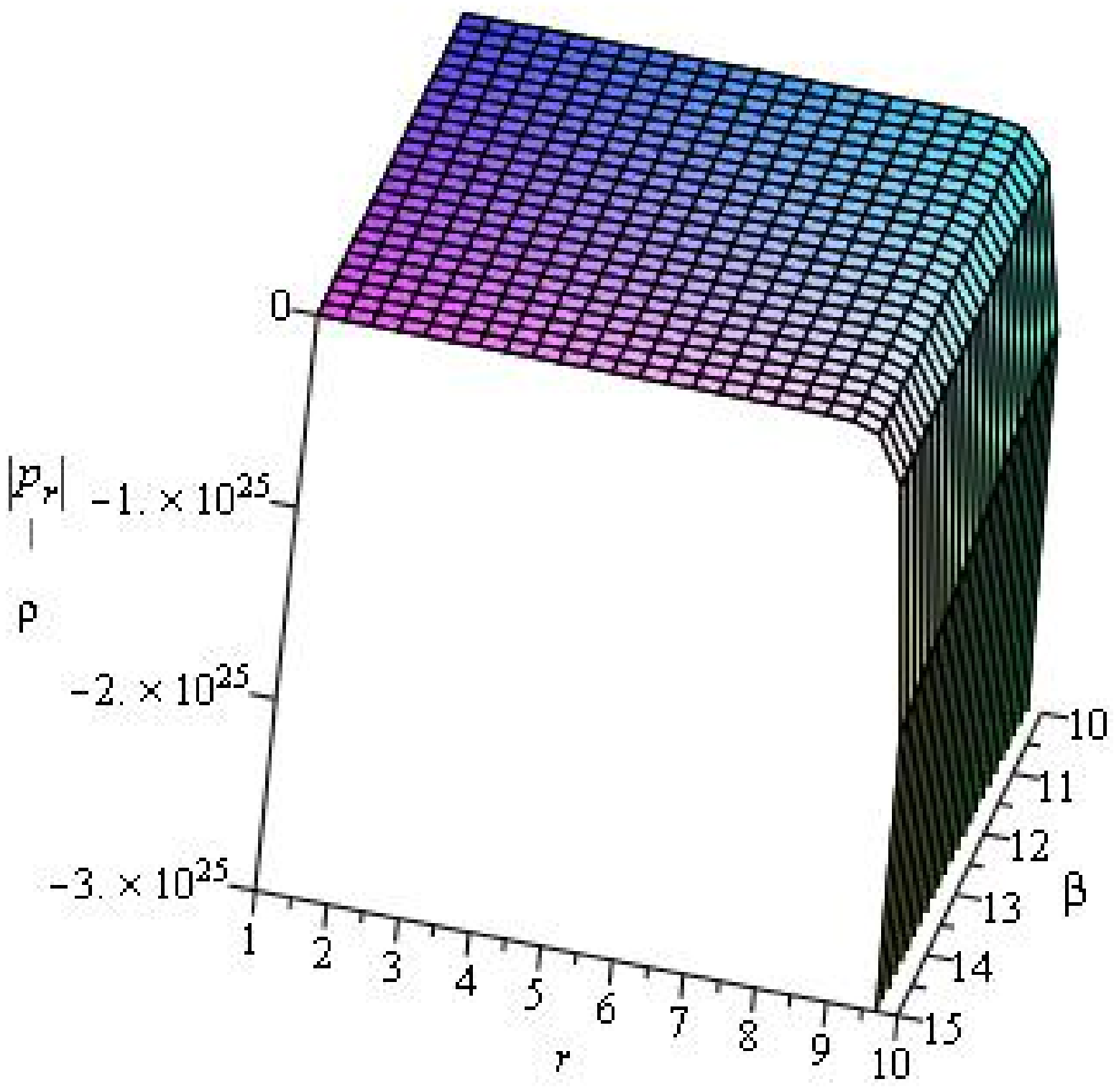}
 	(b)\includegraphics[width=3.5cm, height=4cm, angle=0]{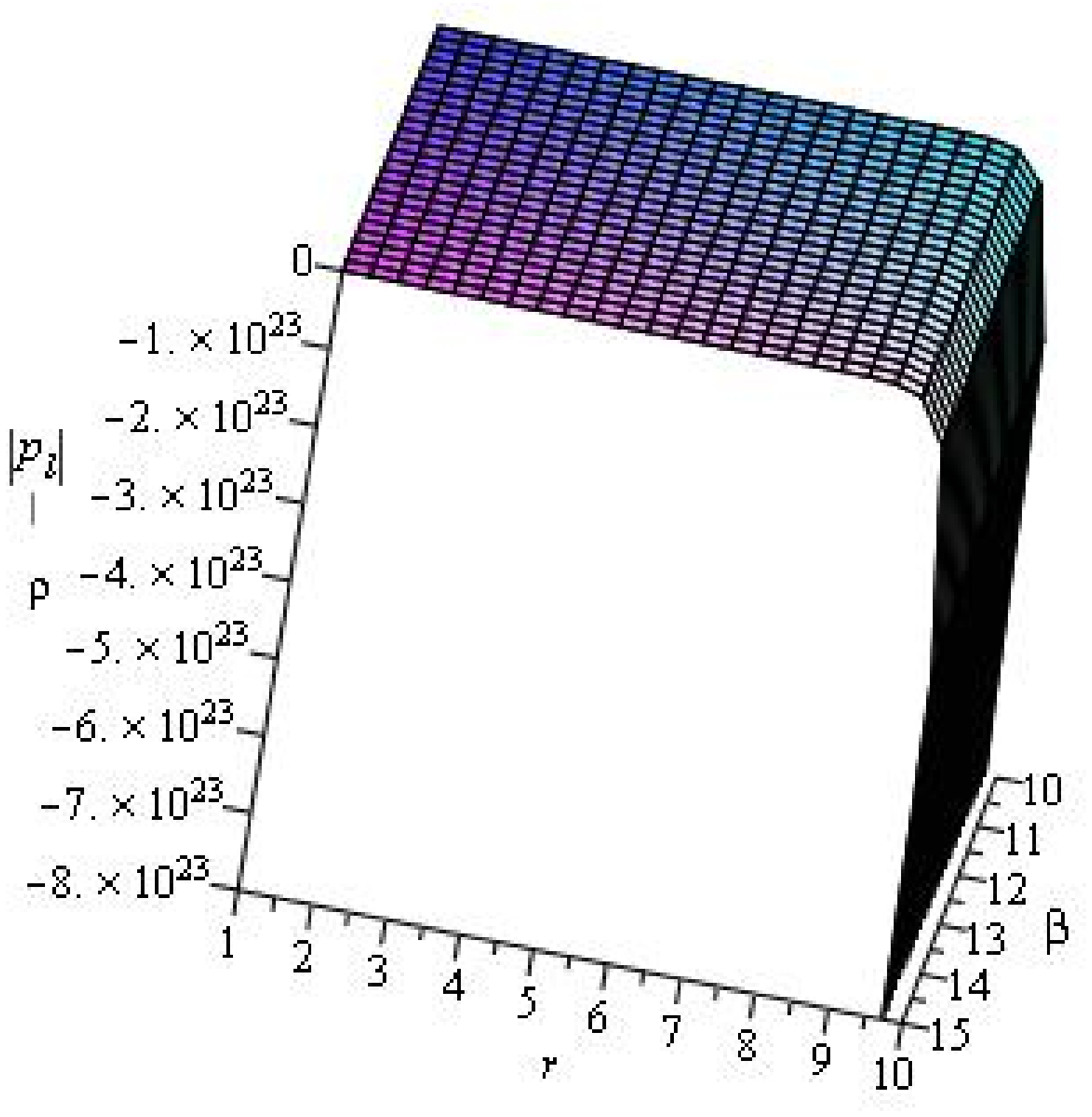}
 	(c)\includegraphics[width=3.5cm, height=4cm, angle=0]{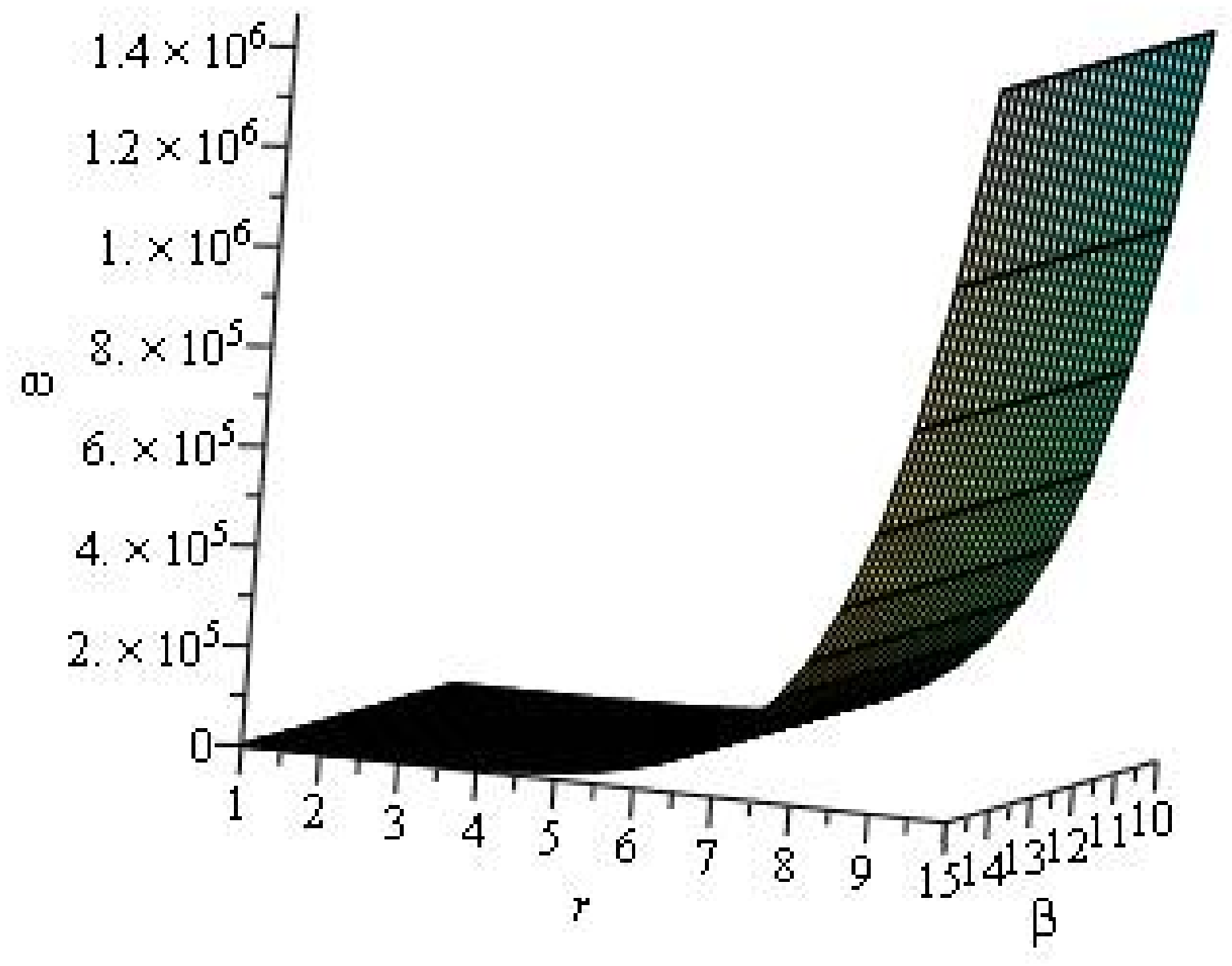}
 	(d)\includegraphics[width=3.5cm, height=4cm, angle=0]{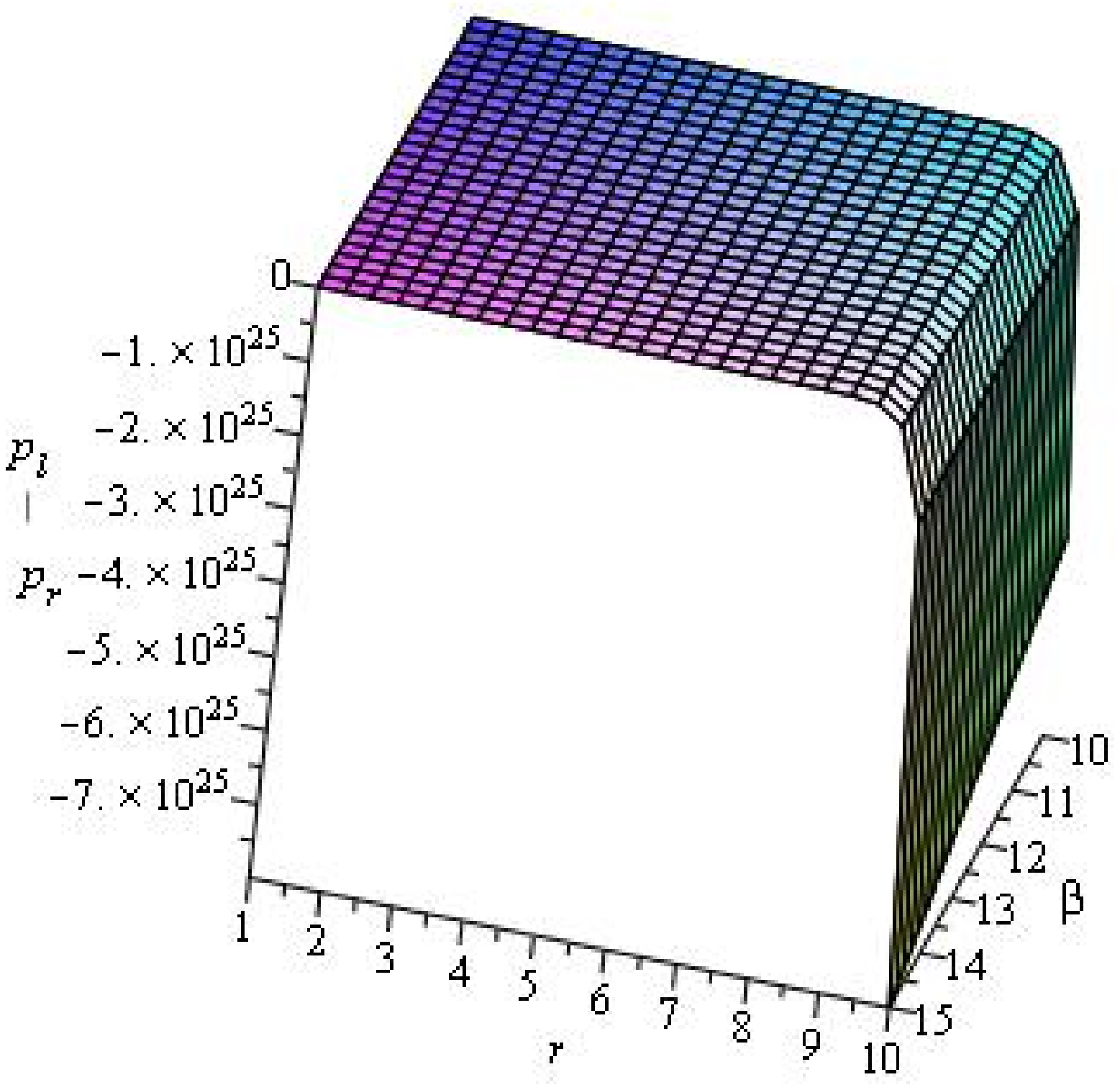}
 	\caption {(a) DEC, $\rho - p_r$ for $\alpha=5, a=0.5, m=2, n=4$, (b) DEC, $\rho - p_l$ for$\alpha=5, a=0.5, m=2, n=4$, (c) EOS,for $\alpha=5, a=0.5, m=2, n=4$ (d)$\bigtriangleup=p_l-p_r$ for $\alpha=5, a=0.5, m=2, n=4$.} 
 \end{figure}
  \begin{figure}
 	\centering 
 	(a)\includegraphics[width=3.5cm, height=4cm, angle=0]{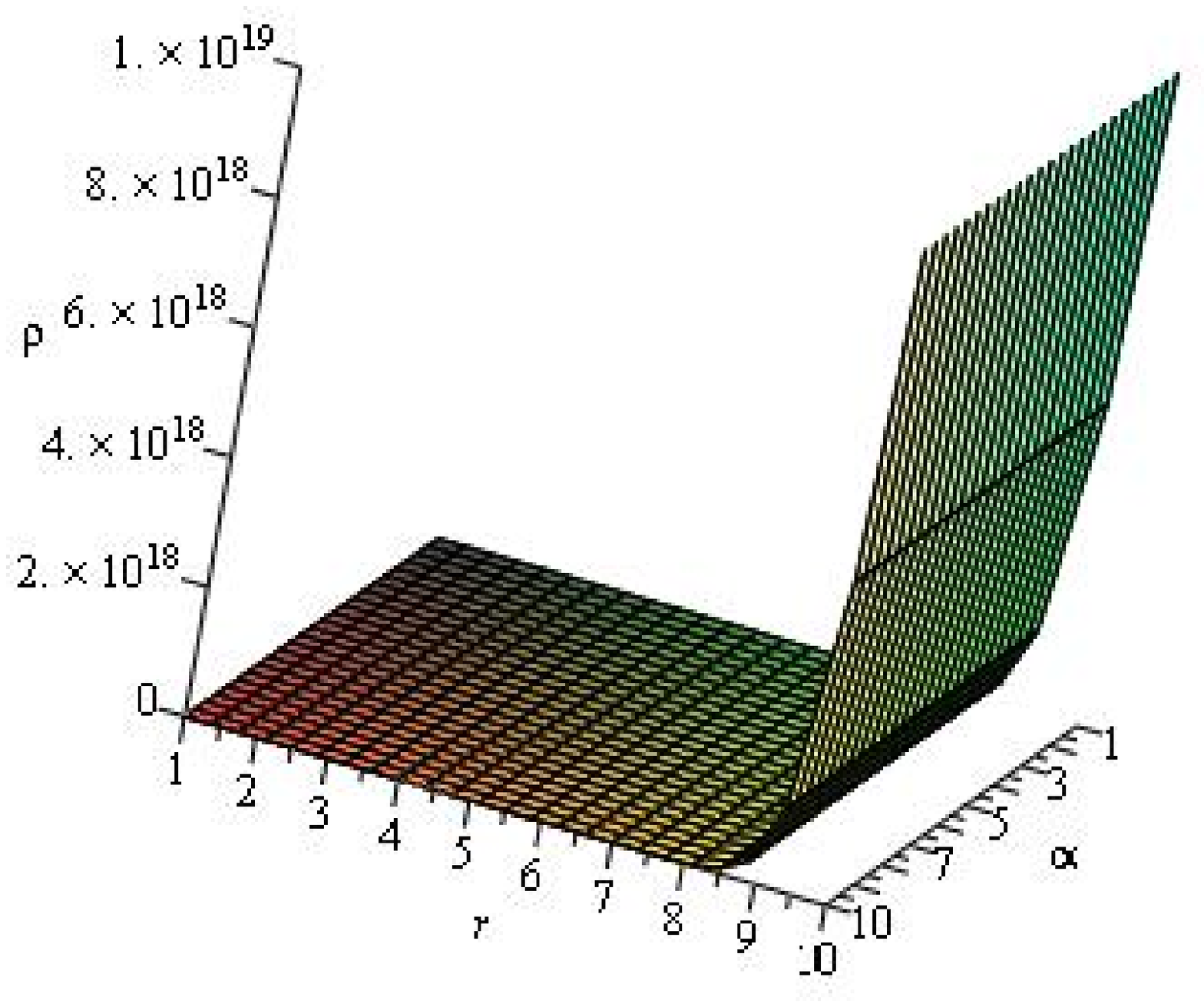}
 	(b)\includegraphics[width=3.5cm, height=4cm, angle=0]{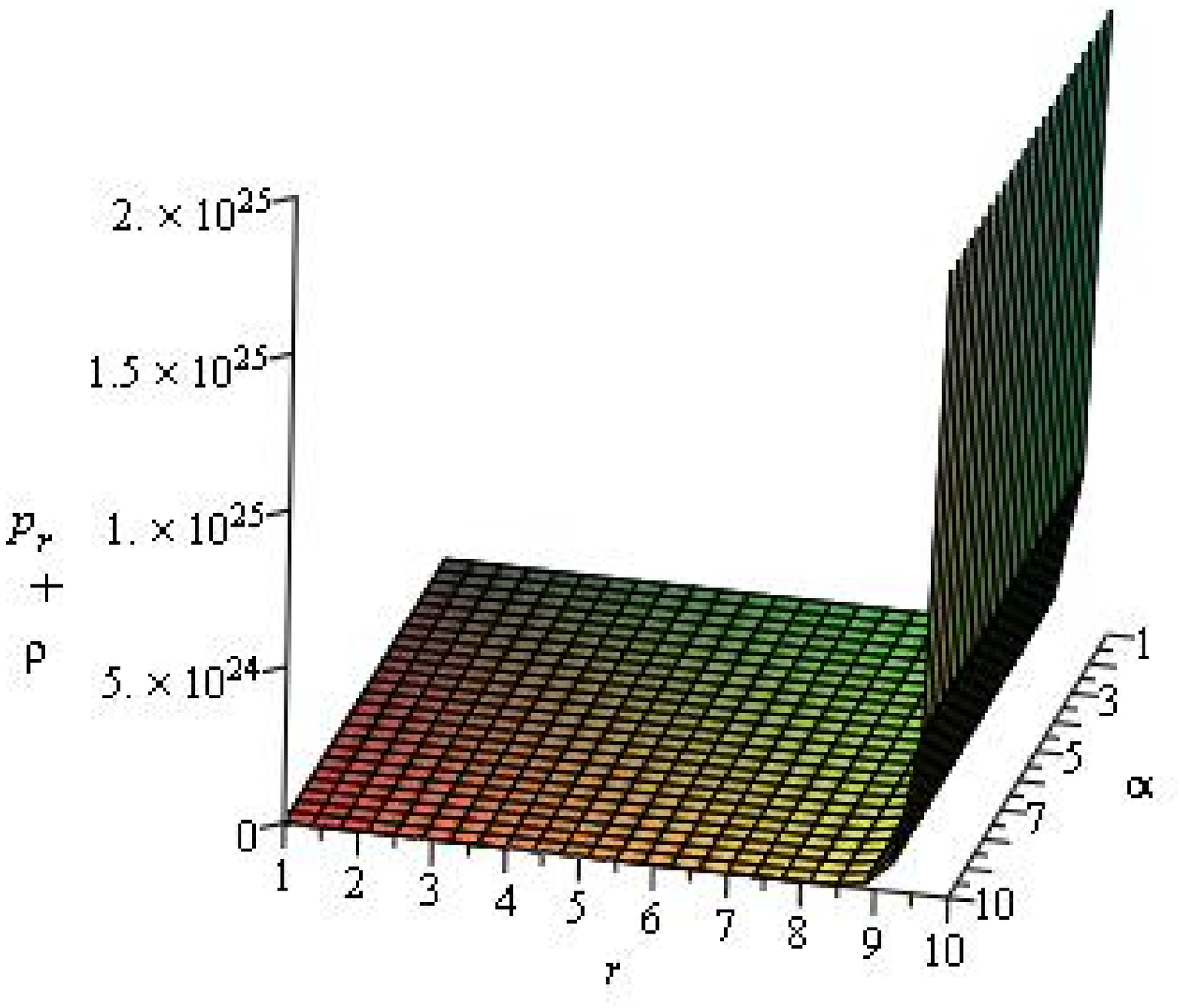}
 	(c)\includegraphics[width=3.5cm, height=4cm, angle=0]{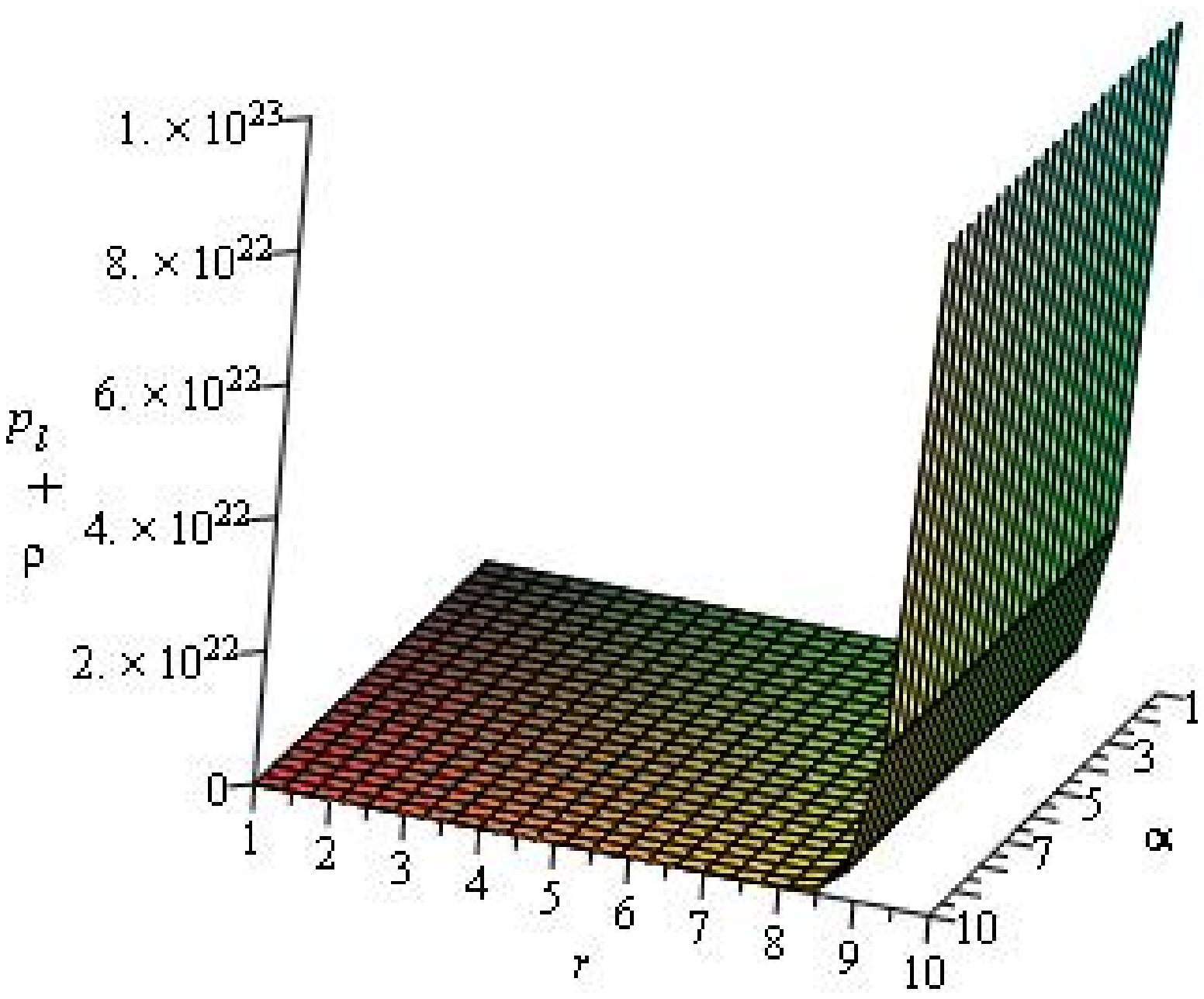}
 	(d)\includegraphics[width=3.5cm, height=4cm, angle=0]{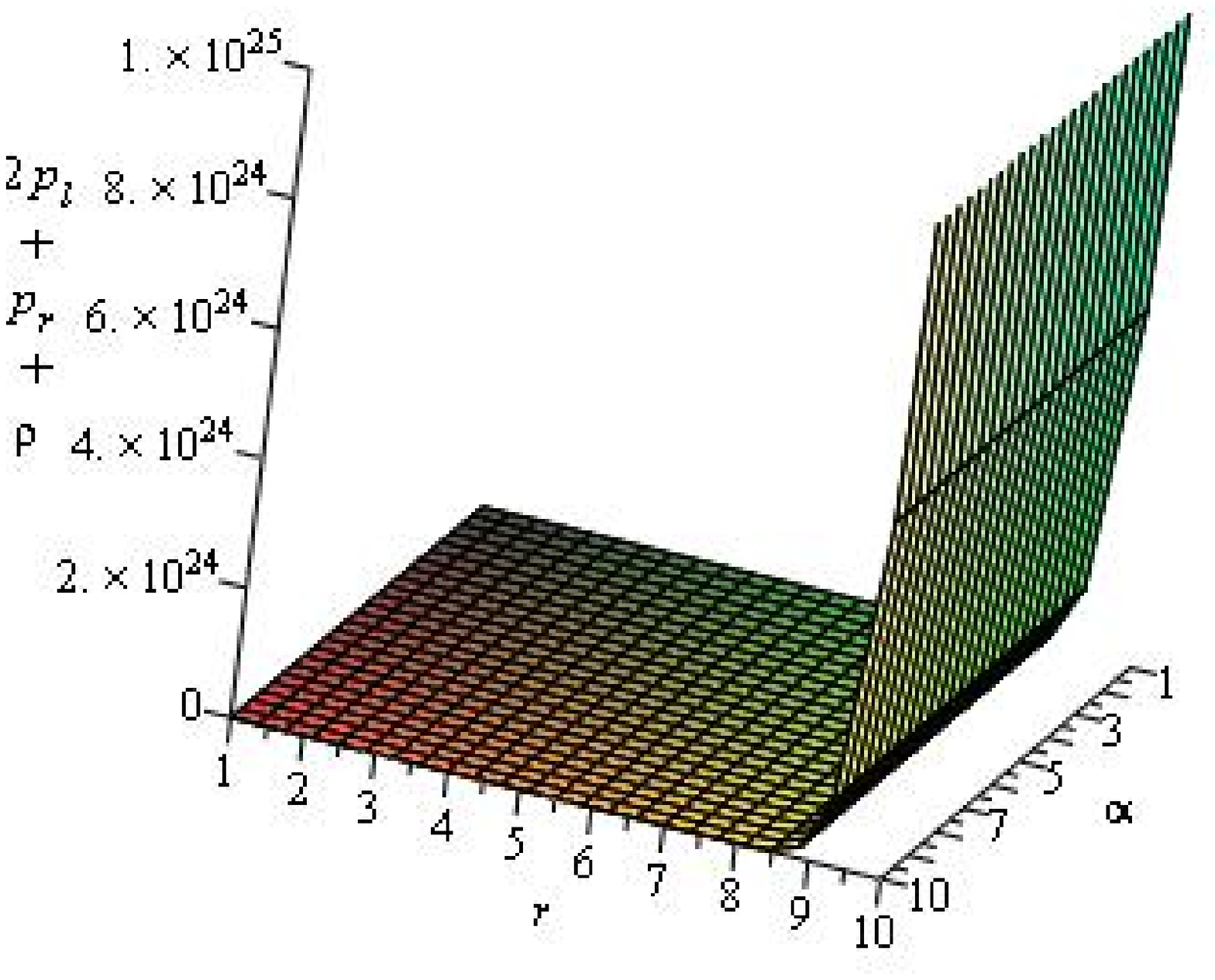}
 	\caption {(a) WEC, $\rho$ for $\beta=5, a=0.5, m=2, n=4$ (b) NEC, $\rho + p_r$ for $\beta=5, a=0.5, m=2, n=4$, (c) NEC, $\rho + p_l$ for$\beta=5, a=0.5, m=2, n=4$,(d) SEC, $\rho + p_r + 2p_l$ for$\beta=5, a=0.5, m=2, n=4$.}
 \end{figure}       
   \begin{figure}
   	\centering 
   	(a)\includegraphics[width=3.5cm, height=4cm, angle=0]{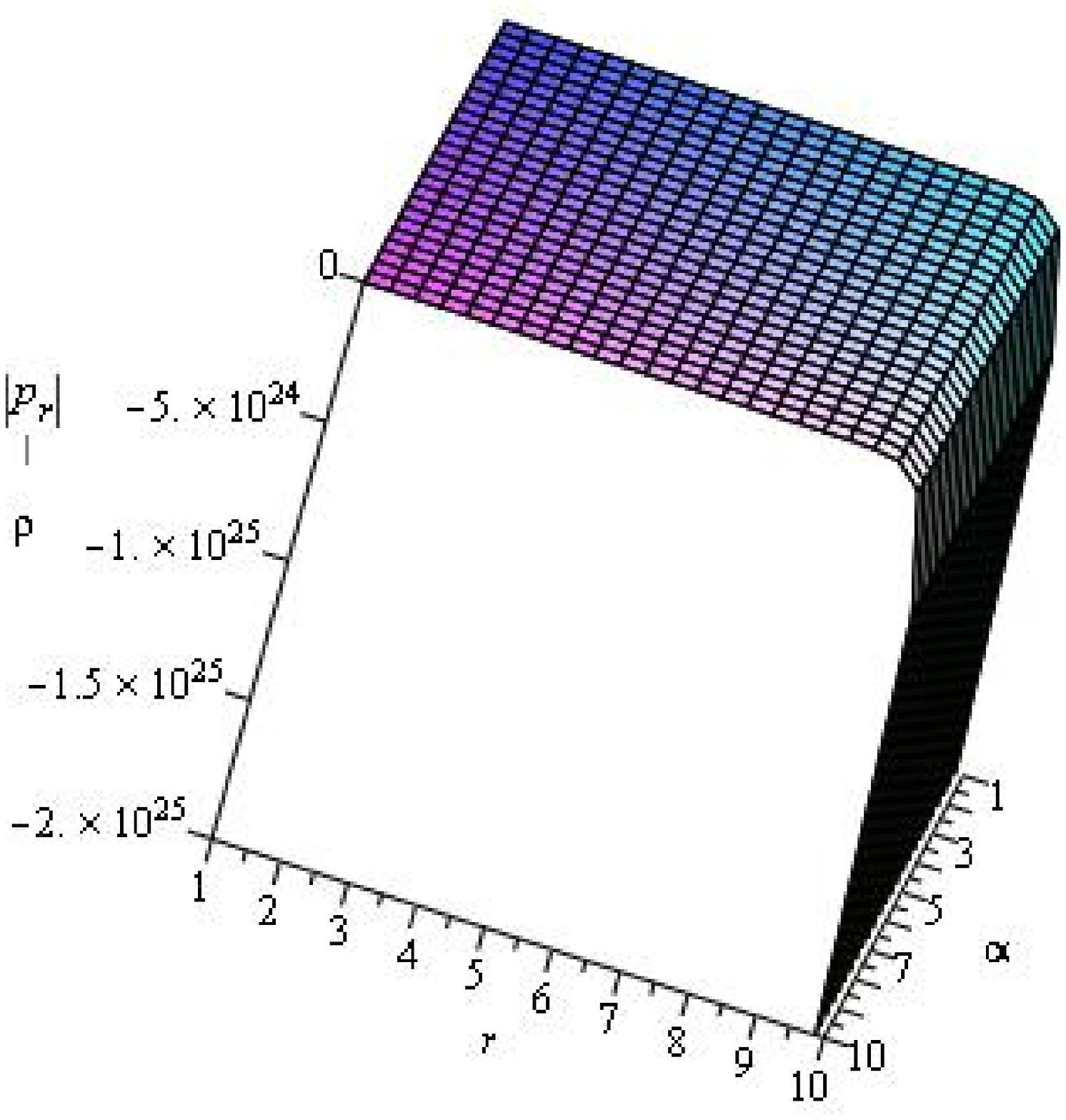}
   	(b)\includegraphics[width=3.5cm, height=4cm, angle=0]{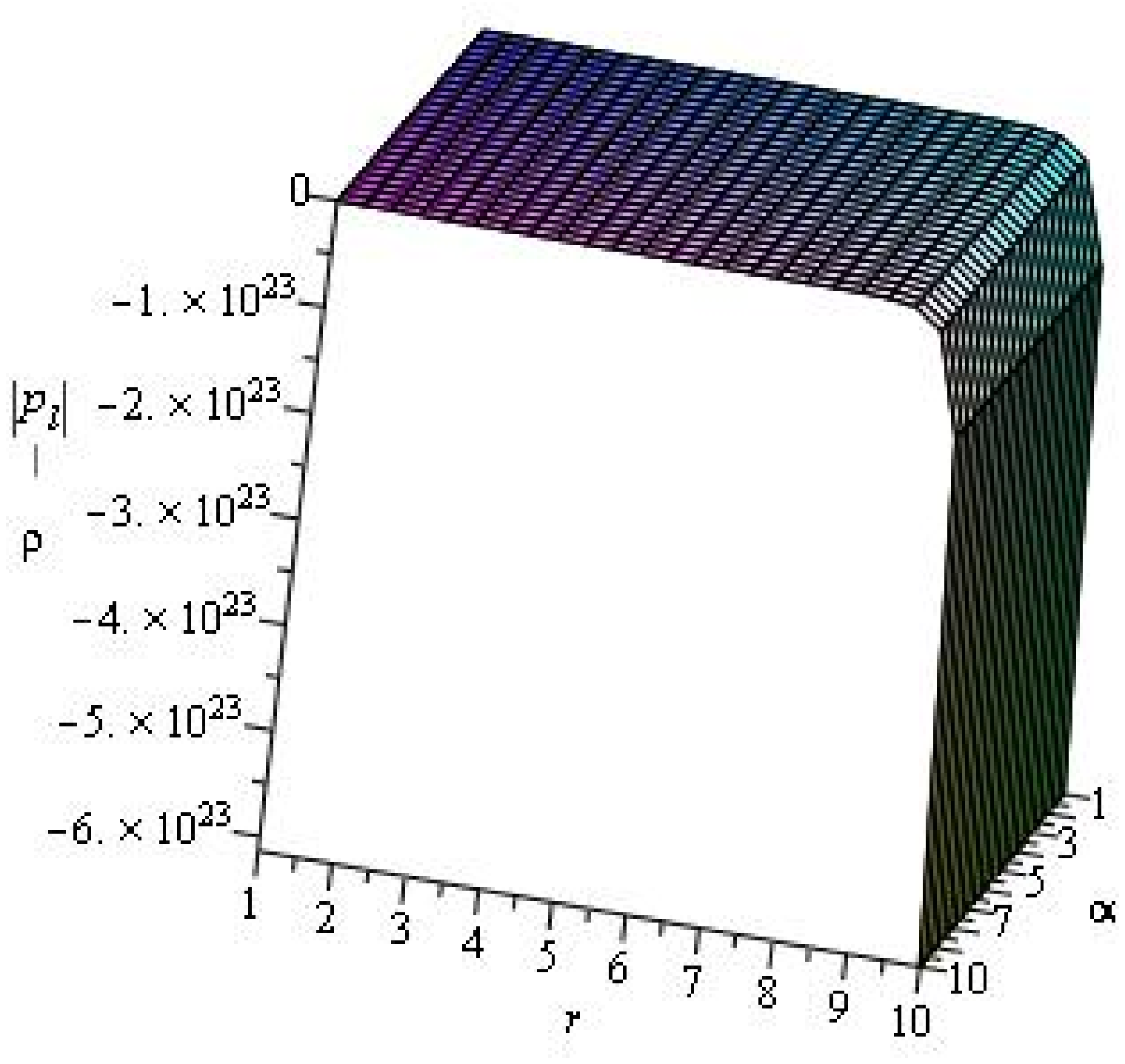}
   	(c)\includegraphics[width=3.5cm, height=4cm, angle=0]{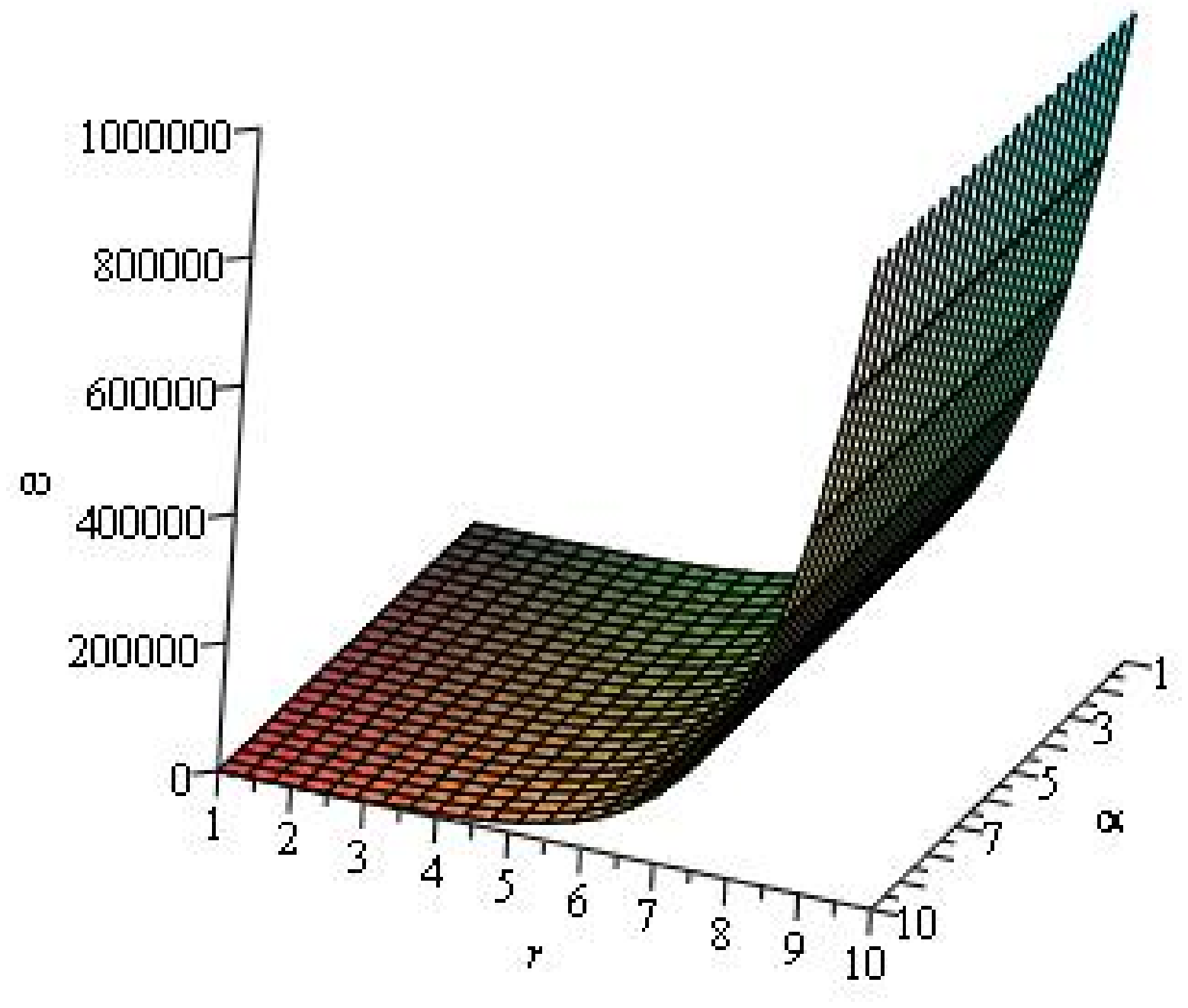}
   	(d)\includegraphics[width=3.5cm, height=4cm, angle=0]{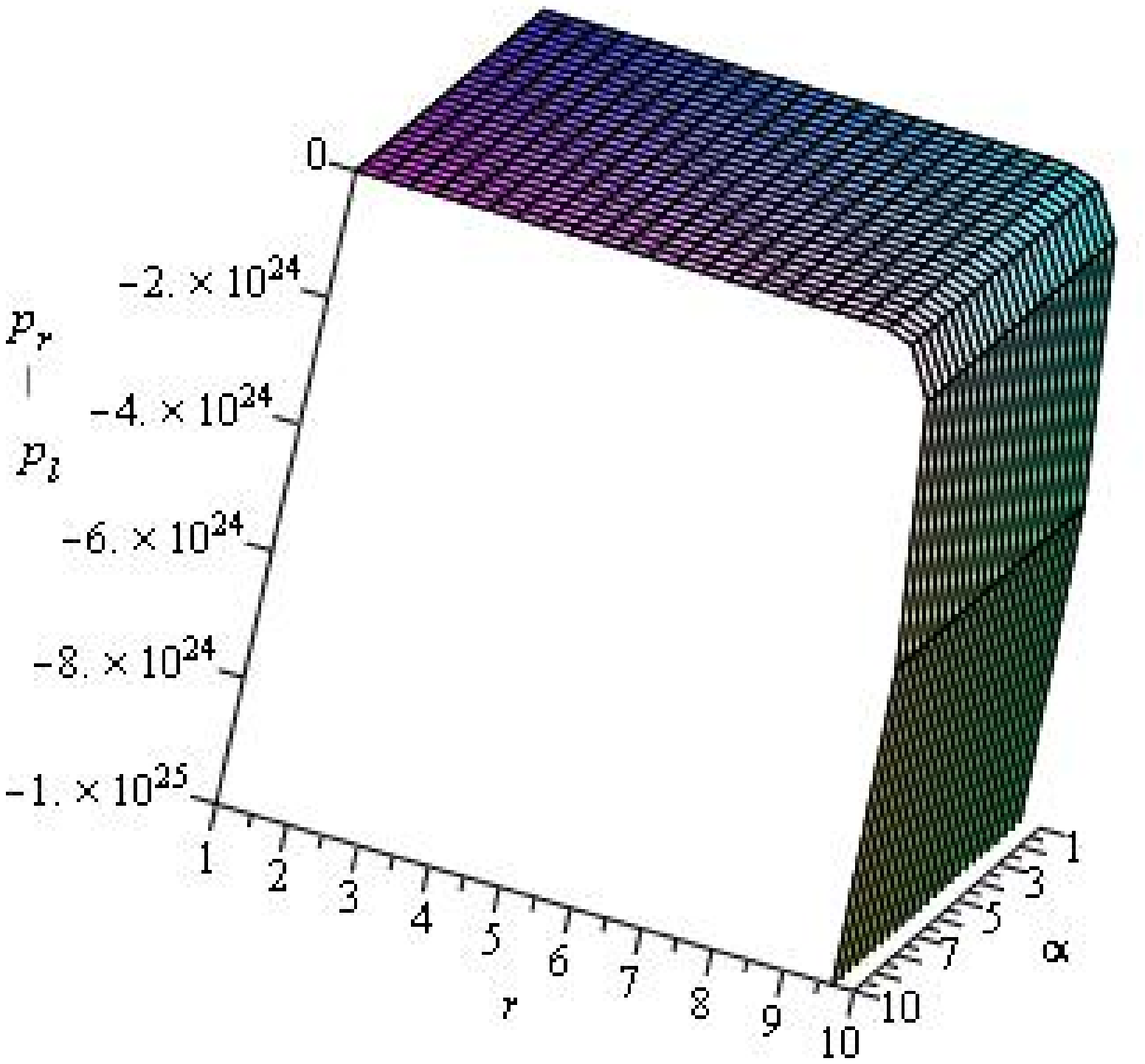}
   	\caption {(a) DEC, $\rho - p_r$ for $\beta=5, a=0.5, m=2, n=4$, (b) DEC, $\rho - p_l$ for$\beta=5, a=0.5, m=2, n=4$, (c) EOS,for $\beta=5, a=0.5, m=2, n=4$ (d)$\bigtriangleup=p_l-p_r$ for $\beta=5, a=0.5, m=2, n=4$.} 
   \end{figure}     
 \section{Results and Discussion}       
        
The wormhole geometries are very much affected on shape function $b(r)$, so in literature various shape functions are defined and wormhole solutions are investigated. Several researchers explored violation or non-violation of energy conditions in the context of wormhole solutions. In the present investigation, the shape function $b(r)= {r_{{0}}}\left(\frac {{a}^{r}}{{a}^{r_{0}}}\right)$ is taken into account and explored the wormhole solutions in the framework of $f(R)$ gravity with $f(R)= R + \alpha R^{m}-\beta R^{-n}$, where $\alpha, \beta, m $ and $n$ are positive real numbers. Here different cases are analyzed for various values of parameters in function $f(R)$ and NEC, WEC, SEC, DEC, EOS (equation of state) and anisotropy parameter are examined.

\subsection{Case 1: General Relativity Case}            

For general relativity case the parameters $\alpha$ and $\beta$ is considered to be zero i.e. $f(R)= R$. The energy density $\rho$ is found to be a negative function of radial coordinate, which indicate the presence of exotic matter in wormhole geometry. In Fig. 2(a) and 2(b), NEC is plotted in terms of radial coordinate and tangential coordinate.  $\rho + p_{r}$ is found to be negative and   $\rho + p_{l}$ is found to be positive. In Fig.2(c), SEC is plotted. It is observed that SEC is also violating in this case. DEC in terms of $\rho-\left| p_{r}\right|$ and $\rho-\left| p_{l}\right|$ is also plotted in Fig.3(a) and 3(b). Both are obtained to be negative. Therefore, for present model NEC, SEC, and DEC are violated. Hence, in general relativity this model strongly support that the wormhole is filled with exotic matter. In Fig.3(c), EOS parameter $\omega$ is plotted, which comes out to be positive. This shows the role of matter in the universe. From Fig.3(d), it is found that anisotropy parameter $\bigtriangleup$ is positive, that supports  the repulsive geometry of universe.

\subsection{Case 2: $f(R)= R + \alpha R^{m}$}             
        
For this case energy density is plotted in Fig.4(a), which comes out to be positive function of $r$. We also found NEC in term of $\rho +p_{r}$ is violating at the throat. The first NEC is plotted in Fig.4(b). The second NEC in term of $\rho + p_{l}$ is plotted in Fig.4(c). It is observed that second NEC is violating for $\alpha > 125$. In Fig.4(d), it is analyzed that SEC is also violating for $ r> r_{0}=1$. The DEC in terms of $\rho - |p_{r}|$ and $\rho - |p_{l}|$ is plotted in Figs.5(a) and 5(b), which are observed to be negative. Therefore, in this model violation of energy conditions confirmed  the presence of exotic matter in wormhole geometry. EOS parameter $\omega$ is plotted in Fig.5(c). It is analyzed that phantom fluid is dominant in wormhole. The anisotropy parameter $\bigtriangleup$ is shown in Fig.5(d), which is found to be positive at throat $r_{0}= 1$. Also anisotropy parameter is observed positive for $r > 2$, that indicate the repulsive nature of the geometry of Universe. 

\subsection{Case 3: $f(R)= R - \beta R^{-n}$}           
 
In this model $\beta$ and $n$ are positive real numbers. In Fig.6(a), energy density is plotted for $a = 0.5$, $n=4$ and $\beta > 20$, which is found to be positive. For the same values of parameters the first and second NEC are plotted in Figs.6(b)and 6(c), which are observed to be positive. The SEC is also analyzed as positive function of $r$, which can be seen in Fig.6(d). However, first and second DEC are violating as both are observed to be negative in Figs.7(a)and 7(b). In Fig.7(c), EOS parameter is plotted, which is found to be positive and anisotropy parameter is found to be negative in Fig.7(d). Therefore, in the present model validation of NEC and SEC indicates presence of non-exotic matter and $w>0$ shows role of non-phantom fluid, $\bigtriangleup <0$ indicated attractive geometry. Thus in this model we conclude that wormhole solution with attractive geometry could exists without violating NEC and SEC with presence of non-phantom fluid.            
  
\subsection{Case 4: $f(R)= R + \alpha R^{m} - \beta R^{-n}$}  
   
In this model $\alpha \neq 0, \beta \neq 0$, $m$ and $n$ are positive real numbers. In the first sub-case energy density is plotted in Fig.8(a) for $a=0.5, \alpha = 5, m = 2, n=4$ and $\beta > 10$. Energy density is found to be positive. For the same values of parameters first and second NEC are analyzed, which are observed positive in Figs.8(b) and 8(c). The SEC is also found to be positive in Fig.8(d). However, both DEC are observed to be negative in Figs.9(a) and 9(b). The EOS parameter is found to be positive function of radial coordinate $r$ in Fig.9(c). However, $\bigtriangleup$ is observed to be negative.
 
In the second sub-case the parameters are taken as $a=0.5, \beta =5, m=2, n=4$ and $\alpha >1$. The energy density is plotted in Fig.10(a), which is found to be positive function of radial coordinate. The first and second NEC are observed to be positive in Figs.10(b) and 10(c). The SEC is also found to be positive in Fig.10(d). However, first and second DECs are analyzed to be negative in Figs. 11(a)and 11(b). The EOS parameter is observed as positive and anisotropy parameter is negative in Figs.11(c) and 11(d). 
Therefore, in both the sub-cases this model indicated the existence of wormhole without violating NECs and SEC. However, DECs are violating. The anisotropy parameter and EOS parameter indicates attractive geometry with non-phantom fluid.

 Thus, we observed that results obtained in viable modified theories of gravity in cases (2-4) are completely different from general relativity case.


\section{Conclusion}
Modified gravitational theories have received considerable attention over the last few decades as a possible alternative to GR. Obviously, this incorporates two stages of universe expansion: rapid expansion (early inflation) and present cosmic acceleration. The present paper is focused on wormhole models using the framework of $f(R)$ gravity by considering a new shape function $b(r)= {r_{{0}}}\left(\frac {{a}^{r}}{{a}^{r_{0}}}\right), \,\,0<a<1.$ The lower bound of radial coordinate $r=r_{0}=1$ defines the throat radius, also there are no black holes/horizons i.e wormhole regions are valid everywhere for $r>r_{0}$. The shape function satisfies all desired condition for wormhole structure like $b'(r_{0})<1$ etc. Here, the function $f(R)$ is taken in well known form as $f(R)= R + \alpha R^{m}-\beta R^{-n}$ (see in \cite{Nojiri2003}) and exact solutions are obtained. On varying the parameters ($\alpha, \beta, m $ and $n$), energy conditions (NEC, SEC, DEC), equation of state parameter and anisotropic parameter are analyzed in four different cases. It is observed that in the case of GR, wormhole is filled with exotic matter, geometry is repulsive, while in other cases of $f(R) = R + R^{m}$, wormhole is found filled with exotic matter with phantom fluid whose geometry is repulsive. In other two case of $f(R)$, it is analyzed that wormhole solutions may exists without violation of NEC and SEC, moreover role of non-phantom fluid is observed in these cases with attractive geometry. In this case wormhole satisfies energy conditions (NEC and SEC) due to matter threading and gravitational fluid is also interpreted by its higher order curvature derivative terms. Therefore, framework of modified gravity found  significant while discussing the existence of wormhole solutions. Thus, $f(R)$ gravity can be a useful framework in wormhole physics in addition to its applications in cosmology.


\section*{Acknowledgments} 
The authors are thankful to GLA University, Mathura, India for providing help and support to do this research work. The authors are also appreciative to A. Pradhan for his fruitful suggestions which improved the paper in present form.

\end{document}